\documentclass[a4paper]{book}
\usepackage{graphicx}
\usepackage{caption}
\usepackage{subcaption}
\usepackage{amsmath,amsthm}
\usepackage{amssymb,amsfonts}
\usepackage{color}
\usepackage[squaren]{SIunits}
\usepackage{epigraph}

\newtheorem{definition}{Definition}

\title{\huge{\bf Quantum coherence\\ and correlations\\ in cold atom systems}}

\author{\Large{Piotr Sza\'{n}kowski}}

\begin{document}


\begin{titlepage}
    \begin{center}
    	{\Large University of Warsaw}\\
    	{\Large Faculty of Physics}\\[.5cm]
    	\includegraphics[scale=.4]{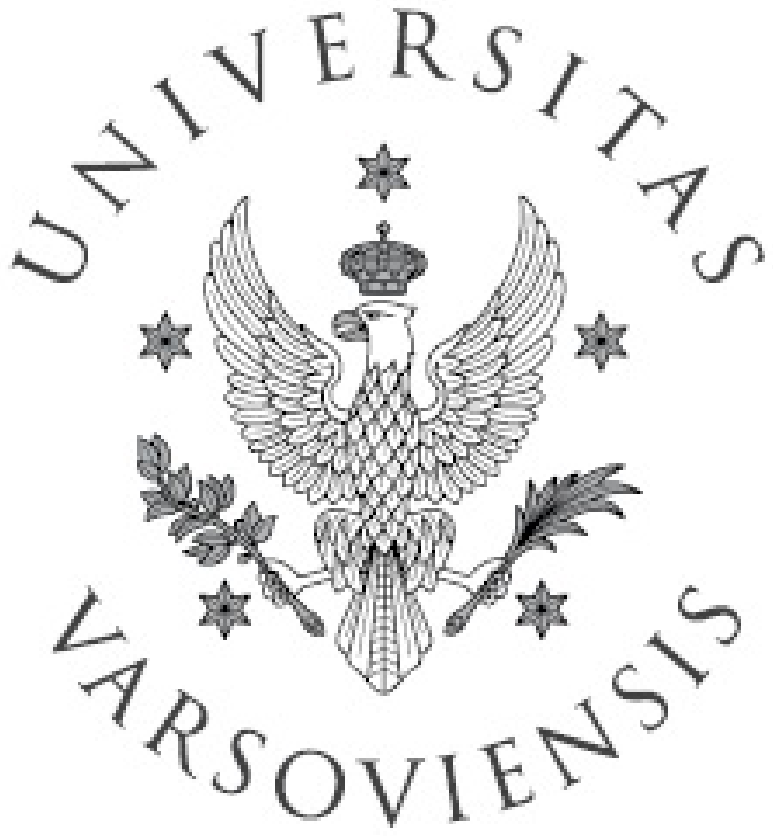}
   \end{center}
   \vspace*{2cm}
   \begin{center}
   	{\Large Piotr Sza\'{n}kowski}\\[1cm]
   	{\LARGE \textbf{Quantum coherence and correlations\\ in cold atom systems}}\\[.5cm]
	{\large \textit{PhD dissertation}}
    \end{center}
    \vspace*{2.5cm}
    \begin{flushright}
    {\Large Supervisor}\\[.5cm]
    {\Large prof. dr hab. Marek Trippenbach}\\[.1cm]
    {\large Faculty of Physics\\ University of Warsaw}
    \end{flushright}
    \vspace*{2cm}
    \begin{center}
    {\large October, 2014}
    \end{center}
\end{titlepage}%

{\thispagestyle{empty}
\newpage
\vspace*{18cm}
\begin{flushright}
{\it Mojemu bratu}
\end{flushright}}

\tableofcontents

\chapter{Introduction}

\epigraph{
I believe in intuition and inspiration.\\
Imagination is more important than knowledge. For knowledge is limited, whereas imagination embraces the entire world, stimulating progress, giving birth to evolution. It is, strictly speaking, a real factor in scientific research.}
{{\it Albert Einstein}}

\epigraph{
The physicist needs a facility in looking at problems from several points of view.}
{{\it Richard P. Feynman}}

\section{Aim of this thesis}

I believe that quantum correlations (also known as {\it quantum entanglement}) acquired a special, almost mystical status in the collective conscious of physics community. My impression is that a lot of physicist tend to think of entanglement as some elusive and incomprehensible property of quantum states. It exists, they believe, in addition to other well understood physical properties of quantum systems, such as coherence or symmetry, but it is always placed on a different level. Such a viewpoint is understandable for various reasons. For example, it is evident form the study of the literature dating before around 1995, that the concept of entanglement have not yet penetrated the vast majority of fields of physics%
 (with the exception of topics revolving around famous EPR paradox and Bell inequalities, see Sec. \ref{sec:non-class_corrs}). Nevertheless, it is clear that the advancements in all those fields where, and still are, unhindered by the lack of explicit considerations of entanglement.
A reasonable explanation for this state of affair is that in fact, the entanglement is not really an issue of its own. Rather, it is a complex, non-uniform in its nature, construct made out of a collective of intertwined ``normal'' physical properties. In other words, by focusing on those standard physical properties one can explain all occurring phenomena and never notice that formally the entanglement was involved. 

Does this mean that entanglement is an empty concept and should be discarded? I believe the answer is negative. One of the greatest struggles with understanding quantum mechanics is its remoteness from the everyday experience of ``classical beings'' such as us. For me and many others, the question which aspects of the theory can be understood in terms of ``semi-classical'' models, and which are inherently ``quantum'' has always been the most interesting one. Originally, the concept of entanglement has been conceived to help making this distinction. Presently, it evolved beyond the scope of this purpose, especially in the field of quantum information, but still I believe it is the best starting point we have for this kind of investigation.

The aim of this thesis is to demystify the entanglement: to find out what it {\it is} in terms of physical properties of the system. I try to steer away from formal mathematical approach which could easily become detached from the physical intuition. Instead, I focus the discussion on building this intuition. The final result is the classical model constructed to represent quantum states, similarly to the celebrated Bohr's model of hydrogen atom. The kinematic properties of the model are given by correlations present in the state. Thus, this approach might provide a new vantage point to examine the entanglement from the perspective of classical concepts we are all accustomed to.
%

This thesis is not meant to simply summarize scientific results that I obtained during my studies; these can already be found in my publications. My understanding of the subject of non-classical correlations matured alongside the various other projects I was involved in. The preparation of the thesis was a great opportunity to collect my thoughts on the subject and organize them into a cohesive whole, which I could now share with the Reader.


\section{Non-classical correlations}\label{sec:non-class_corrs}

Although the foundations of quantum and classical physics are much different, it is often difficult to pinpoint which features of a particular system are intrinsically ``quantum''.
A good example of a ``quantum'' behavior is a wave-particle duality, which is a consequence of the ability of particles to exist in superpositions of quantum states.
On one hand, the wave nature of massive particles is manifested in a Young double-slit experiment \cite{Young_double_slit_88,Young_double_slit_91}, which shows their ability to interfere, just like waves on a surface of a pond.
On the other hand, the photoelectric effect \cite{Einstein_05} is a proof of a particle nature of the electromagnetic field, which is a wave, but also it consists of individual particles -- photons. Even though a system which is a wave-particle hybrid is without precedence in classical world, still the wave or particle side of quantum phenomena on their own are perfectly conceivable in terms of classical physics.

The most clear-cut distinction between ``classical'' and ``quantum'' can be made for systems composed of many particles when the properties of the ensemble are determined by the correlations between the constituents. Among the most important and evident types of quantum correlations are those between identical particles. Classical physics allows for tracking every particle with perfect accuracy without altering the dynamics of the system. Accordingly, even though the particles can be identical, i.e. characterized by the same set of features like mass or charge, they are always distinguishable by the virtue of their trajectories. The situation is dramatically different for identical quantum particles. In quantum world the concept of trajectory simply does not exist due to Heisenberg's uncertainty principle. Even if we were able to determine exactly the initial position of each particle it would be indeterminate in the following instant. Since, the particles cannot be tracked, they cannot be distinguished. Consequently, the physical properties of the system cannot change if any two identical particles are interchanged. Hence, the quantum state of identical particles must be described by a wave function which is symmetric or antisymmetric with respect to permutation of particles. It is well known that this requirement has a profound consequences both for atomic-scale phenomena as well as for our everyday life as we know it. For example, Pauli's exclusion principle, which is a direct consequence of identical fermions, like electrons and quarks, being described by antisymmetric wave functions, is an undergrid of the structure of the periodic table and overall large-scale stability of matter. Another example are bosons, which include photons, helium-4 and Cooper pairs. They are described by symmetric wave functions, and as a consequence, tend to ``bunch'' together in the same quantum state. This bunching leads to phenomena like superfuidity, superconductivity and Bose-Einstein condensation of ultra-cold atoms. Overall, the concept of indistinguishability is unique to quantum mechanics and it has no counterpart in classical physics.


Indistinguishability is not the only type of non-classical correlations. The famous gedankenexperiment proposed by Einstein, Podolski and Rosen (referred to collectively as EPR) revealed, then thought of as paradoxical, feature of quantum mechanics where a pair of particles in a particular quantum state would exhibit non-local properties \cite{EPR_35}. Shortly after, Schr\"{o}dinger made an attempt to extract the essence of non-classicality of EPR-type states, which led him to the concept of {\it entanglement} \cite{Schroedinger_35}, which is the keynote of this thesis. The formal definition of entangled state used commonly nowadays reads:
\begin{definition}\label{def:entanglement}
 The state $\hat\varrho$ of $N$ parties is entangled if it is not {\it separable}, that is, it cannot be written as
\begin{equation}
\hat\varrho = \sum_{i} p_i \,\hat\varrho_i^{(1)}\otimes\hat\varrho^{(2)}_i\otimes\ldots\otimes\hat\varrho^{(N)}_i\label{sep_state}
\end{equation}
where $\hat\varrho^{(n)}_i$ is a density matrix of party $n$ and $p_i\geqslant 1$ for all $i$ with $\sum_i p_i =1$.
\end{definition}
Here ``parties'' may refer to subsystems composed of particles, sets of degrees of freedom (e.g. spin and position of an electron) or even parts of configuration space of a single particle -- so called modes. The physical interpretation of entanglement follows from the definition of separable state. The state which is not entangled, i.e. can be written in the form of RHS of (\ref{sep_state}), can be prepared ``classically''. Each summand of RHS of (\ref{sep_state}) is a product state, which means that it can be initialized by an independent measuring devices operating on each party individually. Suppose that these devices are equipped with a switch with settings $i$, set up in such a way that the device assigned to party $n$ yields a state $\hat\varrho^{(n)}_i$ as a result of the measurement. The set of devices can now be supplemented with random number generator which would supply each of them with a choice of setting $i$ with probability $p_i$. Hence, the state prepared with the use of procedure described above is separable. The correlations between parties are determined only by the random number generator, which can be chosen to be a purely classical device. Thus, any state which could not be prepared by such a procedure has to contain correlations between parties that are not classical and is said to be entangled.

Entanglement, by the virtue of definition, successfully formalize the notion of non-classical correlations. However, this discriminative definition is also the source of the limitations of the concept. We do not know what entanglement {\it is}, we only know what entanglement {\it is not}. It would be very naive to think that entanglement is ``uniform'' and there is no room for different types of quantum correlations that could be classified by their properties. The other problematic issue spawned by the general scope of the definition is an ambiguity of the concept of ``party''. In most cases the physical situation explicitly defines the parties involved. Nevertheless, if the definition is invoked recklessly it may lead to nonsensical conclusions, like for example, equating a superposition with entanglement, which in turn can be destroyed or created by simple rotation of the reference frame\footnote{
Consider two dimensional harmonic oscillator with Hamiltonian
\begin{equation*}
\hat H = \left( \frac{\hat p_x^2}{2m}+\frac{m\omega^2 \hat x^2}2\right)+ \left( \frac{\hat p_y^2}{2m}+\frac{m\omega^2 \hat y^2}2\right)
\end{equation*}
Written in a product basis of $x$ and $y$ directions, the state of single excitation in the direction $x$ is separable
\begin{equation*}
\left|\psi\right\rangle = \hat a_x^\dagger \left|0_x\right\rangle\left|0_y\right\rangle=\left| 1_x\right\rangle\left|0_y\right\rangle
\end{equation*}
Here $\hat a_i=\sqrt{\frac{m\omega}{2\hbar}}\left(\hat x_i +\frac{i}{m\omega}\hat p_i\right)$ is an annihilation operator in the direction $i=x,y$. However, the same state viewed in the reference frame rotated around $z$ axis about $\frac\pi4$ angle is entangled:
\begin{equation*}
\left|\psi\right\rangle = \left| 1_x\right\rangle\left|0_y\right\rangle = \left(\cos\frac\pi 4 \hat a_{x'}-\sin\frac\pi 4\hat a_{y'}\right)\left|0_{x'}\right\rangle\left|0_{y'}\right\rangle 
 =\frac 1{\sqrt 2}\left( \left| 1_{x'}\right\rangle\left| 0_{y'}\right\rangle - \left| 0_{x'}\right\rangle\left| 1_{y'}\right\rangle\right)\,.
\end{equation*}
}. I shall address these two critical points in the following sections.

\section{Entanglement of modes}

According to the definition of entanglement, the system can be partitioned into parties in an arbitrary way, as long as the basis has been chosen so that it has a tensor product structure. Consequently, any superposition of states can be rewritten in such a way that it can be formally considered as an entangled state. To illustrate this statement, consider a two dimensional quantum system. I choose an orthonormal basis $\{ \left|\uparrow\right\rangle, \left|\downarrow\right\rangle \}$ and make a formal assignment mimicking the second quantization formalism
\begin{equation}
\left|\uparrow\right\rangle \leftrightarrow \left| 1_{\uparrow} \right\rangle\left| 0_{\downarrow}\right\rangle \quad {\rm and }\quad \left|\downarrow\right\rangle \leftrightarrow \left|0_{\uparrow}\right\rangle\left| 1_{\downarrow}\right\rangle
\end{equation}
By performing this mathematical trick I managed to turn an arbitrary superposition of the basis states into entanglement of two parties
\begin{equation}
\left|\Psi\right\rangle = \alpha \left|\uparrow\right\rangle + \beta\left|\downarrow\right\rangle \leftrightarrow \alpha \left| 1_{\uparrow} \right\rangle\left| 0_{\downarrow}\right\rangle + \beta\left|0_{\uparrow}\right\rangle\left| 1_{\downarrow}\right\rangle.
\end{equation}
In this case, the entangled parties are ``modes'' defined as an orthonormal states that can be occupied by the particle composing the system. However, such an entanglement can be erased or created at will by a simple change of basis. For example, if I where to change the basis of our system to $\{\left|\Psi\right\rangle, \left|\Psi^\perp\right\rangle\}$ ($\left|\Psi^\perp\right\rangle$ is a state orthogonal to $\left|\Psi\right\rangle$) and make a corresponding mode assignment
\begin{equation}
\left|\Psi\right\rangle \leftrightarrow \left| 1_{\Psi} \right\rangle\left| 0_{\Psi^\perp}\right\rangle \quad {\rm and }\quad \left|\Psi^\perp\right\rangle \leftrightarrow \left|0_{\Psi}\right\rangle\left| 1_{\Psi^\perp}\right\rangle,
\end{equation}
the state $\left|\Psi\right\rangle$ would be a separable state of modes $\Psi$ and $\Psi^\perp$. These observations suggest that the concept of entanglement is merely a mathematical curiosity devoid of any physical significance. Indeed, this is the case if the entanglement of modes is considered in ``vacuum'', without a context. Usually, the physical backdrop of a experimental setting favors particular choice of a modes describing the system, thus eliminating ambiguity in the definition of parties. A perfect example is provided by, the EPR experiment \cite{EPR_35}, which I will now discuss in more detail focusing on the issue of the choice of parties.

Consider a system composed of two qubits, for example, photons which can be vertically and horizontally polarized. The system is initialized in a so called EPR state 
 and then one of the photons is sent to detector located at far left and the other one to the detector located at far right. Thus, the synchronous measurements of polarizations of photons at the left and right site are performed on a state ket given by
\begin{equation}\label{ch1.2:og_EPR_state}
\left|{\rm EPR}\right\rangle =\frac 1{\sqrt 2}\Big( \left| \updownarrow\right\rangle_L\left|\leftrightarrow\right\rangle_R+\left|\leftrightarrow\right\rangle_L\left|\updownarrow\right\rangle_R\Big).
\end{equation}
Here the subscript $L$ ($R$) indicate a state of a particle measured at the left (right) detector and $\left|\updownarrow\right\rangle$ ($\left|\leftrightarrow\right\rangle$) designates vertical (horizontal) polarization of a detected photon. The state (\ref{ch1.2:og_EPR_state}) is entangled without the shadow of doubt. However, it is not clear what exactly are the parties involved, although the answer might seem obvious at the first glance. It is tempting to say that the party described by ket $\left|\varphi\right\rangle_L$ is the photon which went left, and $\left|\psi\right\rangle_R$ is a state of the second photon, the one which went right. This interpretation has to be dismissed immediately, because photons are {\it indistinguishable} and labeling them as ``the one which went left/right'' is meaningless. Such labels can only be attached to modes associated with states of definite values of physical quantities measured by left and right detectors, which as a classical objects, are distinguishable. Hence, Eq.~(\ref{ch1.2:og_EPR_state}) is a proxy, or a shorthand of notation, for a more precise formula
\begin{align}
\left|{\rm EPR}\right\rangle {}&= \frac 1{\sqrt 2}\left( 
  \left| 1_{L,\updownarrow}\right\rangle\left| 0_{L,\leftrightarrow}\right\rangle\left| 0_{R,\updownarrow}\right\rangle\left| 1_{R,\leftrightarrow}\right\rangle+
  \left| 0_{L,\updownarrow}\right\rangle\left| 1_{L,\leftrightarrow}\right\rangle\left| 1_{R,\updownarrow}\right\rangle\left| 0_{R,\leftrightarrow}\right\rangle
\right)=\nonumber\\
&= \frac 1{\sqrt 2}\left( \hat a^\dagger_{L,\updownarrow}\hat a^\dagger_{R,\leftrightarrow} + \hat a^\dagger_{L,\leftrightarrow}\hat a^\dagger_{R,\updownarrow}\right)\left|{\rm VAC}\right\rangle
\end{align}
Here I used second quantization language: $\hat a_{L/R,\updownarrow/\leftrightarrow}$ is a bosonic annihilation operator removing particle from vertical/horizontal mode located at left/right detector and $\left|{\rm VAC}\right\rangle$ is vacuum state for which $\hat a_i \left|{\rm VAC}\right\rangle=0$ for all possible modes $i$. The left and right modes are explicitly chosen by the measuring devices and it has been predicted that the entanglement of these particular modes is a necessary ingredient for observation of non-local properties of the quantum mechanics \cite{Bell_64}. Thought experiment proposed by Einstein, Rosen and Podolsky and the measurements on EPR state has been realized in laboratory \cite{Aspect_81,Aspect_82,Aspect_82_2} and it has been demonstrated that the entanglement of modes can lead to {\it observable} phenomena unconceivable by classical physics. In fact, most of advances in the field of quantum information are based on utilizing the entanglement of modes chosen by a properly designed measurement schemes, including: teleportation, quantum cryptography, and quantum computing algorithms.

\section{Entanglement of particles}

The dependence on a choice of basis has proven to be the main difficulty with physical interpretation of mode entanglement. I argued that the ambiguity of assigning modes as parties can be lifted by a physical context and measurements associated with it. This dilemma cease to exist if particles are designated as parties instead of modes. The Hilbert space of many particle system is a tensor product of subspaces describing each particle. Separable state of parties-particles cannot be made entangled, and {\it vice versa}, by means of the change of basis. Indeed, the choice of basis states cannot compromise or alter the identity of particles, hence the only allowed basis transformations are local to particle subspaces. Consequently, the question whether particles are entangled or not is independent of the choice of basis -- a property very appealing form the physical point of view. 

The entanglement of distinguishable particles is no different then the entanglement of modes. As long as particles are not identical a separable states can be prepared classically, because in principle each party can be measured and initialized independently. Such a  classical procedure of state preparation is impossible to implement for indistinguishable particles, since the measuring devices that could address parties individually cannot exist. Thus, the notion of equating separable states to classical states have to be reconsidered when dealing with identical particles. For starters, it might seem that because of symmetrization/anti-symmetrization condition enforced on a state ket all states of identical particles must be non-separable. In fact, this is the case for fermions but is not for bosons, to which I now turn our discussion. 

I begin by stating that a separable pure state of $N$ identical bosons $\left|\phi;N\right\rangle$ must be a product of $N$ identical single-particle orbitals $\left|\phi\right\rangle$ \cite{CSI}, i.e.
\begin{equation}\label{coh_def}
  \left|\phi;N\right\rangle=\left|\phi\right\rangle^{\otimes N}.
\end{equation}
When the bosonic field operator $\hat\Psi(x)$ acts on the state (\ref{coh_def}), the result is
\begin{equation}\label{coh}
  \hat\Psi(x)\left|\phi;N\right\rangle=\sqrt N\phi(x)\left|\phi;N-1\right\rangle,
\end{equation}
which is a fixed-$N$ counterpart of the property of a coherent state of light $\left|\Phi\right\rangle$  defined by the relation ${\mathcal E}^{(+)}(x)\left|\Phi\right\rangle=\Phi(x)\left|\Phi\right\rangle$, where ${\mathcal E}^{(+)}(x)$ is
the positive-frequency part of the electromagnetic field ${\mathcal E}(x)$.
In Eq.~(\ref{coh}), $\phi(x)$ is a single-particle function determining the spatial properties of the system. Note that an analogical state of fermions is impossible and thus, particle-separable state of fermions cannot exist. In general, the separable state of $N$ identical bosons is a mixture of states (\ref{coh_def}), thus follows the definition of particle entanglement of bosons:
\begin{definition}\label{def:particle_entanglement}
 The state $\hat\varrho$ of $N$ identical bosons is particle entangled if it cannot be written as
\begin{equation}
\hat\varrho = \int\mathcal D\phi \mathcal P(\phi) \left|\phi; N\right\rangle\left\langle \phi;N\right|\label{sep_boson}
\end{equation}
Here $\mathcal D\phi$ denotes the integration over complex field $\phi$ and $\mathcal P(\phi)$ is a probability distribution, i.e. it is normalized and its integral with every $F(\phi)\geqslant 0$ over any volume $\mathcal V$ is non-negative,
\begin{equation}\label{cond_pos}
\int\limits_{\mathcal V}\mathcal D\phi \mathcal P(\phi)F(\phi) \geqslant 0
\end{equation}
\end{definition}
There is a direct analogy between the so-called P-representation of the state of light, where the density matrix is represented as $\hat\varrho_{\rm light} = \int\mathcal D\Phi \mathrm{P}(\Phi) \left|\Phi\right\rangle\left\langle\Phi\right|$. If
functional $\mathrm{P}$ is not a probability distribution (i.e does not satisfy condition similar to (\ref{cond_pos})), the electromagnetic field is considered to be non-classical \cite{CSI,Sudarshan_63,Glauber_63}. Analogically for $N$ indistinguishable bosons, if condition (\ref{cond_pos}) is not fulfilled, the density matrix cannot be written as a statistical mixture of separable coherent state, meaning that correlations between the particles are genuinely quantum.

On the other hand, the separable states of bosons are classical because the symmetrization enforced by the indistinguishability on state (\ref{coh_def}) is only superficial. In principle, identical state can be prepared out of independent, distinguishable particles by initializing each one of them in the same quantum state using a proper set of measuring devices. Hence, the indistinguishability of bosons in separable state is inconsequential and such a state can be simulated by a classical system; at least for as long as the processes which address parties individually are absent.


An entanglement of identical particles is often considered as ``useless'' as opposed to mode entanglement which has proven to be a valuable resource for many applications in the field of quantum information. Many protocols utilizing entanglement are designed under the assumption that each party can be address individually, thus excluding the possibility of using indistinguishable particles. Nevertheless, as I already argued previously, correlations of identical particles are very important and have profound consequences. I shall discuss the usefulness of particle entanglement, which does not have to be considered only in terms of applicability of certain class of protocols, in the upcoming sections.


\section{Useful entanglement}\label{sec:Intro_Useful_Ent}

The difficulty in grasping the nature of entanglement is a direct consequence of a general and discriminative definitions (\ref{def:entanglement}) and (\ref{def:particle_entanglement}). Possible solutions to this problem is to examine the subject form the utilitarian point of view. The idea is that certain tasks, like cryptography or computing, can be performed better when entangled states are used instead of states that are only classically correlated.  This line of reasoning leads to the concept of classification of entanglement by the degree of their usefulness for a given task. Now I proceed to formalize this abstraction.  

Let $\mathcal Q^{\mathcal T}(\hat\varrho)$ be the efficiency at which a given task $\mathcal T$ is performed using state $\hat\varrho$. Since the form of separable state is known it is often possible to find a maximal efficiency achievable with non-entangled states: 
\begin{equation}\label{eq:Intro_UsefulEnt_eff_crit}
\mathcal Q^{\mathcal T}(\hat\varrho_{\rm separable})\leq \mathcal Q^{\mathcal T}_0
\end{equation}
Therefor, if a state allows for efficiency greater then the classical bound $\mathcal Q^{\mathcal T}_0$, then the state has to be entangled:
\begin{equation}\label{eff_crit}
\mathcal Q^{\mathcal T}(\hat\varrho)>\mathcal Q^{\mathcal T}_0 \Rightarrow \text{ $\hat\varrho$ is entangled}.
\end{equation}
The reverse implication is not guaranteed, i.e. not all entangled states are able to outperform classically correlated states. A criterion such as (\ref{eff_crit}) gives means to classify types of entanglement, in this case, in terms of their usefulness for a given task. Moreover, the efficiency $\mathcal Q$ can be treated as an indicator of the degree of non-classical correlations -- the greater the $\mathcal Q$ the more of the useful entanglement is present. 

This approach to entanglement detection is reminiscent of the method of entanglement witnesses \cite{Terhal_00,Toth_09} and criteria such as spin squeezing \cite{Ueda_93,Sorensen_01,Toth_07,Wang_03} and Cauchy-Schwarz criterion \cite{CSI}. The efficiency criterion (\ref{eff_crit}) posses an advantage over other tests that, by its definition, it automatically provides an application for useful entanglement and the physical context. Therefore, the properties of usefully entangled states can be juxtaposed with everything that is know about the task it is useful for, which can turn out to be a valuable source of physical intuition. On the other hand, the entanglement detected by the other mentioned criteria can be characterized only by the fact that it is detected by those criteria. For these reasons I will adopt the notion of useful entanglement and the efficiency criterion as a basis for further discussion on subject of nature of non-classical correlations.

\chapter{Atomic interferometer}\label{ch:Atomic_interferometer}

In previous chapter I argued for the advantages of using the efficiency criterion (\ref{eff_crit}) as a tool for investigating the properties of entanglement. However, in order to implement such approach, first one has to decide which task $\mathcal T$ is to be performed with the help of non-classically correlated states. My task of choice is {\it quantum metrology}, in particular the {\it quantum interferometry}.

\section{Quantum metrology}

Metrology is the science of measurement. Its main goal is to develop methods, both practical and theoretical, for measuring properties of physical systems as precisely as possible. Precise measurements are essential in our everyday life, without it we would not be able to build safe and efficient buildings, machines, medications and so on. Metrology is also of great importance for science itself. In the end any scientific inquire boils down to the measurement and more precise results allow for drawing more decisive conclusions which can help in testing a theory or point us in new directions of research.

Any measurement procedure consists of three steps: the preparation of a probe, its interaction with the system to be measured, and the probe readout. 
%
However, 
this process 
is inevitably affected by statistical or systematic errors. The source of the former can be accidental (e.g. deriving from an insufficient control of the probes or of the measured system) or fundamental (e.g. deriving from the Heisenberg uncertainty relations). Whatever the origin of the error, it can be reduced by repeating the measurement and averaging the resulting outcomes. According to the central-limit theorem, given a large number $N$ of independent measurement results each having a standard deviation $\sigma$, the average error converges (with growing $N$) to a Gaussian distribution with the standard deviation equal to ${\sigma}/{\sqrt{N}}$, so that the resultant error scales as $ 1/{\sqrt N}$. On the other hand, a single measurement realized with a probe composed of $N$ uncorrelated sub-probes (e.g. particles forming a matter-wave) includes the independent repetitions automatically thus yielding the error which also scales as $1/{\sqrt{N}}$. This behavior is referred to as the {\it shot noise limit} (SNL) and I will demonstrate in the upcoming chapter that it is also associated with procedures which do not fully exploit the quantum nature of the probe. This implies that SNL can be surpassed only when one employs quantum effects, such as the entanglement among the probing particles utilized for the measurements. Consequently, the SNL is not a fundamental quantum mechanical bound as it can be overcome by using non-classical strategies. {\it Quantum metrology} studies the fundamental bounds on precision imposed by quantum mechanics and the strategies which allow for attaining them. More generally it deals with measurement and discrimination procedures that receive an enhancement in precision through the use of quantum effects. Therefore, the precision of measurement can serve as a efficiency criterion sensitive to ``quantum effects'' and in particular it can be used for detecting quantum entanglement between particles constituting the probe.

Atomic interferometers \cite{Schmiedmayer_RevModPhys_2009} form a family of devices which exploit the wave nature of matter, providing a unique and powerful tool for modern quantum metrology. They have been used to measure atomic properties \cite{Schmiedmayer_95}, to study quantum degenerate systems \cite{Hoff_08}, for precision measurements \cite{Gring_12}, and are very sensitive probes for inertial effects with applications in gravimeters and gyroscopes \cite{Chu_93,Clauser_88,Chu_99,Gustavson_00}.

Bose-Einstein condensates (BECs) are promising candidates for atom interferometry owing to their macroscopic coherence properties. Following the first observation of the BEC interference in 1997, various building blocks of the BEC interferometers have been realized individually. 
The 
interference experiments with BECs were performed using Bragg beams in ballistic expansion, with freely propagating matter-waves in a guide \cite{Perrin_12,McDonald_13,micrograv_13}. By splitting a single trapped BEC into two separated clouds in a double well, interference was observed after switching off the trapping potential \cite{Shin_04,Schumm_05,Baumg_10}. These and many other examples show that a long-standing goal of realizing a full interferometer with ultra-cold atom systems is in our reach.
 
A fundamental difference between photon and matter-wave optics is the presence of atom-atom interactions which might lead to generation of non-classical correlations. The unprecedented degree of control over these interactions via Feshbach resonance techniques \cite{RMP_feshbach_10} allow for preparation of a BEC probe in strongly entangled states which can beat the shot noise sensitivity of the interferometric measurements. This is the main advantage of atomic interferometry over its optical counter-part: with BECs it is relatively easy to entangle thousands of particles, while with light the current technical limitations only allow for few photon entanglement.

In the following section I describe a few important examples of experimental realizations of atomic interferometers based on BECs and introduce a theoretical tools for describing such systems.

\section{Examples of atomic interferometers}\label{sec:Atom_Inter_Examples}

The basic design of an interferometer involves two ``arms'' through which the probe can travel and the ability to ``mix'' the signals from the two ports in a coherent manner.  In case of atomic interferometers the ``arms'' can be realized by well separated modes; this includes modes in the real space, momentum space or even in the space of internal degrees of freedom such as Zeeman levels. The beam-splitter operation which transfers the particles between modes as well as phase difference imprint can be achieved by properly devised external fields. Having this picture in mind I proceed to describe in more detail some of the most important experimental realizations of atomic interferometers.

\begin{figure}[t]
	\centering
	\includegraphics[clip,width=\textwidth]{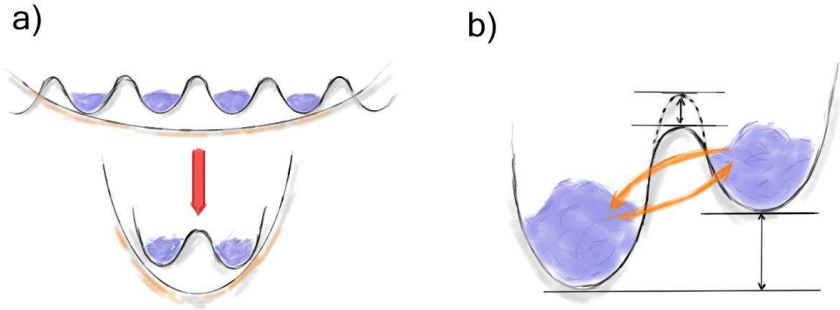}
	\caption{{\it Example of atomic interferometer: double well setup \cite{Oberthaler_08}. a) Initially Bose-Einstein condensate is created in the optical lattice with atoms occupying 2-6 well separated sites. The harmonic potential with tunable frequency is imposed onto the system. The potential is adiabatically ``squeezed'' forcing atoms in two adjacent sites, which define the arms of the interferometer. b) The height of the barrier between potential wells can be manipulated in order to control the rate of coherent tunneling, which allows for realization of beam-splitter operation. The relative depth of the wells can also be manipulated by introducing the energy offset and thus realizing the phase imprint operation.}}\label{fig:double_well_setup}
\end{figure}

The first paradigm is the double well setup achieved with condensate trapped in the optical lattice. In \cite{Oberthaler_08} the particles in BEC of ${}^{87}\mathrm{Rb}$ atoms where distributed over a small number of lattice sites (between two and six) in a one-dimensional optical lattice. The occupation number per site ranges from 100 to 1,100 atoms. The two modes representing arms of the interferometer where the two states of the external atomic motion corresponding to the condensate mean-field wavefunctions in two, well separated adjacent lattice sites (see Fig.~\ref{fig:double_well_setup} (a)). The beam-splitter operation which allows for mixing particles occupying the two modes was realized by coherent tunneling through potential barrier separating the lattice sites. By adjusting the height of the barrier it is possible to control the rate of tunneling or even stop it all together. The imprint of the phase difference between arms comes from the difference in depth of lattice sites (see Fig.~\ref{fig:double_well_setup} (b)). The difference could be caused by the variation of external potential (e.g. gravitation, magnetic field) on the length scale of wells separation thus allowing for high-resolution measurement of this variation. The information about imprinted phase can be inferred from absorption imaging measurements of atom number in each lattice site. The wells of the lattice can be fully resolved thanks to imaging techniques with a resolution of \SIunits{1}{\micro\meter}, which is well below the lattice spacing of\SIunits{ 5.7}{\micro\meter}. This allows for the determination of the atom number in each lattice site by direct integration of the measured atomic density. Local interference measurements after a condensate expansion time short enough that only neighboring sites overlap reveal the phase between these wells, confirming that the coherence of the condensate is preserved. The non-classical correlations between particles where provided by the on-site repulsive atom-atom interactions \cite{Ueda_93}.

 A full interferometric sequence composed of beam-splitter followed by phase imprint followed by another beam-splitter with trapped ${}^{87}\mathrm{Rb}$ BEC confined on an atom chip was demonstrated in \cite{Schmiedmayer_13}. The interferometric scheme relied on the coherent splitting and recombination of a BEC in a tunable magnetic double-well potential, where the matter wave is confined at all times. Thanks to a spatial separation of $\sim$ \SIunits{2}{\micro\meter} between the two wave packets, the geometry was sensitive to accelerations and rotations. By tilting the double well out of the horizontal plane for a variable time, the energy difference was applied and thereby imprint a controlled relative phase between the interferometer arms. A non-adiabatic recombiner translates the relative phase into an atom number difference, which is directly read out using a highly sensitive time-of-flight fluorescence detector. As in the previous case also here the particle interactions in BEC matter waves lead to a nonlinearity which generated the entanglement between atoms.

\begin{figure}[h]
	\centering
	\includegraphics[clip,width=\textwidth]{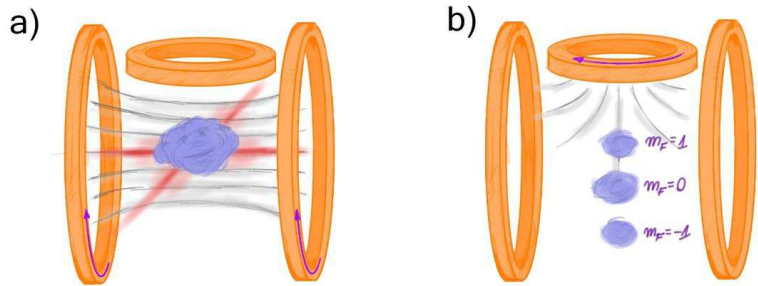}
	\caption{{\it Example of atomic interferometer: internal degrees of freedom \cite{Smerzi_exp_11}. a) The Bose-Einstein condensate of atoms in $F=2$ hyperfine state is confined in optical dipole trap at a homogeneous magnetic field. The spin dynamics in atom-atom collisions lead to creation of pairs in spin up and down state ($m_F=\pm 1$). b) In order to read out the populations of the Zeeman modes, the magnetic trap is turned off and the strong magnetic field gradient is turned on. In the fashion of the Stern-Gerlach device, the internal modes of the atomic cloud become spatially separated and ready for density measurement with standard imagining techniques.}}\label{fig:inner_dof}
\end{figure}

Now I turn to a different paradigm where the arms of the interferometer are realized with modes of internal degrees of freedom. The experiments reported in \cite{Smerzi_exp_11} started by creating a  condensate of $2.8 \times 10^4$ ${}^{87}\mathrm{Rb}$ atoms in the hyperfine state $F = 2$ with horizontal spin orientation (Zeeman substate $m_F = 0$) confined in an optical dipole trap. The spin dynamics in particle collisions where used to create up to $10^4$ paired neutral atoms in spins up and down states ($m_F = \pm 1$). These collisions are bosonically enhanced if the output modes are occupied. Therefore, they act as a parametric amplifier for a finite initial population in $m_F = \pm 1$ or for pure vacuum fluctuations. During the parametric amplification of vacuum, the total number of atoms produced in $m_F = \pm 1$ and its fluctuations increase exponentially with time (see Fig.~\ref{fig:inner_dof} (a)). The conjugate variable of the total number is the sum of the two atomic phases, whose fluctuations are exponentially damped. Furthermore, the number difference between $m_F = \pm 1$ atoms is zero (without fluctuations), and hence the corresponding conjugate variable, the relative phase, is fully undetermined. The underlying physics closely resembles that of optical parametric down-conversion in nonlinear crystals, currently the most important technique to generate non-classical states of light. The spin dynamics where initiated at a magnetic field, where an excited spatial mode is populated and vacuum fluctuations are amplified. The states $F = 2$, $m_F = \pm 1$ are populated by spin dynamics for an optimal duration of \SIunits{15}{\milli\second}. The internal-state beam-splitter was implemented by driving the transition connecting the $F = 2$, $m_F = \pm 1$ states with three resonant microwave pulses. Subsequently, the dipole trap was switched off and all three spin components where recorded by absorption imaging after they where spatially separated by a strong magnetic field gradient (see Fig.~\ref{fig:inner_dof} (b)). 

\begin{figure}[h]
	\centering
	\includegraphics[clip,width=\textwidth]{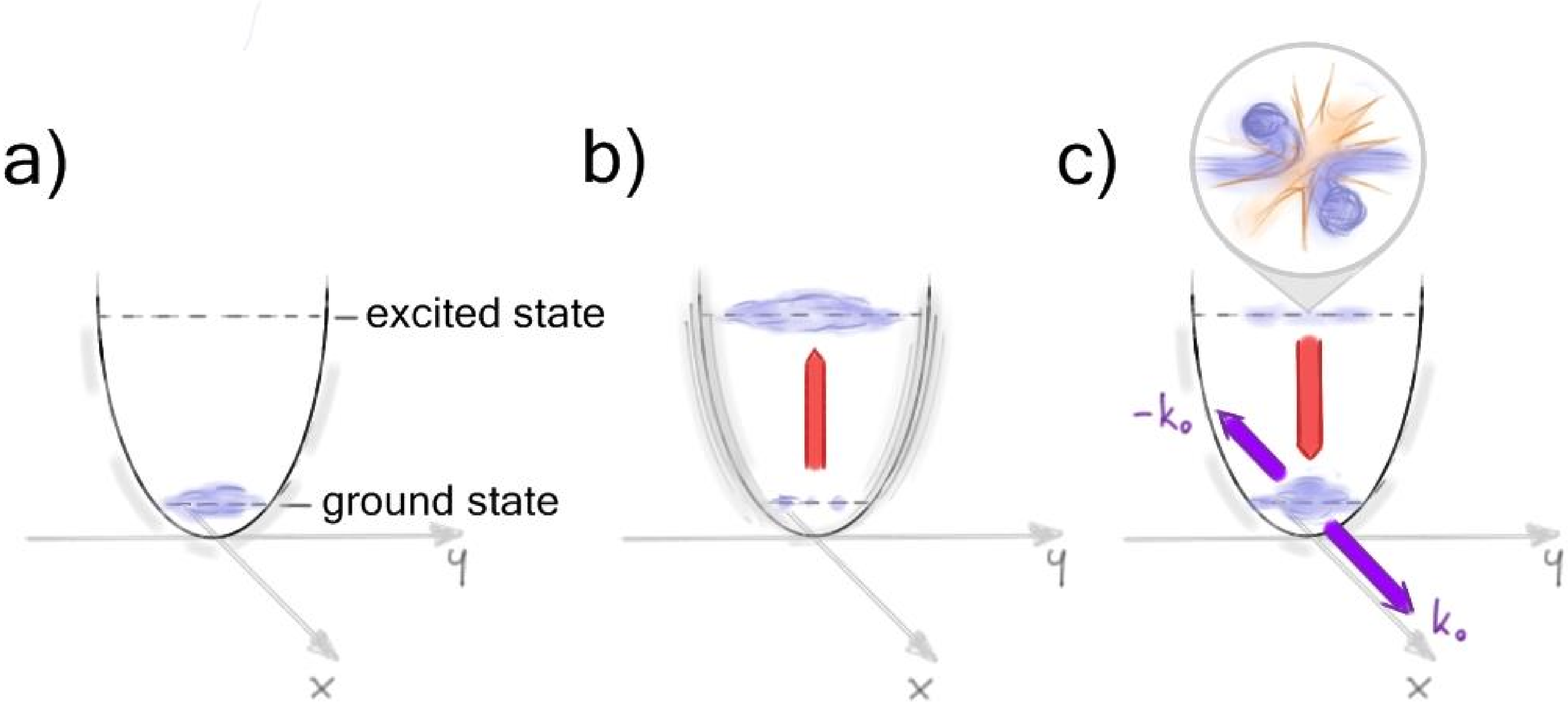}
	\caption{{\it Example of atomic interferometer: twin matter beams \cite{Bucker_11}. a) Special geometry of the trapping potential creates the effective two-level system. b) An optimally controlled shaking of the trap inverts the population from the ground state to excited state with almost 100\% efficiency. c) The excited state is metastable. Due to collisions pairs of atoms get demoted to ground state and the excess of energy is converted into motion with opposite momenta directed along $x$-axis due to appropriately shaped trapping potential.}}\label{fig:twin_beam}
\end{figure}

As a last example I refer to an experiment described in \cite{Bucker_11}, where it was demonstrate how collisional deexcitation of a one-dimensional degenerate Bose gas can be used to efficiently create matter wave beams separated in momentum space. The starting point is a quasi-BEC of ${}^{87}\mathrm{Rb}$ atoms magnetically trapped in a tight waveguide potential with a shallow axial harmonic confinement (say, along $x$-axis
) on an atom chip. The scheme relied on an effective two-level system in the radial vibrational eigenstates of the waveguide (see Fig.~\ref{fig:twin_beam} (a)). This was accomplished by creating unequal level spacings in the radial $y,z$-plane by radio frequency dressing, which introduces anharmonicity and anisotropy. Due to the increasing level spacings, the ground state and the first excited state along $y$, have the lowest energy difference among all possible combinations, establishing a closed two-level system. Having prepared the gas in the ground state, the population was inverted by transferring the atoms almost entirely to the excited state (the efficiency of coherent transfer reached $\sim 97$\%). The transition is driven by shaking the trap along the radial $y$-direction on the scale of the ground state size (\SIunits{$\sim 100$}{\nano\meter}). The trajectory of the shake (total duration \SIunits{5}{\milli\second}) has been optimized by an iterative optimal control algorithm (see Fig.~\ref{fig:twin_beam} (b)). In the experiment, the displacement was achieved by driving a current in an auxiliary chip wire, parallel to the main trapping wire. The population inversion represents a highly non-equilibrium state of the system, analogous to a laser gain medium after a pump pulse. For the ensuing relaxation, due to wave-guide geometry the only allowed channel was a two-particle collisional process, emitting atom pairs with opposite momenta. Within a binary collision, two atoms are scattered from the excited state to the ground state and the excess potential energy is transformed into the kinetic energy of each atom. Due to momentum conservation each one acquires momentum of the same magnitude but opposite direction along the elongated axis of the trapping potential (see Fig.~\ref{fig:twin_beam} (c)). Similarly to the previous example the emission process can be understood as a matter wave analogue of degenerate optical parametric amplifier, where the initially empty twin-modes are seeded by vacuum fluctuations and gain an exponentially growing population if the phase matching conditions are fulfilled. Finally, the beam-splitter can be achieved by applying a Bragg pulse which can coherently bring matter-wave into superposition of counter-propagating beams. The populations of twin-beams was measured with fluorescence images once the trap potential is switched off and the atoms propagate freely separating from the source.

In the last two examples the source of twin-mode matter waves was a collisional phenomena taking place among atoms forming the condensate. Also, due to indistiguishability of atoms and bosonic enhancement, the scattering into well separated modes is responsible for formation of strong non-classical correlation among the particles. Reference \cite{Wasak_twin_beam_14} reviews the Bogoliubov theory in the context of such twin-beam experiments and describes the process of pair generation leading to highly entangled states useful for interferometry. The nature of this entanglement will be discussed in the upcoming chapters.

\section{Pseudo-spin of two-mode atomic interferometer}\label{sec:Dyn_Ent_Pseudo-spin}

According to examples presented in the previous section Bose-Einstein condensate undergoing interferometric measurement can be considered as an ensemble of $N$ {\it qubits} defined by two distinct modes $L$ and $R$ constituting the arms of interferometer. Due to bosonic nature of atoms the state of the system for fixed $N$ can be described in terms of {\it occupation numbers} of each mode. The mode occupation number states form an orthonormal basis in this $N+1$ dimensional space and are defined with the help of  annihilation and creation operators which add and subtract a particle in a given mode:
\begin{equation}
\substack{\text{Basis of}\\\text{mode occupation}\\\text{number states}} \,:\,\,\left\{|n\rangle_L|N-n\rangle_R \equiv \frac{(\hat a_L^\dagger)^n}{\sqrt{n!}}\frac{(\hat a_R^\dagger)^{(N-n)}}{\sqrt{(N-n)!}}|\mathrm{VAC}\rangle \right\}_{n=0,1,\ldots,N}
\end{equation}
Here $\hat a_{L/R}$ are the annihilation operators for mode $L/R$ which satisfy standard bosonic commutation relations ($[\hat a_{L/R} , \hat a_{L/R}^\dagger ] = 1$  and all other commutators equal zero), $|\mathrm{VAC}\rangle$ is a vacuum defined as a state which satisfies $\hat a_{L/R}|\mathrm{VAC}\rangle =0$.

Incidentally, mode occupation number states are also eigenstates of operator
\begin{equation}
\hat J_z{} = \frac{\hat a_L^\dagger\hat a_L-\hat a_R^\dagger\hat a_R}2=\frac{\hat n_L-\hat n_R}2\, .
\end{equation}
This operator, together with
\begin{align}
\hat J_x{}& = \frac{\hat a_L^\dagger\hat a_R+\hat a_R^\dagger\hat a_L}2\\
\hat J_y{}& = \frac{\hat a_L^\dagger\hat a_R-\hat a_R^\dagger\hat a_L}{2i}\\
\end{align}
constitutes a set of three orthogonal components of fictitious spin operator. Straightforward calculation shows that this set of operators indeed satisfy commutation relations of angular momentum algebra $[\hat J_i,\hat J_j]=i\epsilon_{ijk}\hat J_k$. The action of these operators on basis states is given by
\begin{equation}
\left\{\begin{array}{l}
\hat J_z|n\rangle_L|N-n\rangle_R =\left(\tfrac{\hat n_L-\hat n_R}{2}\right)|n\rangle_L|N-n\rangle_R =\tfrac{n-(N-n)}2|n\rangle_L|N-n\rangle_R\\[.3cm]
\left(\hat J_x^2+\hat J_y^2+\hat J_z^2\right)|n\rangle_L|N-n\rangle_R=\\
=\left[\tfrac{(\hat n_L+\hat n_R)^2}{4}+\tfrac{\hat n_L+\hat n_R}2\right]|n\rangle_L|N-n\rangle_R=\tfrac N2\big(\tfrac N2+1\big)|n\rangle_L|N-n\rangle_R\\
\end{array}\right.\,.
\end{equation}
Therefore, the system composed of $N$ bosonic qubits is equivalent to single pseudo-particle with spin $N/2$:
\begin{equation}
\big\{ |n\rangle_L|N-n\rangle_R \big\}_{n=0,\ldots,N}\,\leftrightarrow\, \big\{|J, M\rangle\big\}_{J=N/2,M=-J,\ldots,J}
\end{equation}

This formal equivalence can always be established as long as the two-mode approximation holds. Moreover, pseudo-spin operators can also be ascribed a transparent physical interpretation. The component $\hat J_z$ is an observable associated with {\it population imbalance} between the modes of interferometer. As it was noted in the previous section, the high resolution imaging techniques allow for very accurate evaluation of the number of atoms in each mode and hence the population difference is the main property of the system accessible through direct measurement. The $x$ and $y$ component combine into ladder operator $\hat J_+=\hat J_x+ i\hat J_y=\hat a^\dagger_L\hat a_R$ and $\hat J_-=\hat J_x-i\hat J_y=\hat a_R^\dagger\hat a_L$ which describe a process where a particle is transfered from one mode to the other. These operators are associated with a beam-splitter operation, which is a necessary ingredient of operational atomic interferometer (see previous section).

The equivalence between the two-mode state of bosonic qubits and a pseudo-particle with spin will prove to be very productive in the upcoming chapter where I use it to analyze types of particle entanglement found in ultra-cold atom systems. 

\chapter{Efficiency of interferometer}\label{ch:Eff}

In this chapter I provide a basic introduction to the theory of parameter estimation, on which the efficiency criterion for atomic interferometers is based on.

\section[Distinguishability of quantum states and the problem of parameter estimation]{Distinguishability of quantum states and the problem of parameter estimation%
              \sectionmark{Distinguishability of quantum states and estimation}}\label{sec:Atom_Inter_QFI}
\sectionmark{Distinguishability of quantum states and estimation}

The principle of operation of the atomic interferometer is based on detecting changes in a state of the probe induced by the interaction with the measured  system. The evolution of the probe state obviously depends on the properties of the system; in particular it depends on the unknown value of the parameter $\theta$ characterizing its certain aspect. The goal is to estimate the exact value of this parameter. The efficiency at which this task is performed is tied to the precision of the estimation procedure.

Once the interaction with the system is concluded we end up with the output state which has some information about the parameter imprinted onto it,
\begin{equation}
\hat\varrho_{\mathrm{in}}\xrightarrow{\substack{\text{interaction with} \\ \text{the measured system}}} \hat\varrho_{\mathrm{out}}(\theta)\,
\end{equation}
To gain access to this information a measurement (in a sense of quantum mechanics) has to be performed on $\hat\varrho_{\mathrm{out}}$. Preferably, the measurement should be chosen so that it addresses those properties of the output state which where most affected by the parameter imprint. In the final step of the procedure the data acquired from the measurement is used to estimate the real value of $\theta$.

\begin{figure}[h]
\centering
\includegraphics[clip,width=\textwidth]{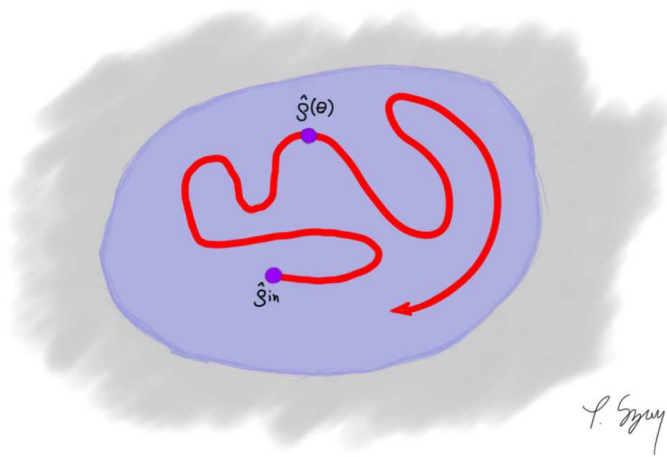}
\caption{{\it The evolution of probe state under the influence of the measured system viewed as a $\theta$-parametrized path in the space of density matrices. The ability to locate an output state $\hat\varrho_\mathrm{out}=\hat\varrho(\theta)$ on the path is required for estimating the value of $\theta$.}}\label{fig:trajectory_in_state_space}
\end{figure}

\begin{figure}[h]
\centering
\includegraphics[clip,width=\textwidth]{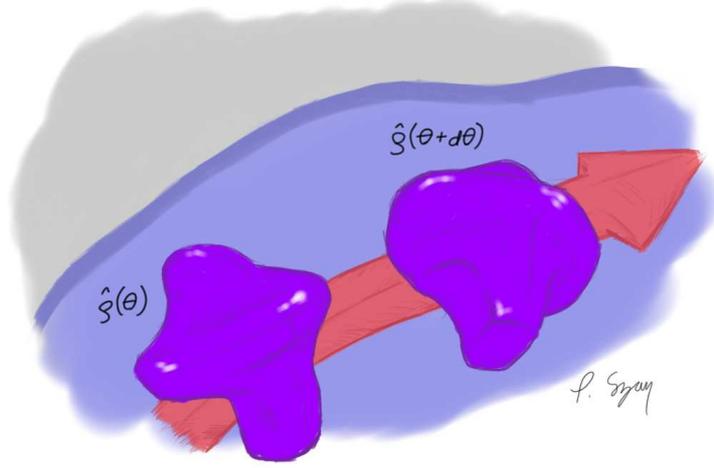}
\caption{{\it States which are more susceptible to the influence of the measured system can undergo a significant change and cover grater distances in the density matrix space even for very small variations of parameter $\theta$. For such states it is easier to distinguish $\hat\varrho(\theta)$ from the neighboring $\hat\varrho(\theta+d\theta)$ which allows to locate the state on the path with greater reliability. This translates directly to precision of parameter estimation.}}\label{fig:distinguishability}
\end{figure}

An alternative point of view is that the value of $\theta$ parametrizes a path through the space of quantum states followed be the probe under the influence of the system (see Fig.~\ref{fig:trajectory_in_state_space}). Therefore, estimation of the value of $\theta$ is directly related to the ability to pinpoint the position of the state on this path. Then, the precision of the estimation relays on the capability for distinguishing the neighboring quantum states occupying the path. The precision can, thus, be increased by preparing the probe in the state which is most susceptible to the evolution driven by the interaction with the system. Indeed, if the state undergoes a significant change when exposed to the system it should be easier to determine a small variation of the parameter. In other words, the state which moves along the path with greater ``speed'' will travel a greater distance even for small increments of $\theta$ (see Fig.~\ref{fig:distinguishability}).

This point of view was adopted in \cite{Caves_QFI} where the authors formulated the problem of parameter estimation in terms of distinguishing neighboring quantum states in a space of density matrices. They have found measurements which optimally resolve neighboring states, and characterized their degree of distinguishability in terms of a Riemannian metric, increasing distance corresponding to more reliable distinguishability. These considerations allowed to establish a type of uncertainty principles which allowed to relate the precision of estimation with the susceptibility of the probe state to change under process of parameter imprint. Below I present a brief summary of the derivation of this result.

\section{Distinguishability metric}\label{sec:Eff_Dist_metric}

I begin by reviewing a derivation of the distinguishability metric for probability distributions \cite{Wootters_81}. After drawing $\nu$ samples from a probability distribution, one can estimate the probabilities $p(\xi)$ as the observed frequencies $f(\xi)$, The probability for the frequencies is given by a multinomial distribution, which for large $\nu$ is proportional to a Gaussian $\propto\exp\big[-(\nu/2)\big(f(\xi) -p(\xi)\big)^2/p(\xi)\big]$. A nearby distribution $\widetilde{p}(\xi)$ can be reliably distinguished from $p(\xi)$ if the Gaussian $\exp\big[-(\nu/2)\big(\widetilde{p}(\xi) -p(\xi)\big)^2/p(\xi)\big]$ is small. Thus the quadratic form $\big(\widetilde{p}(\xi) -p(\xi)\big)^2/{p(\xi)}$ provides a natural distinguishability metric on the space of probability distributions (PD), called the {\it statistical distance}:
\begin{equation}\label{eq:Atom_Inter_s_PD}
ds^2_{\mathrm{PD}}\big(p(\xi)\big) \equiv \int d\xi \frac{\big[dp(\xi)\big]^2}{p(\xi)}=\int d\xi\, p(\xi) \big[d \log p(\xi)\big]^2 \,.
\end{equation}
The notion of the statistical distance has been generalized by the authors to mixed quantum states and thus obtain a natural Riemannian geometry on the space of density operators. 


Consider now a curve $\hat\varrho(\theta)$ on the space of density matrices. Performing a measurement on the state is the only way available in quantum mechanics to distinguish $\hat\varrho(\theta)$ from the neighboring matrices $\hat\varrho(\theta+d\theta)$. In general quantum mechanical measurements are described by a set of non-negative, Hermitian operators $\hat\Pi_\xi$ which are complete in the sense that
\begin{equation}
\int d\xi\, \hat \Pi_\xi = \hat{\mathbf 1}\,.
\end{equation}
Such a set is refered to as {\it positive operator valued measure} (POVM). The quantity $\xi$ labels the ``results'' of the measurement; although written here as a single continuous real variable, it could be discrete or multivariate. The probability density for result $\xi$, given the parameter $\theta$, is
\begin{equation}
p(\xi |\theta) = \mathrm{Tr}\big(\hat \Pi_\xi\hat\varrho(\theta)\big)\,.
\end{equation}
Thus, the density matrix $\hat\varrho(\theta)$ has been mapped onto probability distribution which can be treated with classical statistical distance $ds_{\mathrm{PD}}^2$:
\begin{align}
&ds_{\mathrm{PD}}^2 \big(p(\xi|\theta)\big) = \int d\xi \frac{\big[\partial_\theta p(\xi|\theta)\big]^2}{p(\xi|\theta)}=\int d\xi\,p(\xi|\theta)\big[ \partial_\theta\log p(\xi|\theta)\big]^2=\nonumber\\
& \equiv F\big(\hat\varrho(\theta),\hat\Pi_\xi\big)d\theta^2\,,
\end{align}
where the quantity $F$ is called the {\it classical Fisher information} (CFI).

The statistical distance depends on the choice of POVM. This dependence is to be removed by optimization over all possible quantum measurements. Therefore, the problem of finding quantum analog of classical statistical distance is equivalent to the problem of maximizing the Fisher information over all POVMs, i.e., symbolically
\begin{equation}\label{eq:Atom_Inter_quantum_metric}
ds^2_Q \big(\hat\varrho(\theta)\big)= d\theta^2 \max_{\{\hat\Pi_\xi\}} F\big(\hat\varrho(\theta),\hat\Pi_\xi\big)\,.
\end{equation}
The subscript $Q$ reminds one that this is a metric on a space of density matrices of quantum states.

Derived in \cite{Caves_QFI} the upper bound of Fisher information, called the {\it quantum Fisher information} (QFI) is given by
\begin{align}
F_Q(\hat\varrho_\mathrm{out}(\theta)) {}&\equiv 2\sum_{i,j}\frac{|\langle i^\mathrm{out}|\partial_\theta \hat\varrho_\mathrm{out}(\theta)|j^\mathrm{out}\rangle|^2}{p_i^\mathrm{out}+p_j^\mathrm{out}}=\nonumber\\
&=\sum_i \frac{\big(\partial_\theta p_i^\mathrm{out}\big)^2}{p_i^\mathrm{out}} 
	+ 2\sum_{i,j}\frac{(p_i^\mathrm{out}-p_j^\mathrm{out})^2}{p_i^\mathrm{out}+p_j^\mathrm{out}}|\langle i^\mathrm{out}| \hat h|j^\mathrm{out}\rangle|^2\nonumber\\
&\geqslant F\big(\hat\varrho_\mathrm{out}(\theta),\hat\Pi_\xi\big)\,,\label{eq:Atom_Inter_QFI_bounds_Fisher}
\end{align}
where $|i^\mathrm{out}\rangle$ are the eigenstates of $\hat\varrho_\mathrm{out}(\theta)$ with corresponding eigenvalues $p_i^\mathrm{out}$ and the hermitian operator $\hat h$ generates the infinitesimal unitary basis transformation
\begin{equation}\label{eq:Atom_Inter_generator_h}
| i^\mathrm{out}(\theta+d\theta)\rangle = e^{i d\theta \hat h}|i^\mathrm{out}(\theta)\rangle = |i^\mathrm{out}(\theta)\rangle + i d\theta \hat h|i^\mathrm{out}(\theta)\rangle\, .
\end{equation}
When QFI is supplemented with an appropriate interpretation and physical context it becomes a powerful tool for investigating the structure of particle entanglement, as it will be shown in the upcoming chapters.

The general formula for $F_Q$ is quite intimidating and can be difficult to work with. Fortunately, this is not always the case. For example, when the interaction of the prob and the measured system results in unitary evolution, $\hat\varrho_\mathrm{out}(\theta)=e^{-i \theta\hat h}\hat\varrho_{\mathrm{in}}e^{i\theta\hat h}$, the eigenvalues of input density matrix $p_i^\mathrm{in}$ remain unaltered. Corresponding output eigenstates are related to input eigenstates via $|i^\mathrm{out}(\theta)\rangle =e^{-i\theta\hat h}|i^\mathrm{in}\rangle$. In this simple case the QFI is given by
\begin{equation}
F_Q\big(e^{-i \theta\hat h}\hat\varrho_{\mathrm{in}}e^{i\theta\hat h}\big) = 2\sum_{i,j}\frac{\big(p_i^\mathrm{in}-p_j^\mathrm{in}\big)^2}{p_i^\mathrm{in}+p_j^\mathrm{in}}\big|\langle i^\mathrm{in}|\hat h|j^\mathrm{in}\rangle\big|^2\,.
\end{equation}
This expression simplifies even further when the input state is pure%
\footnote{{Note that the sum over eigenvalues in Eq.~(\ref{eq:Atom_Inter_QFI_bounds_Fisher}) also includes cases when one of $p_i$-s is zero.}}, $\hat\varrho_\mathrm{in}=|\Psi\rangle\langle\Psi|$
\begin{equation} 
F_Q\big(e^{-i\theta\hat h}|\Psi\rangle\big) = 4\left[\langle\Psi|\hat h^2|\Psi\rangle -\big(\langle\Psi|\hat h|\Psi\rangle\big)^2\right] = 4\langle \Delta^2\hat h\rangle_\Psi\,.
\end{equation}
Hence for pure states, QFI is simply proportional to the variance of the transformation generator $\hat h$.

I conclude the discussion on distinguishability metric by noting that upper bound (\ref{eq:Atom_Inter_QFI_bounds_Fisher}) is achievable, i.e. for given $\hat\varrho(\theta)$ there always exists an optimal POVM $\{\hat\Pi^{(\mathrm{opt})}_\xi\}$, so that $F\big(\hat\varrho(\theta),\hat\Pi_\xi^{(\mathrm{opt})}\big) =F_Q(\hat\varrho(\theta))$ \cite{Caves_QFI}. Therefore, the distinguishability metric (\ref{eq:Atom_Inter_quantum_metric}) on density operators becomes
\begin{equation}\label{eq:Atom_inter_Q_metric_final}
ds^2_Q = F_Q\big(\hat\varrho(\theta)\big)d\theta^2 \,.
\end{equation}
The density matrix metric (\ref{eq:Atom_inter_Q_metric_final}) also appears in another context. For example, a distance between density operators was defined in \cite{Bures_69,Uhlmann_76,Hubner_92}:
\begin{equation}
d(\hat\varrho_1,\hat\varrho_2)^2 = 2\left[ 1 - \mathrm{Tr}\left( \sqrt{\sqrt{\hat\varrho_1}\,\hat\varrho_2\,\sqrt{\hat\varrho_1}}\right)\right]
\end{equation}
In quantum information theory this quantity is interpreted as a {\it fidelity} -- the measure of the ``closeness'' of two quantum states. It can be shown that for neighboring density matrices it reduces to 
\begin{equation}
d(\hat\varrho,\hat\varrho+d\hat\varrho)^2 = \frac 14 ds_Q^2+o(d\hat\varrho^3)\,.
\end{equation}

\section{Precision of estimation}

As I noted above, the problem of precise estimation of $\theta$ is equivalent to the problem of distinguishing density matrices along the trajectory $\hat\varrho(\theta)$. Consider now the following procedure: a series of quantum measurements $\hat\Pi_\xi$ (not necessary the optimal ones) are repeated $\nu$ times, yielding the results $\xi_1,\ldots,\xi_\nu$. These results are then used for estimation the value of parameter $\theta$ via a function called {\it estimator}: $\theta_{\mathrm{est}}=\theta_{\mathrm{est}}(\xi_1,\ldots,\xi_\nu)$. The variance of any unbiased estimator, i.e., functions which satisfy condition $\overline{\theta_{\mathrm{est}}}=\theta$ for all values of $\theta$ (the average $\overline{(\ldots)}$ is taken with probability distributions $p(\xi_i|\theta)$ ) is bounded from below by the so-called Cram\'{e}r-Rao lower bound \cite{Cramer_book}:
\begin{equation}\label{eq:Atom_Inter_CRLB}
\overline{ \Delta^2\theta_{\mathrm{est}}} \geqslant \frac{1}{\nu}\frac 1{F\big(\hat\varrho_\mathrm{out}(\theta),\hat\Pi_\xi\big)}\,.
\end{equation}
According to Fisher theorem \cite{Fisher_1925}, for given probability distribution $p(\xi|\theta)$ and sufficiently large number of repetitions $\nu$ this bound is achievable by unbiased {\it maximum-likelihood estimator} \cite{Braunstein_92,Braunstein_92_2}.

The bound (\ref{eq:Atom_Inter_CRLB}) is known from the classical estimation theory. It can now be combined with the quantum result of optimization over POVMs (\ref{eq:Atom_Inter_QFI_bounds_Fisher}). This brings us to the final conclusion that the error of estimation of parameter $\theta$ for an arbitrary choice of estimator {\it and} quantum measurements performed on the output state of the probe is bounded by QFI
\begin{equation}\label{eq:Quantum_CRLB}
\Delta^2\theta \geqslant \frac{1}{\nu}\frac{1}{F_Q\big(\hat\varrho_\mathrm{out}(\theta)\big)} = \frac 1\nu \frac 1{ds_Q^2/d\theta^2}\,.
\end{equation}

Aside of its utility for atomic interferometry and quantum metrology as a whole, the bound (\ref{eq:Quantum_CRLB}) also leads to formulation of a generalization of uncertainty principle. Note that QFI can be bounded by a following expression (I drop subscript ``out'' for convenience)
\begin{align}
F_Q\big(\hat\varrho(\theta)\big){}& = \left(\sum_i \frac{(\partial_\theta p_i)^2}{p_i} + 2\sum_{i,j}\frac{(p_i-p_j)^2}{p_i+p_j}|\langle i| \hat h|j\rangle|^2\right)\\
&\leqslant \sum_i \frac{(\partial_\theta p_i)^2}{p_i} + 4\langle\Delta^2\hat h\rangle_{\hat\varrho(\theta)}\, .
\end{align}
Here $\hat h$ was introduced in (\ref{eq:Atom_Inter_generator_h}) and $\langle\Delta^2\hat h\rangle_{\hat\varrho} = \mathrm{Tr}(\hat h^2\hat\varrho)-(\mathrm{Tr}(\hat h\hat\varrho))^2$ is a variance on state $\hat\varrho$. When this bound is combined with (\ref{eq:Quantum_CRLB}) we obtain
\begin{equation}
\nu\,\Delta^2\theta\,\left(\sum_i \frac{(\partial_\theta p_i)^2}{p_i} + 4\langle\Delta^2\hat h\rangle_{\hat\varrho(\theta)}\right) \geqslant 1
\end{equation}
The above inequality transition between a ``classical'' uncertainty principle, when $\hat h = 0$, which limits distinguishability of probability distributions, and a ``quantum'' uncertainty principle, when $\partial_\theta p_i=0$ for all $i$, which involves the generator $\hat h$ ``conjugated'' to phase $\theta$.

In summary, I showed that the precision of parameter estimation is tied to QFI which quantifies the susceptibility of the probe state to change induced by the interaction with measured system. This result is a starting point for the upcoming chapters where I describe how the efficiency of metrological task can be used for investigation of properties of quantum states.

\chapter{Dynamical entanglement}

In this chapter I discuss the role of quantum Fisher information as a criterion for useful particle entanglement and introduce the concept of dynamical entanglement.

\section[Quantum Fisher information as an efficiency criterion for entanglement detection]{Quantum Fisher information as an efficiency criterion for entanglement detection%
              \sectionmark{QFI as an efficiency criterion}}\label{sec:Dyn_Ent_QFI_is_crit}
\sectionmark{QFI as an efficiency criterion}

In this section I aim to show that quantum Fisher information (QFI) associated with {\it coherent transformation}, i.e. transformation which acts on each party in the same way, can serve as an entanglement criterion \cite{Smerzi_QFI}. Moreover, I will demonstrate that the value of QFI strongly relies upon the correlations between the parties forming the system, thus it can also be used as an indicator of the degree of non-classical correlations.

Consider a $N$-parties system that has been initialized in a separable state (as in Eq.~(\ref{sep_state})
\begin{equation}\label{eq:DynEnt_QFIasCrit_in_state}
\hat\varrho_{\mathrm{in}}=\hat\varrho_{\mathrm{separable}}=\sum_k\,P_k\,\hat\varrho^{(1)}_k\otimes\ldots\otimes\hat\varrho^{(N)}_k,
\end{equation}
where $0\leqslant P_k\leqslant1$ and $\sum_k P_k =1$. I assume that parties are copies of the same subsystem described by a Hilbert space of finite dimension. By doing so I can consider a scenario where parties are added or removed from the system. In such a case $N$ can be treated as a resource utilized for performing a given task. 

The system undergoes an interferometric transformation generated by {\it coherent Hamiltonian} $\hat H= \sum_{n=1}^N \hat h^{(n)}$, where $\hat h^{(n)}=\hat{\mathbf{1}}^{\otimes(n-1)}\otimes\hat h\otimes\hat{\mathbf{1}}^{\otimes(N-n)}$ are operators acting only in the subspace of party $n$. The transformation imprints information about the phase $\theta$ onto the initial state
\begin{equation}\label{eq:DynEnt_QFIasCrit_out_state}
\hat\varrho_{\mathrm{in}}\to \hat\varrho_{\mathrm{out}}(\theta) =e^{-i\theta \hat H}\hat\varrho_{\mathrm{in}}e^{i\theta\hat H}\, .
\end{equation}
Quantum Fisher information associated with this process is given by (see Eq.~(\ref{eq:Atom_Inter_QFI_bounds_Fisher}))
\begin{equation}\label{eq:DynEnt_QFIasCrit_QFI_def}
F_Q(\hat\varrho_{\mathrm{in}},\hat H) = 2\sum_{i,j}\frac{\left|\left\langle i^{\mathrm{out}}\right|\partial_\theta \hat\varrho_{\mathrm{out}}(\theta)\left|j^{\mathrm{out}}\right\rangle\right|^2}{p^{\mathrm{out}}_i+p^{\mathrm{out}}_j}\,,
\end{equation}
where $\left|i^{\mathrm{out}}\right\rangle$ and $p_i^{\mathrm{out}}$ are eigenstates and corresponding eigenvalues of $\hat\varrho_{\mathrm{out}}$ (note that the sum also includes cases when one of the eigenvalues is zero). The second argument of $F_Q$ reminds us that the transformation of input state is generated by the Hamiltonian $\hat H$. I proceed by finding the upper bound on the value of $F_Q(\hat\varrho_{\mathrm{separable}},\hat H)$ in order to establish a relation of type (\ref{eq:Intro_UsefulEnt_eff_crit}) which is a necessary condition for the QFI to be considered as a criterion for entanglement detection.

In section \ref{sec:Eff_Dist_metric} I showed that the quantum Fisher information is obtained by optimizing classical Fisher information (CFI) over all possible measurements. Consequently QFI inherits certain properties of CFI, including convexity as a function of density matrix. Therefore, when the input state is separable, i.e. it is a convex combination of product density matrices, as in Eq.~(\ref{eq:DynEnt_QFIasCrit_in_state}), I get a following inequality
\begin{align}
&F_Q(\hat\varrho_{\mathrm{separable}},\hat H) \leqslant \sum_k P_k F_Q(\hat\varrho^{(1)}_k\otimes\ldots\otimes\hat\varrho^{(N)}_k,\hat H)=\nonumber\\
& = 2\sum_k P_k \sum_{i,j}\frac{|\langle i^{\mathrm{out}}_k|\partial_\theta(e^{-i\theta\hat H}\hat\varrho^{(1)}_k\otimes\ldots\otimes\hat\varrho^{(N)}_ke^{i\theta\hat H}) |j^{\mathrm{out}}_k\rangle|^2}{p^k_i+p^k_j}\, .\label{eq:DynEnt_QFIasCrit_convex_bound}
\end{align}
Note that the unitary interferometric transformation does not change probabilities $P_k$ and eigenvalues $p_i^k$ of product density matrices . The derivative of the output state with respect to parameter $\theta$ is given by $\partial_\theta(e^{-i\theta \hat H} \hat\varrho\, e^{i\theta\hat H}) = -i[ \hat H , e^{-i\theta \hat H} \hat\varrho\, e^{i\theta\hat H} ]$. Substituting this result into Eq.~(\ref{eq:DynEnt_QFIasCrit_convex_bound}) I get
\begin{align}
F_Q{}&\leqslant 2\sum_k P_k\sum_{i,j}\frac{|\langle i^{\mathrm{out}}_k|[\hat H, e^{-i\theta\hat H}\hat\varrho^{(1)}_k\otimes\ldots\otimes\hat\varrho^{(N)}_ke^{i\theta\hat H}]|j_k^{\mathrm{out}}\rangle|^2}{p^k_i+p^k_j}=\nonumber\\
&=2\sum_k P_k\sum_{i\neq j}\frac{\left(p_i^k-p_j^k\right)^2}{p_i^k+p_j^k}|\langle i_k^{\mathrm{out}}|\hat H|j_k^{\mathrm{out}}\rangle|^2\leqslant\nonumber\\
&\leqslant 2\sum_k P_k\sum_{i\neq j}\left(p_i^k+p_j^k\right)|\langle i_k^{\mathrm{out}}|\hat H|j_k^{\mathrm{out}}\rangle|^2
=4\sum_k P_k\sum_{i}p_i^k\langle\Delta^2 \hat H\rangle_{i_k^{\mathrm{out}}}
\end{align}
where $\langle\Delta^2 \hat H\rangle_{i_k^{\mathrm{out}}}=\left\langle i_k^{\mathrm{out}}\right|\hat H^2\left|i_k^{\mathrm{out}}\right\rangle-(\left\langle i_k^{\mathrm{out}}\right|\hat H\left|i_k^{\mathrm{out}}\right\rangle)^2$ is a variance of $\hat H$ on state $\left|i_k^{\mathrm{out}}\right\rangle$. However, unitary transformation generated by coherent Hamiltonian do not entangle parties and state $\hat\varrho^{(1)}_k\otimes\ldots\otimes\hat\varrho^{(N)}_k$ remains separable. It follows that its eigenstates are product states of form $\left| i_k^{\mathrm{out}}\right\rangle=| \phi^{(1)}_{ik}\rangle\otimes\ldots\otimes| \phi^{(N)}_{ik}\rangle$ and the variance of Hamiltonian $\hat H$ breaks up into sum of variances of single-party Hamiltonians $\langle\Delta^2 \hat H\rangle_{i_k^{\mathrm{out}}}=\sum_{n=1}^N \langle\Delta^2 \hat h\rangle_{\phi^{(n)}_{ik}}$. The variance itself is bounded by the difference of extreme values of operator spectrum, namely $\langle\Delta^2\hat h\rangle \leqslant (\lambda_{\mathrm{max}}-\lambda_{\mathrm{min}})^2/4$. The bound on variance of single-party Hamiltonian together with the conditions $\sum_i p^{k}_i =1$ and $\sum_k P_k =1$ yields the inequality which ends the inquiry
\begin{equation}\label{eq:general_SNL}
F_Q(\hat\varrho_{\mathrm{separable}},\hat H) \leqslant N(\lambda_{\mathrm{max}}-\lambda_{\mathrm{min}})^2\, .
\end{equation}
I proceed to show that non-separable states can give QFI greater than the bound above. 

Suppose that the input state is now arbitrary. I start with Eq.~(\ref{eq:DynEnt_QFIasCrit_QFI_def}) and note that this expression can be bounded by the variance calculated with the whole density matrix
\begin{align}
F_Q(\hat\varrho_{\mathrm{in}},\hat H){}&=2\sum_{i,j}\frac{|\langle i^{\mathrm{out}}|\partial_\theta\hat\varrho_{\mathrm{out}}|j^{\mathrm{out}}\rangle|^2}{p^{\mathrm{out}}_i+p^{\mathrm{out}}_j}
=2\sum_{i\neq j}\frac{\left(p_i^{\mathrm{out}}-p_j^{\mathrm{out}}\right)^2}{p_i^{\mathrm{out}}+p_j^{\mathrm{out}}}|\langle i^{\mathrm{out}}|\hat H|j^{\mathrm{out}}\rangle|^2\nonumber\\
&\leqslant 2\sum_{i\neq j}\left(p_i^{\mathrm{out}}+p_j^{\mathrm{out}}\right)|\langle i^{\mathrm{out}}|\hat H|j^{\mathrm{out}}\rangle|^2
=4\sum_{i}p_i^{\mathrm{out}}\langle\Delta^2 \hat H\rangle_{i^{\mathrm{out}}}=\nonumber\\
&=4\mathrm{Tr}\Big( \hat H^2 \sum_i p_i^{\mathrm{out}} |i^{\mathrm{out}}\rangle\langle i^{\mathrm{out}}|\Big)
-4\sum_i p_i^{\mathrm{out}}\mathrm{Tr}\Big(\hat H|i^{\mathrm{out}}\rangle\langle i^{\mathrm{out}}|\Big)^2\nonumber\\
&\leqslant 4 \mathrm{Tr}(\hat H^2\hat\varrho_{\mathrm{out}}) - 4 \mathrm{Tr}(\hat H\hat\varrho_{\mathrm{out}})^2 = 4\langle\Delta^2 \hat H\rangle_{\varrho_{\mathrm{out}}}\label{eq:DynEnt_QFIasCrit_bound_by_var}
\end{align} 
In turn, the variance is bounded by the difference of maximal and minimal eigenvalues of $\hat H$ which are simply $N\lambda_{\mathrm{max}}$ and $N\lambda_{\mathrm{min}}$. Thus I obtain the ultimate bound on the QFI which cannot be surpassed by any input state
\begin{equation}\label{eq:general_HL}
F_Q(\hat\varrho_{\mathrm{in}},\hat H ) \leq N^2(\lambda_{\mathrm{max}}-\lambda_{\mathrm{min}})^2
\end{equation}
Note the difference between bounds (\ref{eq:general_SNL}) and (\ref{eq:general_HL}): separable states cannot exceed QFI that is proportional to number of parties $N$ while the ultimate bound scales with $N^2$ -- the square of number of parties. This shows that indeed there might exist some states which give QFI greater then separable bound.

As a last step, I identify the family of states which saturate the ultimate bound to be of ``Schr\"{o}dinger cat'' type
\begin{equation}\label{eq:general_Cat}
\left|\mathrm{Cat}_\varphi\right\rangle =\frac{|\lambda_{\mathrm{max}}\rangle^{\otimes N}+e^{i\varphi}|\lambda_{\mathrm{min}}\rangle^{\otimes N}}{\sqrt 2},
\end{equation}
where $\varphi$ is an arbitrary phase and $|\lambda_{\mathrm{max}/\mathrm{min}}\rangle$ are the eigenstates of single-party Hamiltonians $\hat h^{(n)}$. Indeed, the QFI for such a pure state yields
\begin{align}
F_Q{}&(|\mathrm{Cat}_\varphi\rangle,\hat H)= 4\langle\Delta^2\hat H\rangle_{\mathrm{Cat}_\varphi} =4\langle\mathrm{Cat}_\varphi|\hat H^2|\mathrm{Cat}_\varphi\rangle -4\langle\mathrm{Cat}_\varphi|\hat H|\mathrm{Cat}_\varphi\rangle^2=\nonumber\\
&= 2(N^2\lambda_{\mathrm{max}}^2 +N^2\lambda_{\mathrm{min}}^2) -(N\lambda_{\mathrm{max}}+N\lambda_{\mathrm{min}})^2=\nonumber\\
&=N^2\lambda_{\mathrm{max}}^2+N^2\lambda_{\mathrm{min}}^2-2N^2\lambda_{\mathrm{max}}\lambda_{\mathrm{min}}=N^2(\lambda_{\mathrm{max}}-\lambda_{\mathrm{min}})^2\label{eq:Cat_is_max_ent}
\end{align}
This concludes the proof but also is a pleasing result in itself. Schr\"{o}dinger cat states, which in case of systems composed of qubits are also called NOON states or GHZ states for $N=3$, are considered as maximally entangled states (see Ref. \cite{Plenio_ent_measures} for extensive discussion on the problem of multipartite entanglement measures). Therefore, since maximally entangled state such as cat states can yield a tremendous gain of order $N$ over separable states, QFI is also a good indicator for a degree of non-classical correlations between parties.

The relation established by the Cram\'{e}r-Rao lower bound (see Eq.~(\ref{eq:Atom_Inter_CRLB})) asserts that QFI is a well defined efficiency criterion associated with task of parameter estimation in interferometric transformation generated by $\hat H$. In this context the efficiency introduced in Sec. \ref{sec:Intro_Useful_Ent} is the inverse of the error of the estimation divided by the number of experiment repetitions $\nu$
\begin{equation}
\mathcal Q^{(\text{estimation of $\theta$})}(\hat\varrho_{\mathrm{in}}) = \sqrt{\frac{\nu}{\overline{\Delta^2\theta}}} \leqslant \sqrt{F_Q(\hat\varrho_{\mathrm{in}},\hat H )}\,.
\end{equation}
For a separable state the efficiency is no greater then $\sqrt N |\lambda_{\mathrm{max}}-\lambda_{\mathrm{min}}|$, or the error scales as $\Delta\theta \sim 1/{\sqrt{N}}$ -- the precision reaches the shot noise limit. Certain entangled states allow for beating this limit and for maximally entangled Schroedinger cat states the improvement is of order $\Delta\theta \sim 1/ N$, which is particularly lucrative for large number of parties. The ultimate bound (\ref{eq:general_HL}) which defines the maximal achievable precision is called the \emph{Heisenberg limit} (HL).

Quantum Fisher information backed up with the context of atomic interferometry will now be the centerpiece of the upcoming discussions. I will use this powerful concept to analyze and classify non-classical correlations present in ultra-cold atom systems.

\section{Introducing dynamical entanglement}\label{sec:Dyn_Ent_Introducing_dyn_ent}

The troublesome aspect of using the QFI as a measure of entanglement is the requirement for associating it with some transformation which imprints parameter to be estimated. Of course, the context of atomic interferometry provides us with the choice of transformation. Nevertheless, still it would be desirable to make an attempt to find some generic type (or types) of transformation which would justify the use of QFI on its own.

In chapter \ref{ch:Atomic_interferometer} I argued that treating atoms as qubits provides a satisfactory description of ultra-cold atom systems used for ongoing research on quantum entanglement. By restricting the discussion to qubit system I will be able to detach QFI from the context of interferometric experiment by removing the ambiguity of choice for coherent transformation. Then, the QFI becomes a function of the state only and it can be treated as a quantity characterizing its properties, including non-classical correlations. However, the relation between QFI and atomic interferometry is too valuable and it would be unwise to discard it all together. As I will argue, the direct correspondence between this generic QFI and arbitrary interferometric transformation can be easily restored.

From now on I shall assume that parties constituting the system of interest are particles living in a two dimensional Hilbert spaces. This restriction simplify the problem significantly. In case of qubit systems single-particle transformations, which build up
coherent transformations, are mapped onto rotations of spin $1/2$ system generated by triple of spin operators $\hat J^{(n)}_i=\hat\sigma^{(n)}_i/2$, with $i=x,y,z$ and $\hat\sigma^{(n)}_i$ are Pauli matrices of $n$-th qubit. Any given unitary transformation of a qubit can be characterized by an angle $\theta$ and the unit vector ${\mathbf{n}}$ parallel to the axis of rotation in a following way: $e^{-i\theta\, {\mathbf{n}}\cdot\hat{\mathbf{J}}}$. However, the reference frame can be changed at will and it can always be chosen in such a way that the new $z$-axis coincides with the axis of rotation ${\mathbf{n}}$. Therefore, each transformation can now be considered as a family of rotations about the $z$ axis characterized by a single parameter -- the angle $\theta$. Now imagine that such a generic transformation describes an interferometer while the angle $\theta$ is an unknown parameter to be estimated. The quantum Fisher information associated with the precision of this fictional estimation task is given by
\begin{align}
F_Q(\hat\varrho_{\mathrm{in}},\hat J_z){}&=2\sum_{i,j}\frac{|\langle i^{\mathrm{out}}|[\hat J_z,\hat\varrho_{\mathrm{out}}]|j^{\mathrm{out}}\rangle|^2}{p^{\mathrm{out}}_i+p^{\mathrm{out}}_j}
= 2\sum_{i\neq j}\frac{(p_i-p_j)^2}{p_i+p_j}|\langle i|\hat J_z|j\rangle|^2 =\nonumber\\
&= 4\sum_i p_i\langle\Delta^2 \hat J_z\rangle_i -8\sum_{i\neq j}\frac{p_ip_j}{p_i+p_j}|\langle i|\hat J_z|j\rangle|^2\,,
\end{align}
where $\hat\varrho_{\mathrm{out}}=e^{-i\theta\hat J_z}\hat\varrho_{\mathrm{in}}e^{i\theta\hat J_z}$ and $p^{\mathrm{out}}_i$ are its eigenvalues with corresponding eigenstates $| i^{\mathrm{out}}\rangle$. In order to obtain this equation I used the fact that unitary transformation does not change eigenvalues of density matrix ($p_i^{\mathrm{out}}=p_i$) and $\langle i^{\mathrm{out}}|\hat J_z|j^{\mathrm{out}}\rangle = \langle i|e^{i\theta\hat J_z}\hat J_z e^{-i\theta  \hat J_z}|j\rangle = \langle i|\hat J_z|j\rangle$, where $|i\rangle$ is an eigenstate of $\hat\varrho_{\mathrm{in}}$ corresponding to $p_i$. As a result, $F_Q$ \emph{does not} depend on the value of $\theta$ and the explicit form of the transformation $e^{-i\theta \hat J_z}$ becomes superfluous. Therefore, I introduce new function of state inspired by QFI, which stands on its own, not relaying on a context of interferometric experiment
\begin{equation}\label{eq:dyn_susceptibility}
\mathcal F (\hat\varrho)=4\sum_i p_i\langle\Delta^2 \hat J_z\rangle_i - 8\sum_{i\neq j}\frac{p_i p_j}{p_i+p_j}|\langle i|\hat J_z|j\rangle|^2\,.
\end{equation}
I shall refer to $\mathcal F$ as \emph{dynamical susceptibility}. The name is inspired by the interpretation of QFI as a susceptibility of the state to change due to driving by the system Hamiltonian, as it was discussed in Ch.~\ref{ch:Eff}.

Since $\mathcal F$ equals QFI associated with interferometer generated by $e^{-i\theta \hat J_z}$ all the properties derived in previous section carry over. Hence the dynamical susceptibility can also serve as a criterion for entanglement detection. Since spin operator $\hat\sigma_z/2$ has only two eigenvalues $+1/2$ and $-1/2$ a general SNL bound (\ref{eq:general_SNL}) reduces to
\begin{equation}
\mathcal F(\hat\varrho_{\mathrm{separable}})\leqslant N\,.
\end{equation}
The ultimate HL bound (\ref{eq:general_HL}) in this case is
\begin{equation}
\mathcal F(\hat\varrho) \leqslant N^2\, .
\end{equation}
Similarly to criterion based on QFI, dynamical susceptibility will serve as the mean to analyze and quantify non-classical correlations in ultra-cold atoms systems for which qubit approximation is valid. I will say that a state $\hat\varrho$ which gives $\mathcal F(\hat\varrho)>N$ is \emph{dynamically entangled} to the degree $\mathcal F(\hat\varrho)$.

\section[Relation between useful and dynamical entanglement]{Relation between useful and dynamical entanglement%
              \sectionmark{Useful and dynamical entanglement}}\label{sec:Dyn_Ent_QFI_vs_dyn_sus}
\sectionmark{Useful and dynamical entanglement}

The aim of this section is to establish a procedure to relate the dynamical susceptibility with an interferometric experiment. To achieve this goal I need to be able to express the QFI associated with a task of estimating a value of phase imprinted by an arbitrary interferometric transformation in terms of $\mathcal F$.

A general (unitary) interferometric transformation can be parametrized by a sequence of rotations with one of the angles being unknown $\theta$
\begin{align}
\hat U_\theta {}&= e^{-i\boldsymbol{\alpha}_1\cdot\hat{\mathbf J}}\ldots e^{-i\boldsymbol{\alpha}_k\cdot\hat{\mathbf J}}e^{-i \theta\, {\mathbf{n}}\cdot\hat{\mathbf J}}
e^{-i\boldsymbol{\beta_1}\cdot\hat{\mathbf{J}}}\ldots e^{-i\boldsymbol{\beta_m}\cdot\hat{\mathbf{J}}} =\nonumber\\
&=e^{-i\boldsymbol{\alpha}\cdot\hat{\mathbf J}}e^{-i \theta\, {\mathbf{n}}\cdot\hat{\mathbf J}}e^{-i\boldsymbol{\beta}\cdot\hat{\mathbf{J}}}\equiv \hat U_\alpha e^{-i \theta\, {\mathbf{n}}\cdot\hat{\mathbf J}}\hat U_\beta
\end{align}
Here $\hat U_{\alpha/\beta}$ is a composition of rotations (therefore being rotation itself) preceding/following the phase imprint. The task is to estimate the value of $\theta$. As usual, the precision of estimation is bounded by QFI, which in this case reads
\begin{align}
F_Q(\hat\varrho_{\mathrm{in}},\hat U_\theta) {}&= 2\sum_{i,j}\frac{|\langle i^{\mathrm{out}}|\partial_\theta \hat\varrho_{\mathrm{out}}(\theta)|j^{\mathrm{out}}\rangle|^2}{p_i+p_j}
=2\sum_{i,j}\frac{|\langle i^{\mathrm{out}}|[{\mathbf{n}}'\cdot\hat{\mathbf{J}},\hat\varrho_{\mathrm{out}}]|j^{\mathrm{out}}\rangle|^2}{p_i+p_j}\nonumber\\
&=2\sum_{i\neq j}\frac{(p_i-p_j)^2}{p_i+p_j}|\langle i^{\mathrm{out}}|{\mathbf{n}}'\cdot\hat{\mathbf{J}}|j^{\mathrm{out}}\rangle|^2
\end{align}
Here ${\mathbf{n}}'=\mathbf{R}^{-1}_\alpha{\mathbf{n}}$ is a unit vector ${\mathbf{n}}$ rotated counter-clockwise by angle $|\boldsymbol\alpha|$ about the axis $\boldsymbol\alpha$ by means of rotation matrix $\mathbf R_\alpha$. This relation follows from\footnote{
One can consider even more general interferometric transformation, where the phase $\theta$ is imprinted simultaneous with another rotation, i.e. 
\begin{align}
\hat U_\theta{}&=\hat U_\alpha \mathcal T\!\exp\!\left[ -i \!\!\int_0^\tau \!\!\!\!ds \left(\frac \theta\tau \mathbf{n} + \boldsymbol{\gamma}(s)\right)\!\cdot\hat{\mathbf J}\right]\hat U_\beta = 
\hat U_\alpha \hat U_\gamma \mathcal T\!\exp\!\left(\!\!-i \theta\!\! \int_0^1 \!\!\!\!ds \,\hat U_\gamma(\tau s)(\mathbf{n}\cdot\hat{\mathbf{J}})\hat U_\gamma^\dagger(\tau s)\!\right)\hat U_\beta\nonumber
\end{align}
where $\mathcal T\exp$ is a time-ordered exponential and $\hat U_\gamma = \mathcal T\exp\left(\!-i\int_0^\tau ds\, \boldsymbol{\gamma}(s)\cdot\hat{\mathbf{J}}\right)$. In such a case we have $\partial_\theta\hat\varrho_{\mathrm{out}}=-i[(\mathbf{R}^{-1}_\alpha\mathbf{R}^{-1}_\gamma\widetilde{\mathbf{n}})\cdot\hat{\mathbf{J}},\hat\varrho_{\mathrm{out}} ]\nonumber
$, with $\widetilde{\mathbf{n}}\cdot\hat{\mathbf{J}}=\int_0^1 ds\, \hat U_\gamma(\tau s)(\mathbf{n}\cdot\hat{\mathbf{J}})\hat U_\gamma^\dagger(\tau s)$.
}
\begin{align}
\partial_\theta \hat\varrho_{\mathrm{out}}{}&=(\partial_\theta  \hat U_\theta)\hat\varrho_{\mathrm{in}}\hat U_\theta^\dagger+\hat U_\theta\hat\varrho_{\mathrm{in}}(\partial_\theta  \hat U_\theta^\dagger)=\nonumber\\
&=(-i\, {\mathbf{n}}\cdot(\hat U_\alpha  \hat{\mathbf J}\hat U_\alpha^\dagger))\hat\varrho_{\mathrm{out}}+\hat\varrho_{\mathrm{out}}(i \,{\mathbf{n}}\cdot(\hat U_\alpha  \hat{\mathbf J}\hat U_\alpha^\dagger))=\nonumber\\
&=-i [ \,(\mathbf{R}^{-1}_\alpha{\mathbf{n}})\cdot\hat{\mathbf J},\hat\varrho_{\mathrm{out}} ] = -i [ {\mathbf{n}}'\cdot\hat{\mathbf{J}}, \hat\varrho_{\mathrm{out}} ]\, .
\end{align}
Further manipulations confirm that $F_Q$ is independent of $\theta$ 
\begin{align}
F_Q &=2\sum_{i\neq j}\frac{(p_i-p_j)^2}{p_i+p_j}|\langle i^{\mathrm{out}}|{\mathbf{n}}'\cdot\hat{\mathbf{J}}|j^{\mathrm{out}}\rangle|^2 =
2\sum_{i\neq j}\frac{(p_i-p_j)^2}{p_i+p_j}|\langle i|{\mathbf{n}}'\cdot(\hat U_\beta^\dagger\hat{\mathbf{J}}\hat U_\beta)|j\rangle|^2\nonumber\\
&=2\sum_{i\neq j}\frac{(p_i-p_j)^2}{p_i+p_j}|\langle i|(\mathbf R_\beta{\mathbf{n}}')\cdot\hat{\mathbf{J}}|j\rangle|^2 \equiv 
2\sum_{i\neq j}\frac{(p_i-p_j)^2}{p_i+p_j}|\langle i|{\mathbf{n}}''\cdot\hat{\mathbf{J}}|j\rangle|^2\label{eq:DynEnt_DynEnt_QFI_to_susc}
\end{align}
Where $\mathbf R_\beta$ is the clockwise rotation about axis $\boldsymbol\beta$ by an angle $|\boldsymbol \beta|$.The final step before I can relate $F_Q$ to dynamical susceptibility is to introduce a rotation $\hat R$ such that $\hat R({\mathbf{n}}''\cdot\hat{\mathbf J})\hat R^\dagger=\hat J_z$. This transformation always exist and can be applied in (\ref{eq:DynEnt_DynEnt_QFI_to_susc})
\begin{align}
&F_Q(\hat\varrho_{\mathrm{in}},\hat U_\theta) =2\sum_{i\neq j}\frac{(p_i-p_j)^2}{p_i+p_j}|\langle i|\hat R^\dagger\hat R({\mathbf{n}}''\cdot\hat{\mathbf{J}})\hat R^\dagger\hat R|j\rangle|^2=\nonumber\\
&=2\sum_{i\neq j}\frac{(p_i-p_j)^2}{p_i+p_j}|(\langle i|\hat R^\dagger)\hat J_z(\hat R|j\rangle)|^2 =\nonumber\\
&=4\sum_i p_i \langle\Delta^2\hat J_z\rangle_{\hat R|i\rangle}-8\sum_{i\neq j}\frac{p_ip_j}{p_i+p_j}|(\langle i|\hat R^\dagger)\hat J_z(\hat R|j\rangle)|^2 =\mathcal F(\hat R\hat\varrho_{\mathrm{in}}\hat R^\dagger)\, .\label{eq:Dyn_Ent_QFI_is_susc}
\end{align}
Hence, any usefully entangled state can be transformed by means of rotation (which \emph{does not} introduce any entanglement) into state that is dynamically entangled to equal degree. The particular transformation is uniquely determined by the interferometric sequence in question.

The reciprocal relation also exist but is not unique. Working out the derivation backwards I obtain the following relation valid for arbitrary rotation $\hat R$
\begin{equation}\label{eq:Dyn_Ent_QFI_v_susc_reciprocal}
\mathcal F(\hat\varrho) =F_Q(\hat R\hat\varrho\hat R^\dagger, \hat U_\alpha \hat U_n e^{-i\theta\hat{J}_z}\hat U^\dagger_n \hat U_\beta )\, .
\end{equation}
Rotations $\hat U_\alpha$ and $\hat U_\beta$ which precede and follow imprinting of the phase along axis set by $\hat U_n$ satisfy condition $\hat R=\hat U_n^\dagger\hat U_\alpha^\dagger\hat U_\beta$. That is, these transformations are decomposition of $\hat R$ into sequence of rotations. Such a decomposition can be done in infinite number of ways. The only constrains can be imposed by a practical concerns laid down by a context of experimental setup.

Finally, I consider a case of non-unitary interferometric transformation. Such transformation not only rotates the eigenstates of density matrix but also manipulates its eigenvalues which describe classical ignorance of the observer regarding the preparation of the state. In principle, the information about the parameter to be estimated can be drawn form both of these processes as it is demonstrated by the QFI for a general non-unitary transformation which depends on the parameter $\theta$
\begin{equation}\label{eq:DynEnt_DynEnt_non-uni_QFI}
F_Q^{\mathrm{non-unitary}} = 2\sum_{i\neq j}\frac{(p^{\mathrm{out}}_i-p^{\mathrm{out}}_j)^2}{p^{\mathrm{out}}_i+p^{\mathrm{out}}_j}|\langle i^{\mathrm{out}}|\hat h|j^{\mathrm{out}}\rangle|^2 +
 \sum_i \frac{(\partial_\theta p^{\mathrm{out}}_i)^2}{p^{\mathrm{out}}_i}
\end{equation}
where the matrix elements of hermitian operator $\hat h$ describing the ``unitary'' part is defined as a generator of the infinitesimal transformation $|i^{\mathrm{out}}(\theta+d\theta)\rangle = e^{-i d\theta \hat h}|i^{\mathrm{out}}(\theta)\rangle$ (as in Eq~(\ref{eq:Atom_Inter_generator_h})). In general $F_Q^{\mathrm{non-unitary}}$ is not a proper efficiency criterion since relation such as (\ref{eq:general_SNL}) does not exist. This is because only the ``unitary'' part depends on non-classical correlations present in the system. Therefore, QFI associated with transformations that are not unitary can be biased by the additional ``classical'' part. The dynamical susceptibility is not burdened by this flaw, as by its definition, it ``picks'' only the unitary part of the QFI. 
By choosing $\hat R$ such that the generator $\hat h$ is transformed into $\hat J_z$ we obtain the following relation
\begin{equation}
F_Q^{\mathrm{non-unitary}} = \mathcal F(\hat R\hat\varrho_{\mathrm{out}}\hat R^\dagger) + \sum_i\frac{(\partial_\theta p^{\mathrm{out}}_i)^2}{p^{\mathrm{out}}_i}\, .
\end{equation}
We see that the dynamical susceptibility indeed describes the ``unitary'' part of $F_Q^{\mathrm{non-unitary}}$. Note that $\mathcal F$ no longer depends on the input state $\hat\varrho_{\mathrm{in}}$ only. Instead it is a function of the output state $\hat\varrho_{\mathrm{out}}$ which means that it depends on the type of transformation, the duration of the experiment as well as the initial state. This is not a surprise because in general non-unitary interferometric transformations include the process of decoherence which destroys correlations within the system. Therefore, the degree of entanglement is no longer conserved and this is reflected in the dependence of dynamical susceptibility on the details of the process.

Relation between dynamical susceptibility and the QFI associated with a given transformation discussed above can be understood as follows. The interferometric experiment consists of three stages: state preparation, phase imprint and measurement. So far we focused on the phase imprint stage represented by the interferometric transformation $\hat U_\theta = \hat U_\alpha \hat U_n e^{-i\theta\hat J_z}\hat U_n^\dagger \hat U_\beta$. We took it for granted that the preparation stage already took place and the result was the input state $\hat\varrho_{\mathrm{in}}$. Also we were never concerned with measurements which follow the imprint because QFI is optimized over all possible realization of this stage. However, the boundaries between the stages are not clear-cut. Formally one can regard the transformations $\hat U^\dagger_n\hat U_\beta$ and $\hat U_\alpha\hat U_n$  which precede and follow the phase imprint  as a part of state preparation and a measurement stage, then $e^{-i\theta\hat J_z}$ becomes the new interferometer. In fact, the relation (\ref{eq:Dyn_Ent_QFI_is_susc}) is an example of this formal division. Indeed, dynamical susceptibility is equivalent to the QFI associated with a generic interferometer $e^{-i\theta\hat J_z}$ and the transformation $\hat R$ represents the net result of transformations preceding and following the phase imprint.

\section{Dynamical entanglement and spin squeezing}

The {\it spin squeezing} is another efficiency criterion for entanglement which is related to precision of two-mode interferometer. Spin squeezing parameter $\xi_S$ is defined as the ratio of the error of the phase estimator derived from the population imbalance measurement in the Mach-Zehnder interferometer to the SNL precision. Before I compare criterion based on QFI and the spin squeezing I will review the derivation of $\xi_S$.


In the language of pseudo-spin, the Mach-Zehnder interferometer is represented by the following sequence of rotations
\begin{equation}
\hat U^{(\mathrm{MZ})}_\theta = e^{i\frac \pi 2 \hat J_x}e^{-i\theta \hat J_z}e^{-i\frac \pi 2 \hat J_x}
\end{equation}
and as usual $\theta$ is an unknown phase to be estimated. The rotations about $x$-axis which precede and follow the phase imprint represent beam-splitter operations which enable ``mixing'' of atoms occupying the modes of interferometer. The value of $\theta$ can now be inferred from the oscillatory behavior of population imbalance between the two modes of the output state $\hat\varrho_\mathrm{out}=\hat U^{(\mathrm{MZ})}_\theta\hat\varrho_\mathrm{in}\hat U^{(\mathrm{MZ})\dagger}_\theta$. According to what I showed in Sec.~\ref{sec:Dyn_Ent_Pseudo-spin}, the $z$ component of the spin operator is an observable associated with the difference of mode populations. We can easily verify that the expectation value of $\hat J_z$ indeed depends on the parameter $\theta$
\begin{align}
\langle \hat J_z\rangle_\mathrm{out} {}&= \mathrm{Tr}\big[\hat J_z (\hat U^{(\mathrm{MZ})}_\theta\hat\varrho_\mathrm{in}\hat U^{(\mathrm{MZ})\dagger}_\theta)\big] 
= \mathrm{Tr}\big[(\hat U^{(\mathrm{MZ})\dagger}_\theta\hat J_z\hat U^{(\mathrm{MZ})}_\theta)\hat\varrho_\mathrm{in}\big]=\nonumber\\
&=\cos\theta\langle\hat J_z\rangle_\mathrm{in}+\sin\theta\langle\hat J_x\rangle_\mathrm{in}\equiv f(\theta)
\end{align}
The estimator for parameter $\theta$ can be chosen so that $\overline{\theta_\mathrm{est}} = f^{-1}(\langle\hat J_z\rangle_\mathrm{out})$. The uncertainty of this estimator is determined by the variance of the population imbalance and it can be calculated using the error propagation formula:
\begin{align}
\overline{\Delta^2\theta_\mathrm{est}}{}& = \frac 1\nu \left|\frac{\partial f^{-1}}{\partial \langle\hat J_z\rangle_\mathrm{out}}\right|^2 \langle\Delta^2\hat J_z\rangle_\mathrm{out}
=\frac 1\nu\frac{\langle\Delta^2\hat J_z\rangle_\mathrm{out}}{\left|\partial \langle\hat J_z\rangle_\mathrm{out}/\partial\theta\right|^2} =\nonumber\\
&=\frac 1\nu \frac{\langle\Delta^2\hat J_z\rangle_\mathrm{out}}{\left| -\sin\theta\langle\hat J_z\rangle_\mathrm{in}+\cos\theta\langle\hat J_x\rangle_\mathrm{in}\right|^2}\,,
\end{align}
where $\nu$ is a number of experiment repetitions. The spin squeezing parameter is defined as
\begin{equation}\label{eq:Dyn_Ent_OtherCrit_squeezing_def}
\xi_S^2(\hat\varrho_\mathrm{in}) \equiv \nu\left(\frac{\lim_{\theta\to0}\overline{\Delta^2\theta_\mathrm{est}}}{1/N}\right) =  N\frac{\langle\Delta^2\hat J_z\rangle_\mathrm{in}}{|\langle\hat J_x\rangle_\mathrm{in}|^2}
\end{equation}
Spin squeezing is also a criterion for particle entanglement \cite{Sorensen_01,Toth_07,Wang_03}:
\begin{equation}
\frac 1{\xi_S^2(\hat\varrho_\mathrm{separable})} \leqslant 1\quad,\quad \frac 1{\xi_S^2(\hat\varrho)} \leqslant N
\end{equation}

The main advantage of the spin squeezing is how relatively easy it is to assess $\xi_S$ in most experimental settings. The variance of population imbalance ($\langle \Delta^2\hat J_z\rangle$) can be deduced from standard particle number measurements which are always setup so that they can resolve between different modes of the interferometer. The denominator, $|\langle\hat J_x\rangle|$ measures the coherence between two modes. It can be obtained by examining the visibility of interference fringes observed after the ``mixing'' of matter waves from each mode. For example, this can be achieved by releasing the BEC from the trapping potential and allowing the atomic clouds to freely expand and overlap thus forming the interference pattern. High resolution imagining techniques allow for very precise measurements of the structure of this pattern \cite{Oberthaler_08}. In contrast, extracting QFI is also possible but it requires highly sophisticated methods \cite{Smerzi_measured_QFI}.

The ``standard'' choice of axes defining $\xi_S$ came out naturally as a direct result of particular estimation strategy adopted for the Mach-Zehnder interferometer. The first generalization of the spin squeezing parameter is to replace the denominator with the square of the length of projection of average spin vector onto the plane perpendicular to the $z$ axis:
\begin{equation}
\widetilde{\xi}_S \equiv \sqrt{N\frac{\langle\Delta^2\hat J_z\rangle_{\mathrm{in}}}{|\langle\hat J_x\rangle_\mathrm{in}|^2 + |\langle\hat J_y\rangle_\mathrm{in}|^2}}\,,
\end{equation}
A more general definition which explicitly points out the flexibility in rearranging the directions of squeezing is given by
\begin{equation}
\xi'_S(\hat\varrho,\mathbf{n}) \equiv \sqrt{N\frac{\langle \Delta^2(\mathbf{n}\cdot\hat{\mathbf{J}})\rangle_{\hat\varrho}}{||\langle\hat{\mathbf{J}}\rangle_\perp||^2}} \,.
\end{equation}
Here $||\langle\hat{\mathbf{J}}\rangle_\perp||$ denotes the length of vector $\langle\hat{\mathbf{J}}\rangle_\perp$ which is a projection of the average spin vector $\langle \hat{\mathbf{J}}\rangle_{\hat\varrho}$ onto plane perpendicular to $\mathbf{n}$. Similarly to relations between dynamically and usefully entangled states, a state which is spin squeezed in respect to one direction $\mathbf{n}_1$ is related to state squeezed in direction $\mathbf{n}_2$ through unitary rotation
\begin{equation}
\xi_S'(\hat\varrho,\mathbf{n}_1) =\xi_S'(\hat R\hat\varrho\hat R^\dagger,\mathbf{n}_2)\,,
\end{equation}
where $\hat R(\mathbf{n}_1\cdot\hat{\mathbf{J}})\hat R^\dagger=\mathbf{n}_2\cdot\hat{\mathbf{J}}$. In particular $\widetilde\xi_S(\hat\varrho) = \xi'_S(\hat\varrho,\mathbf{e}_z)=\xi_S(\hat\varrho)$ if the axes of reference frame are chosen so that $\mathbf{e}_x \parallel \langle\hat{\mathbf{J}}\rangle_\perp$. These relations are analogical to relations I have established between QFI and dynamical susceptibility which is essentially the QFI for ``standard'' interferometer. However, the components of pseudo-spin do not refer to direction in real space and $\hat J_z$ is in fact the population imbalance operator while $\hat J_x$ is related to visibility of interference fringes. Passing to the rotated frame of reference might sabotage the main advantage of spin squeezing parameter: the ability to measure it in experiment. Therefore, in almost all circumstances the states which are spin squeezed according to ``standard'' parameter $\xi_S$ (i.e. such $\hat\varrho$ that $\xi_S^2(\hat\varrho)<1$) are most desirable for practical uses. 

From theoretical point of view, dynamical susceptibility (and QFI) is more attractive then spin squeezing. Since $\xi_S$ is a variance of particular estimator it is bounded from below by the inverse of QFI associated with Mach-Zehnder interferometer
\begin{equation}
\overline{\Delta^2\theta_\mathrm{est}} =\frac1\nu\frac {\xi_S^2(\hat\varrho_\mathrm{in})}{N} \geqslant \frac 1\nu\frac 1{F_Q(\hat\varrho_\mathrm{in},\hat U^{(\mathrm{MZ})}_\theta)}\Rightarrow \frac{F_Q(\hat\varrho_\mathrm{in},\hat U^{(\mathrm{MZ})}_\theta)}{N} \geqslant \frac 1{\xi^2_S(\hat\varrho_\mathrm{in})}
\end{equation}
Therefore, QFI, and by extension dynamical susceptibility, detects more types of entangled states then spin squeezing \cite{Smerzi_measured_QFI}. The on-site atom-atom interaction (as it is the case for atomic interferometers realized in double well setup) is the natural source of spin squeezed states (i.e. states for which $\xi_S^2 < 1$ is satisfied). Also this type of states have the greatest overlap with usefully and dynamically entangled state detected by QFI or dynamical susceptibility. Nevertheless the overlap is not perfect. In \cite{Smerzi_QFI} authors carry out a detailed analysis of this family of states and showed that indeed QFI criterion is able to detect more entangled states then spin squeezing.

The advantage of methods for entanglement classification and analysis based on QFI is most explicit when one considers states created in atomic collisions (as in the case of twin-beam type of atomic interferometers). The characteristic feature of this type of states is the vanishing of the average spin vector $\langle \hat{\mathbf{J}}\rangle$ \cite{Wasak_twin_beam_14}. This leads to zero visibility of interference fringes which renders the spin squeezing parameter undetermined because the denominator in Eq.~(\ref{eq:Dyn_Ent_OtherCrit_squeezing_def}) equals zero. Similar problem is encountered in case of Schr\"{o}dinger cat states. This is indeed a serious drawback since these states are considered to be very strongly entangled. It seems reasonable to stipulate that the analysis of this type of non-classical correlations might be the key to understand the nature of entanglement.

\chapter{Physical interpretation of dynamical entanglement}
\chaptermark{Physical interpretation of entanglement}

In this chapter, I employ the dynamical susceptibility, its interpretation as a measure of the susceptibility of state to change, and the equivalence between bosonic qubit- and spin-systems to discuss the possible physical interpretation of particle entanglement encountered in cold-atom systems.

\section{The role of particle indistinguishability}\label{sec:Dyn_Ent_Indistinct}

The key aspect of ultra-cold atom system is the indistinguishability of bosons which constitute it. It is of great importance to analyze this contribution to overall non-classical correlations present in these systems. To this end I will exploit the correspondence between system composed of $N$ qubits and a system composed of a single pseudo-particle with spin degree of freedom. Initially I will not assume that qubits are identical bosons and I will investigate the changes in the degree of dynamical entanglement when the indistinguishability is imposed. By doing so I will establish important relation between the spin of pseudo-particle, entanglement and the indistinguishability of particles.

In Sec. \ref{sec:Dyn_Ent_QFI_is_crit} I showed that no state can surpass the Heisenberg limit (HL) of dynamical entanglement. I have also found the class of Schr\"{o}dinger cat states which reach this limit. Now I shall examine the role of particle indistinguishability in attaining the ultimate HL. I start with the following bound on dynamical susceptibility
\begin{align}
\mathcal F (\hat\varrho) {}&= 4\sum_ip_i\langle\Delta^2\hat J_z\rangle_i-8\sum_{i\neq j}\frac{p_ip_j}{p_i+p_j}|\langle i|\hat J_z|j\rangle|^2 \nonumber\\
&\leqslant 4\sum_ip_i\langle\Delta^2\hat J_z\rangle_i \leqslant \sum_i p_i \langle i|\hat J_z^2|i\rangle\,.\label{eq:DynEnt_Indist_dyn_sus}
\end{align}
With this bound I can utilize the equivalence between the qubit- and the spin-system. On one hand, eigenstates $|i\rangle$ can be written as a superposition of products of single-qubit basis states, i.e. $|i\rangle = \sum c^{(i)}_{\sigma_1\ldots\sigma_N}|\sigma_1\rangle\otimes\ldots\otimes|\sigma_N\rangle$, where $\sigma_n =\uparrow,\downarrow$.  On the other hand, each qubit is equivalent to spin $1/2$ particle and the basis states correspond to spin eigenstates: $\left|\uparrow/\downarrow\right\rangle \leftrightarrow |s=\frac 12, m=\pm\frac 12\rangle$. Since states $|i\rangle$ are superpositions of outer products of $N$ spin eigenstates they can be expanded in the basis of irreducible representation of the rotation group, i.e. the basis of total angular momentum, 
\begin{align}
  |i\rangle {}&= \!\!\sum_{m_1,\ldots,m_N}\!\! c^{(i)}_{m_1\ldots m_N}\left|\tfrac 12, m_1\right\rangle\otimes\ldots\otimes\left|\tfrac 12,m_N\right\rangle =\nonumber\\
  &=\sum_{ J=0 }^{\frac N 2} \sum_{\nu}\sum_{M=-J}^J \langle J,M;\nu\,|i\rangle   | J,M;\nu\rangle\equiv \sum_{ J =0 }^{\frac N2}  C_{J}^{(i)}  | \psi_{J}^{(i)} \rangle\,.\label{eq:DynEnt_Indist_J_decomposition}
\end{align}
States $| \psi_{J}^{(i)} \rangle$ are normalized projections of $|i\rangle$ on the subspace with the total angular momentum $J$, while $\nu$ is a degeneracy index labeling representations with the same $J$. Note that in general the sum over total angular momenta $J$ ranges from $0$ to $N/2$. 

Subspaces with definite $J$ are invariant under rotations, thus
the expectation value of $\hat{J}_z^2$ on the state $|i\rangle$ is equal to the sum of expectation values on each projection 
$|\psi_J^{(i)}\rangle$, i.e. $\langle i|\hat J_z^2|i\rangle = \sum_{J} |C_J^{(i)}|^2 \langle\psi^{(i)}_J|\hat J_z^2|\psi^{(i)}_J\rangle$. 
In the final step, I note that $\langle\psi^{(i)}_J|\hat J_z^2|\psi^{(i)}_J\rangle\leqslant J^2$, so the bound for dynamical susceptibility from Eq.~(\ref{eq:DynEnt_Indist_dyn_sus}) is
\begin{equation}\label{eq:DynEnt_Indist_the_bound}
  \mathcal F(\hat\varrho)\leqslant4\sum_{J=0}^{\frac N2} J^2\sum_i p_i|C_J^{(i)}|^2=4\sum_{J=0}^{\frac N2} J^2\,\mathrm{Tr}\left(\hat\Pi_J \hat\varrho\right)\,.
\end{equation}
Here $\hat \Pi_J = \sum_{\nu}\sum_{M=-J}^J\left| J,M;\nu\right\rangle\left\langle J,M;\nu\right|$ are projectors onto subspace of total angular momentum $J$. This bound is more strict then the ultimate HL -- it takes into account the symmetry properties of the state through probabilities of finding the system in subspaces of total angular momentum
\begin{equation}
P_J\equiv\mathrm{Tr}\left(\hat\Pi_J\hat\varrho\right)\,\Rightarrow\, P_J\geqslant 0\,\text{  and  }\,\sum_{J=0}^{\frac N2}P_J = 1\,.
\end{equation}
Bound (\ref{eq:DynEnt_Indist_the_bound}) is an average square of total angular momentum $J^2$ weighted by probabilities $P_J$. If the state of the system is spread among wide range of total angular momentum subspaces it is less entangled then the state which is distributed only among small number of subspaces with the highest $J$-s. In extreme case, the entanglement of a state confined to $J=N/2$ subspace is bounded by the HL of $N^2$. Moreover, the projectors $\hat\Pi_J$ are rotationally invariant, i.e. for any rotation $\hat R$ we have $[\hat \Pi_J , \hat R] =0$ and hence $P_J=\mathrm{Tr}(\hat\Pi_J \hat\varrho) = \mathrm{Tr}(\hat\Pi_J\hat R^\dagger\hat R\hat\varrho)=\mathrm{Tr}(\hat \Pi_J(\hat R\hat\varrho\hat R^\dagger))$. Therefore, the efficiency of the parameter estimation in interferometric experiment is also bounded by the decomposition of the state into total angular momentum subspaces since according to Eq.~(\ref{eq:Dyn_Ent_QFI_is_susc}) we have $F_Q(\hat\varrho,\hat U_\theta) =\mathcal F(\hat R\hat\varrho\hat R^\dagger)\leqslant 4\sum_{J=0}^{\frac N 2}P_J J^2$.

The sum over $J$ in Eq.~(\ref{eq:DynEnt_Indist_J_decomposition}) results from the rules of addition of angular momenta of the pseudo-spins of qubits constituting the system. Now I will review how the total angular momentum eigenstates $|J,M;\nu\rangle$ are constructed out of product states of qubits' pseudo-spin eigenstates. By examining this process we shall gain a necessary insight to answer the question what role is played by the indistinguishability of qubits.

I start with the even set of  $N$ qubits (the case of odd $N$ is solved analogically).  I take a symmetric superposition of first $J-M$ qubits in state $|\frac 12,-\frac 12\rangle$ and $J+M$ qubits in state $|\frac 12, \frac 12\rangle$ then multiply it by $\frac N2-J$ anitsymmetrized pairs of qubits that remain. Such a state is a total angular momentum eigenstate:
\begin{equation}\label{eq:DynEnt_Indist_construction}
|J,M;\nu=1\rangle=\mathcal S\left[ |\tfrac 12,-\tfrac 12\rangle^{\otimes J-M}|\tfrac 12,\tfrac 12\rangle^{\otimes J+M}\right]|\Psi_-\rangle^{\otimes \frac N2-J}\, .
\end{equation}
Here $\mathcal S$ is a symmetrization operator and 
\begin{equation}
|\Psi_-\rangle = \frac {|\tfrac 12,\tfrac 12\rangle|\tfrac 12,-\tfrac 12\rangle - |\tfrac 12,-\tfrac 12\rangle|\tfrac 12,\tfrac 12\rangle}{\sqrt 2}
\end{equation}
is the antisymmetric singlet state. The value assigned to the degeneracy index $\nu$ indicates that this particular choice of which qubits are to be antisymmetrized is one of many possible.\footnote{{It can be show that the number of distinguished choices equals $d_{J<\frac N2} = {N \choose N/2-J}-{N \choose N/2-J-1}$ and $d_{\frac N2} = 1$.}} From this construction we see that the total angular momentum is set by the number of symmetrized qubits forming the state. Indeed, when a total angular momentum operator, $\hat J_k =\sum_{n=1}^N\hat J_k^{(n)}$, acts on this state, the singlet pairs of qubits never contribute:
\begin{align}
\hat J_k& \left(\mathcal S\left[ |\tfrac 12,-\tfrac 12\rangle^{\otimes J-M}|\tfrac 12,\tfrac 12\rangle^{\otimes J+M}\right]|\Psi_-\rangle^{\otimes \frac N2-J} \right)=\nonumber\\
=&\left(\sum_{n=1}^{2J} \hat J^{(n)}_k \mathcal S\left[ |\tfrac 12,-\tfrac 12\rangle^{\otimes J-M}|\tfrac 12,\tfrac 12\rangle^{\otimes J+M}\right]\right)
|\Psi_-\rangle^{\otimes\frac N2-J}+\nonumber\\
&+\mathcal S\left[ |\tfrac 12,-\tfrac 12\rangle^{\otimes J-M}|\tfrac 12,\tfrac 12\rangle^{\otimes J+M}\right]\left(\sum_{n=2J+1}^{N}\hat J_k^{(n)}|\Psi_-\rangle^{\otimes\frac N2-J}\right)\label{eq:DynEnt_Indist_costruction_explained}
\end{align}
and now I can rearrange terms in the sum of the second line of the above equation
\begin{align}
&\sum_{n=2J+1}^{N}\hat J_k^{(n)}|\Psi_-\rangle^{\otimes\frac N2-J} =\nonumber\\
&= \sum_{n=2J+1}^{N-1}|\Psi_-\rangle^{\otimes n -2J}\left[\left(\hat J_k^{(n)}+\hat J_k^{(n+1)}\right)|\Psi_-\rangle\right]|\Psi_-\rangle^{\otimes \frac N2 + J - n -1} = 0\,.
\end{align}
The last equality follows from the fact that singlet state vanishes under action of any global angular momentum operator:
\begin{align}
&\left(\hat J_z^{(n)}+\hat J_z^{(n+1)}\right)|\Psi_-\rangle
=\left(+\tfrac 12 - \tfrac 12\right)\frac {|\tfrac 12,\tfrac 12\rangle|\tfrac 12,-\tfrac 12\rangle - |\tfrac 12,-\tfrac 12\rangle|\tfrac 12,\tfrac 12\rangle}{\sqrt 2}=0\,,\nonumber\\
&\left(\hat J_\pm^{(n)}+\hat J_\pm^{(n+1)}\right)|\tfrac 12,\pm\tfrac 12\rangle|\tfrac 12,\mp\tfrac 12\rangle = 0\,.
\end{align}
Hence, all that remains from Eq.~(\ref{eq:DynEnt_Indist_costruction_explained}) is the action of the operator on the symmetrized part. An elementary calculations show that
\begin{align}
\sum_{n=1}^{2J}\hat J_z^{(n)}&\mathcal S\left[ |\tfrac 12,-\tfrac 12\rangle^{\otimes J-M}|\tfrac 12,\tfrac 12\rangle^{\otimes J+M}\right] = 
M\mathcal S\left[ |\tfrac 12,-\tfrac 12\rangle^{\otimes J-M}|\tfrac 12,\tfrac 12\rangle^{\otimes J+M}\right]\,,\nonumber\\
\sum_{n=1}^{2J}\hat J_\pm^{(n)}&\mathcal S\left[ |\tfrac 12,-\tfrac 12\rangle^{\otimes J-M}|\tfrac 12,\tfrac 12\rangle^{\otimes J+M}\right] = \nonumber\\
&=\sqrt{(J\pm M)(J\pm M-1)}\,\mathcal S\left[ |\tfrac 12,-\tfrac 12\rangle^{\otimes J-M\mp 1}|\tfrac 12,\tfrac 12\rangle^{\otimes J+M\pm 1}\right]\,,
\end{align}
which confirms that Eq.~(\ref{eq:DynEnt_Indist_construction}) is correct.

Recall that the state describing a system of identical bosons has to be symmetric in respect to qubit permutations. Suppose that $N'<N$ qubits are indistinguishable while the remaining $N-N'$ particles remain distinguishable. In that case the minimal value of the total angular momentum that can be attained by the system is restricted to $J\geqslant \frac {N'}2$, because any possible state has to be symmetrized in respect to at least $N'$ qubits. Therefore, when all $N$ qubits are identical, the bound (\ref{eq:DynEnt_Indist_the_bound}) grows to become the Heisenberg limit $\mathcal F(\hat\varrho)\leqslant N^2$ since only maximal $J=N/2$ is allowed, which implies that $P_{J=N/2}=1$ and all other $P_J$ are zero. Hence, the indistinguishability of qubits  ``automatically'' confines all possible states into subspace of maximal total angular momentum which enables potentially highest degree of entanglement. If particles where not identical, it would require a tremendous amount of effort and ingenuity to be able to perform an experiment with a large ensemble of qubits prepared in a symmetrized state. For example, Schr\"{o}dinger cat states (see also Eq.~(\ref{eq:general_Cat}))
\begin{equation}\label{eq:Symmetric_Cat}
|\mathrm{Cat}_\varphi\rangle = \frac{|\tfrac 12,\tfrac 12\rangle^{\otimes N}+e^{i\varphi}|\tfrac 12,-\tfrac 12\rangle^{\otimes N}}{\sqrt 2}=\frac{ |\tfrac N2,\tfrac N2\rangle+e^{i\varphi}|\tfrac N2,-\tfrac N2\rangle}{\sqrt 2}\, ,
\end{equation}
yield HL degree of entanglement and are explicitly symmetric in respect to qubit permutations even if the particles are distinguishable. However, it is well known that in practice it is very difficult to prepare such a state for a large number of particles. 

The most prominent example of ``experimental friendly'' state is a so called {\it twin Fock state} \cite{Oberthaler_08}. In two mode approximation such a state is often described as a separable in modes ket $|\frac N2\rangle_L|\frac N2\rangle_R$ (here $L/R$ indicate the two modes of the system). In the language of pseudo-spin it is given by a symmetric total angular momentum eigenstate
\begin{equation}\label{eq:TF}
|\tfrac N2\rangle_L|\tfrac N2\rangle_R =\mathcal S\big[ |\tfrac 12,-\tfrac 12\rangle^{\otimes \tfrac N2}|\tfrac 12,\tfrac 12\rangle^{\otimes \tfrac N2}\big] = |\tfrac N2, 0\rangle\,.
\end{equation}
Such a state is relatively easy to obtain in various experimental setups. For example, the twin Fock state is obtained when the potential trapping Bose-Einstein condensate is adiabatically brought into a double well trap \cite{Oberthaler_08}. Due to repulsive interaction between bosons, the condensate is split evenly between the wells.

Although separable in modes, the twin Fock state is usefully entangled to a very high degree since QFI associated with Mach-Zehnder interferometer $\hat U^{(\mathrm{MZ})}_\theta = e^{i\frac \pi 2 \hat J_x}e^{-i\theta \hat J_z}e^{-i\frac \pi 2 \hat J_x}$ for this state is of order of half HL. This interferometer utilizes the beam-splitter operation $e^{-i\frac \pi2\hat J_x}$ to prepare the state, therefore the state corresponding to Twin Fock which posses equal degree of dynamical entanglement is $|\mathrm{TF}\rangle=e^{i\frac \pi2\hat J_x}|\frac N2,0\rangle$:
\begin{equation}\label{eq:Dyn_ent_Indist_TF_MZI}
\mathcal F( |\mathrm{TF}\rangle)=F_Q\big(|\tfrac N2,0\rangle,\hat U^{(\mathrm{MZ})}_\theta\big) =N\left(\frac N2 +1\right)\,. 
\end{equation}
The entanglement between particles forming this state can be ascribed solely to indistinguishability of qubits. Indeed, consider an experiment where instead of adiabatic splitting the double well setup is created by bringing together independently prepared condensates. If the number of particles in both wells is \emph{exactly} the same, say $N/2$, and the two-mode approximation holds, then the state of the system is again $|\mathrm{TF}\rangle$. This result might be counter-intuitive, since one might expect a separable state because the contents of each well never interacted with each other. However, the particles forming both condensates are identical bosons and the state has to be symmetric in respect to qubit permutations, therefore it cannot be separable. To see this lets denote by $\left|\uparrow/\downarrow\right\rangle$ the state of a particle being localized in left/right well. If particles were not identical the state of the whole system would be a product $\left|\uparrow\right\rangle^{\otimes\frac N2}\left|\downarrow\right\rangle^{\otimes\frac N2}$, but for identical bosons the state is symmetrized%
\footnote{{
In realistic circumstances it is impossible to predict how many particles will condense in a given run of the experiment. As a result the state of the whole system is a statistical mixture of form
\begin{align}
\hat\varrho_{\mathrm{two}}{}&=\sum_{N_\uparrow,N_\downarrow}p(N_\uparrow)p(N_\downarrow)\mathcal S[\left|\uparrow\right\rangle^{\otimes N_\uparrow}\left|\downarrow\right\rangle^{\otimes N_\downarrow}]\mathcal S[\left\langle\uparrow\right|^{\otimes N_\uparrow}\left\langle\downarrow\right|^{\otimes N_\downarrow}]\nonumber
\end{align}
Here $p(N)$ is the probability of creating a condensate of $N$ particles. Here I assume that the probability distribution $p$ is flat on interval $[N_\mathrm{min},N_\mathrm{max}]$. The dynamical susceptibility and QFI of this state are
\begin{equation}
\mathcal F(e^{i\frac\pi 2\hat J_x}\hat\varrho_{\mathrm{two}}e^{-i\frac \pi2\hat J_x}) =F_Q(\hat\varrho_{\mathrm{two}},\hat U^{(\mathrm{MZ})}_\theta)=\overline{N}\left(\tfrac{\overline{N}}2+1\right)\Big[1\!-\!\big(\tfrac{\Delta N}{\Delta N+2}\big)^2\Big]-\tfrac 12\big(\tfrac{\Delta N}{\Delta N+2}\big)^2\geqslant \overline{N}\nonumber
\end{equation}
Here $\overline{N}=\frac{N_\mathrm{max}+N_\mathrm{min}}2$ and $\Delta N = N_\mathrm{max}-N_\mathrm{min}$. Since $\mathcal F$ (and $F_Q$) surpass SNL the state is almost always entangled (with the exception of maximal $\Delta N=2\overline{N}$ when $\mathcal F=\overline{N}$). For example, when $\Delta N =25\% \overline{N}$ and $\overline{N}=1000$ we have $\mathcal F\approx 8 \overline{N}$.
}}: $\mathcal S[\left|\uparrow\right\rangle^{\otimes\frac N2}\left|\downarrow\right\rangle^{\otimes\frac N2}]=\mathcal S[ |\tfrac 12,-\tfrac 12\rangle^{\otimes \frac N2}|\tfrac 12,\tfrac 12\rangle^{\otimes \frac N2}]=|\frac N2,0\rangle=e^{-i\frac \pi2\hat J_x}|\mathrm{TF}\rangle$. I again underline that the beam-splitter $e^{i\frac\pi2\hat J_x}$ is a coherent transformation and it does not introduce any entanglement between particles. The only reason why dynamical susceptibility surpasses shot noise limit is due to symmetrization enforced by indistinguishability of bosons. 

The physical interpretation of this seemingly unnerving result is as follows. Although each condensate is prepared far apart from each other, say one was created on the Moon while the other remained on Earth, it is incorrect to assume that identical bosons forming them where not correlated. Since these particles are indistinguishable it is impossible to tell which atoms where on the Moon and which on the Earth, therefore the correlation always exists. It follows that the symmetrization is enforced even if the two condensates where never brought together. Now the question is whether the system of Moon and Earth bound condensates is entangled. Formally one can define rotation operator $e^{i\frac \pi 2\hat J_x}$ representing the beam-splitter and calculate dynamical susceptibility for this Moon-Earth double well system. However, it is not correct to claim that the state is dynamically entangled. The distinction between artificial entanglement such as this, and the proper, physically meaningful entanglement can be made by invoking the context of interferometry. Dynamical susceptibility for this system is meaningless since it is impossible to create an interferometer operating on remote condensates and the relation such as (\ref{eq:Dyn_ent_Indist_TF_MZI}) between QFI and $\mathcal F$ cannot be established. Therefore, it is necessary to bring the condensates together and enable them to ``mix'' with each other, otherwise the non-classical correlations due to indistinguishability are impossible to observed and to utilize.

\section{Symmetry and the dynamical entanglement}

In the previous section I have established a relation between the total angular momentum of a multi-qubit system and the maximal degree of dynamical entanglement this system can achieve (see Eq.~(\ref{eq:DynEnt_Indist_the_bound})). I have found that only completely symmetric states, i.e. states with maximal total angular momentum, are able to reach the ultimate Heisenberg limit of entanglement. The key question is which properties of the state allow for tapping into this potential entanglement that can be detected by a criterion such as the dynamical susceptibility. Once I identify this property, I will be able to make a positive statement: we will know what entanglement \emph{is}, not just what it \emph{is not}. 

I begin by revisiting the example of Twin Fock state, which written in the basis of pseudo-spin eigenstates is given by
\begin{equation}\label{eq:Sym_vs_Ent_sup_of_kittens}
|\mathrm{TF}\rangle =e^{i\frac \pi2\hat J_x}|\tfrac N2,0\rangle =C_0|\tfrac N2 , 0\rangle+ \sum_{m=1}^{\frac N4}C_{2m}\frac{|\tfrac N2,2m\rangle +|\tfrac N2,-2m\rangle}{\sqrt 2}\, .
\end{equation}
Here $C_M =\sqrt 2D^{\frac N2}_{M,0}(-\tfrac \pi 2,-\tfrac\pi 2,\tfrac \pi 2)$ is an element of Wigner D-matrix.\footnote{{
	A rotation operator can be written as
	\begin{equation}
	\hat R =e^{-i\alpha\hat J_x}e^{-i\beta\hat J_y}e^{-i\gamma\hat J_z}\,,
	\end{equation}
	where $(\alpha,\beta,\gamma)$ are Euler angles. The Wigner D-matrix is a square matrix of dimension $2j+1$ with general element
	\begin{equation}
	D^j_{m,m'}(\alpha,\beta,\gamma)=\langle j,m|\hat R|j,m'\rangle=\langle j,m|e^{-i\alpha\hat J_z}e^{-i\beta\hat J_y}e^{-i\gamma\hat J_z}|j,m'\rangle\,.
	\end{equation}
 \label{foot:wigner_D}}} We see that the state $|\mathrm{TF}\rangle$ is a superposition of ``Schroedinger kitten'' states of form $(|\frac N2,M\rangle+|\frac N2,-M\rangle)/{\sqrt 2}$. Similarly to Schr\"{o}dinger cat states, the dynamical susceptibility of the kitten state is determined by the difference of angular momentum projections of eigenstates forming the superposition,
\begin{equation}
\mathcal F\left(\frac{|\frac N2,M\rangle+|\frac N2,-M\rangle}{\sqrt 2}\right)=4M^2\,.
\end{equation}
The dynamical susceptibility of superposition of kitten states (\ref{eq:Sym_vs_Ent_sup_of_kittens}) is
\begin{equation}\label{eq:Sym_vs_Ent_susc_TF}
\mathcal F\left(|\mathrm{TF}\rangle\right)=4\sum_{M=0}^{\frac N2}M^2 |C_{M}|^2\,.
\end{equation}
A simple intuitive picture can be draw to interpret this result. The density matrix of a kitten state when written in the basis of angular momentum eigenstates is of form
\begin{align}
\frac{|\tfrac N2,M\rangle+|\tfrac N2,-M\rangle}{\sqrt 2}{}&\frac{\langle\tfrac N2,M|+\langle\tfrac N2,-M|}{\sqrt 2} =\nonumber\\
&=\frac 12\left(\begin{array}{ccccccccc}
\ddots&&\vdots& &\ddots& & \vdots&&\\
&0&0& &&0 & 0&&\\
\ldots&0&1& 0&\ldots& 0& 1&0&\ldots\\
&&0 & 0 & & &0&0&\\
\ddots&&\vdots & & \ddots & &\vdots&&\ddots\\
&0&0 &  & &0 &0&&\\
\ldots&0 &1& 0&\ldots& 0& 1&0&\ldots\\
&&0&0 && & 0&0&\\
&&\vdots& &\ddots& & \vdots&&\ddots\\
\end{array}\right)
\end{align}
The only non-zero elements are found on the diagonal and in the band located $M$ units away from the diagonal. The dynamical susceptibility can now be broken down as follows: $\mathcal F=4(2\times M^2\times \frac 12)$. It is equal four times square of distance from diagonal to upper matrix element ($M$) times the magnitude of this element ($1/2$) plus the same for lower matrix element. This formula is analogical to the classical {\it moment of inertia} of two points of mass $1/2$ positioned at distances $M$ away from the rotation axis. This analogy can even be extended to the state $|\mathrm{TF}\rangle$, since Eq.~(\ref{eq:Sym_vs_Ent_susc_TF}) resembles a formula for the moment of inertia of a collection of pairs of points corresponding to kitten states with masses given by probabilities $|C_M|^2/2$ and the distances given by the projections $M$. Although the elegant form of Eq.~(\ref{eq:Sym_vs_Ent_susc_TF}) results from the a particular form of $|\mathrm{TF}\rangle$ being a pure superposition of kitten states, the analogy between the moment of inertia of density matrix elements and the dynamical susceptibility is generally valid. Below I show how it comes by and for that I shall start by introducing a language of {\it spherical tensor operators}. This language enables a convenient way for quantification of the symmetry properties of a state which, as it will turn out, is the main ingredient of the dynamical entanglement.

\subsection{Basis of spherical tensor operators}

In order to properly describe the rotational symmetry of a state its density matrix has to be decomposed into parts which transform independently under rotations. For example, recall that the density matrix of a single qubit, which is equivalent to spin $1/2$ system, can be written as a combination of spin operators and the unit matrix: 
\begin{equation}\label{eq:Sym_vs_Ent_qubit_rho}
\hat\varrho_{\text{single qubit}} = \frac 12 \hat{\mathbf 1} + 2\langle\hat{\mathbf{J}}\rangle\cdot\hat{\mathbf{J}}\,,
\end{equation}
where $\hat J_i = \frac 12 \hat \sigma_i$ and $\langle \hat J_i\rangle = \mathrm{Tr}\left( \hat J_i \hat\varrho_{\text{single qubit}}\right)$. 
An arbitrary rotation can only transform spin operator into combination of spin operators, while the unit matrix is always unaffected. 
Therefore, (\ref{eq:Sym_vs_Ent_qubit_rho}) is the desired decomposition, where the part spanned by triple of $J_i$-s is independent of part spanned by $\mathbf 1$. Although the decomposition (\ref{eq:Sym_vs_Ent_qubit_rho}) might seem to trivially result from the properties of Pauli matrices $\hat\sigma_i$, in fact it is a consequence of a more general principle. In the language of group theory the spin operators are vectorial objects. To be more precise, they are proportional to spherical tensor operators \cite{N-qubit_noise,2-qubits_noise,sakurai} with angular momentum $j=1$ and magnetic number $m$ taking on values of $-1$, $0$ and $1$:
\begin{align}
&\hat T_{j=1,m=0}=\hat J_z\label{eq:Sym_vs_Ent_def_10_tensor}\\
&\hat T_{j=1,m=\pm 1} = \mp \frac 1{\sqrt{2}}\left(\hat J_x\pm i \hat J_y\right)\equiv \mp\frac 1{\sqrt 2}\hat J_\pm\label{def_11_tensor}
\end{align}
The spherical tensor operators $\hat T_{jm}$ are an operator analog of the angular momentum eigenstates $\left|j,m\right\rangle$. For example, one way of defining the spherical tensor with quantum numbers $j$ and $m$ is to require for it to satisfy a following set of equations
\begin{align}
&[ \hat J_z , \hat T_{jm}] = m \hat T_{jm}\label{eq:Sym_vs_Ent_def_tensor_m}\\
&[ \hat J_\pm, \hat T_{jm}] = \sqrt{\left(j\pm m\right)\left(j\pm m +1\right)}\hat T_{jm\pm 1}\label{eq:Sym_vs_Ent_def_tensor_l}\,,
\end{align}
which imply
\begin{equation}
\sum_{i=x,y,z}[\hat J_i, [\hat J_i , \hat T_{jm} ] ] =j(j+1)\hat T_{jm}\,.
\end{equation}
These equations are in the direct correspondence with a similar set satisfied by the angular momentum eigenstates: 
\begin{align}
&\hat J_z\left|j,m\right\rangle = m\left| j,m\right\rangle\,,\\ 
&\hat J_\pm\left| j,m\right\rangle = \sqrt{(j\pm m)(j\pm m +1)}\left| j,m\pm 1\right\rangle\,,\\
&\hat{\mathbf{J}}^2|j,m\rangle = j(j+1)|j,m\rangle\,.
\end{align}
Since the action of an arbitrary rotation on an operator is determined by commutators such as (\ref{eq:Sym_vs_Ent_def_tensor_m}) and (\ref{eq:Sym_vs_Ent_def_tensor_l}) it follows that the transformation properties of spherical tensors and angular momentum eigenstates are the same. One of the consequences is that any rotation transforms a spherical tensor with angular momentum $j$ into a combination of spherical tensors with the same $j$.

The set of spherical tensor operators forms an orthogonal basis with respect to the scalar product $( \hat A | \hat B ) \equiv \mathrm{Tr}( \hat A^\dagger \hat B)$ in the space of density matrices of systems with fixed angular momentum. In particular, the density matrix $\hat \varrho_{J}$ of system with angular momentum $J$, operating in $2J+1$ dimensional Hilbert space, can be decomposed into linear combination of spherical tensor operators with angular momenta ranging from $j=0$ to $j=2J$:
\begin{equation}\label{eq:Sym_vs_Ent_tensor_decomp}
\hat\varrho_J = \sum_{j=0}^{2J}\sum_{m=-j}^j \tau_{jm}\hat T_{jm},
\end{equation}
Here $\tau_{jm}=(\hat T_{jm}|\hat T_{jm})^{-1}( \hat T_{jm} | \hat\varrho_J)$ are the expectation values of the spherical tensor operators. The parts of $\hat\varrho_J$ spanned by spherical tensors with fixed $j$ have different symmetry and they transform independently. Therefore, decomposition such as this provides a proper quantification of symmetry properties of a state described by $\hat\varrho_J$. Since the density matrix of $N$ bosonic qubits is equivalent to a matrix of a system with total angular momentum $J=N/2$, this decomposition provides an excellent tool for carrying out my analysis.

\subsection{Quantum moment of inertia}

In order to formalize the analogy with the moment of inertia I shall utilize the decomposition (\ref{eq:Sym_vs_Ent_tensor_decomp}) in conjunction with a theorem which relates dynamical susceptibility with a variance of spin operators. 

For an arbitrary pure-state ensemble $\{ p_k , |\phi_k\rangle\}$ of a density matrix $\hat\varrho$ (i.e. $\hat\varrho=\sum_k p_k |\phi_k\rangle\langle\phi_k|$ and $p_k\geqslant 0$, $\sum_k p_k =1$) the dynamical susceptibility is bounded by the {\it average variance} \cite{Toth_Petz_2013,QFI_is_a_roof}:
\begin{equation}
\mathcal F(\hat\varrho) \leqslant 4\sum_k p_k \langle\Delta^2\hat J_z\rangle_{\phi_k}\,.
\end{equation}
We have already seen an example of this relation when the ensemble in question was composed of orthogonal eigenstates of $\hat\varrho$, for example in Eq.~(\ref{eq:DynEnt_QFIasCrit_bound_by_var}). The theorem states that among all possible ensembles there exists a special one which minimizes the average variance which coincides with the dynamical susceptibility \cite{QFI_is_a_roof}, therefore
\begin{equation}
\mathcal F(\hat\varrho) = \min_{\{p_k,|\phi_k\rangle\}} 4\sum_k p_k\langle\Delta^2\hat J_z\rangle_{\phi_k}\,.
\end{equation}
Note that the variance of $\hat J_z$ on any pure state can be written in terms of expectation value of the commutator with the density matrix of the state
\begin{align}
\mathcal F(\hat\varrho) {}&= \min_{\{p_k,|\phi_k\rangle\}} 4\sum_k p_k\langle\Delta^2\hat J_z\rangle_{\phi_k} =\nonumber\\
&=\min_{\{p_k,|\phi_k\rangle\}} 2\sum_k p_k\mathrm{Tr}\left( |\phi_k\rangle\langle\phi_k|\,\big[\hat J_z,\big[\hat J_z , |\phi_k\rangle\langle\phi_k|\, \big]\big]\right)\,.
\end{align}
Each pure-state density matrix can now be decomposed according to Eq.~(\ref{eq:Sym_vs_Ent_tensor_decomp}), $|\phi_k\rangle\langle\phi_k| = \sum_{j=0}^{N}\sum_{m=-j}^j \tau^{(\phi_k)}_{jm}\hat T_{jm}$. Using the fact that spherical tensors are eigenoperators of $\hat J_z$ [see Eq.~(\ref{eq:Sym_vs_Ent_def_tensor_m})] we get
\begin{align}
\mathcal F(\hat\varrho) {}&=\min_{\{p_k,|\phi_k\rangle\}} 2\sum_k p_k \sum_{j=0}^{N}\sum_{m=-j}^j m^2 \tau^{(\phi_k)}_{jm}\mathrm{Tr}(|\phi_k\rangle\langle\phi_k|\hat T_{jm}) =\nonumber\\
& =\min_{\{p_k,|\phi_k\rangle\}} 2\sum_k p_k \sum_{j=0}^{N}2\sum_{m=1}^j m^2 \mathrm{Tr}(\hat T_{jm}^\dagger\hat T_{jm})|\tau^{(\phi_k)}_{jm}|^2 =\nonumber\\
&=\min_{\{p_k,|\phi_k\rangle\}} 2\sum_{j=1}^{N}\left(2\sum_{m=1}^{j} m^2 \sum_k p_k \frac{|\langle \phi_k|\hat T_{jm}|\phi_k\rangle|^2}{|| \hat T_{jm}||^2}\right)\,,
\end{align}
where $||\hat A||^2 =(\hat A |\hat A)$. The function min renders the above equation highly impractical, however a useful bound can be found by recalling that $p_k$ defines a probability distribution which implies that 
\begin{equation}
\sum_k p_k |\langle\phi_k|\hat T_{jm}|\phi_k\rangle|^2 \geqslant \big|\sum_k p_k\langle \phi_k|\hat T_{jm}|\phi_k\rangle\big|^2 = |\mathrm{Tr}(\hat\varrho\,\hat T_{jm})|^2\,.
\end{equation}
Therefore, we can drop the min function and get the final result
\begin{equation}\label{eq:QMI}
\mathcal F(\hat\varrho)  \geqslant 2\sum_{j=1}^{N} \left(2\sum_{m=1}^j m^2 \frac{|\langle\hat T_{jm}\rangle|^2}{||\hat T_{jm}||^2}\right)\equiv 2 \mathcal I(\hat\varrho)\,,
\end{equation}
where $\langle\hat A\rangle = \mathrm{Tr}(\hat\varrho\,\hat A)$ and the inequality is saturated only for pure states. Thus we have identified new quantity $\mathcal I(\hat\varrho)$ (I will refer to it as {\it quantum moment of inertia} (QMI)) which can serve as a criterion for dynamical entanglement. Indeed, according to inequality (\ref{eq:QMI}) QMI is bounded for separable states: 
\begin{equation}
\mathcal I(\hat\varrho_{\mathrm{separable}})\leqslant \frac 12\mathcal F(\hat\varrho_{\mathrm{separable}}) \leqslant \frac N2\, ,
\end{equation}
and it reaches the maximal value for Schr\"{o}dinger cat states
\begin{equation}
\mathcal I(\hat\varrho)\leqslant \frac 12\mathcal F(\hat\varrho)\leqslant \frac{N^2}2\text{ , }\mathcal I(|\mathrm{Cat}_\varphi\rangle) = \frac 12 \mathcal F(|\mathrm{Cat}_\varphi\rangle) =\frac {N^2}2.
\end{equation}
The drawback of $\mathcal I(\hat\varrho)$ is that it is not an efficiency criterion, since this quantity is not directly related to efficiency of any particular task. However, it is related to dynamical susceptibility, therefore QMI is indirectly related to atomic interferometry.

The QMI generalizes the analogy between dynamical entanglement and pseudo moment of inertia of the matrix elements of Schr\"{o}dinger kitten states. According to the Wigner-Eckhart theorem \cite{sakurai} the matrix element of a spherical tensor operator is given by 
\begin{equation}\label{eq:Phys_Inter_Symm_v_Ent_Wigner-Eckhart}
\langle J,M|\hat T_{jm}|J,M'\rangle = \langle J,M' ; j,m| J, M \rangle\frac{\langle J|| T_{j}||J\rangle}{\sqrt{2J+1}}
\end{equation}
where the reduced matrix element $\langle J|| T_{j}||J\rangle$ is independent of $m$, $M$ and $M'$. The Clebsch-Gordan coefficient $\langle J,M' ; j,m| J, M \rangle$ does not vanish if and only if the condition $M'+m = M$ is satisfied. Hence, the matrix of $\hat T_{jm}$ is sparse and the non-zero elements are located only in a band positioned $m$ units away from the diagonal. It follows that the expectation value $\langle \hat T_{jm}\rangle = \mathrm{Tr}(\hat\varrho\,\hat T_{jm})$ is not zero only when the density matrix has a non-vanishing matrix element located in this band. For example, the density matrix of a kitten state $(|\tfrac N2, M\rangle + |\tfrac N2, -M\rangle)/{\sqrt 2}$ decomposes into tensors with $m=\pm M$ and $m=0$ which span the diagonal. 

The quantities $|\langle \hat T_{jm}\rangle|^2/||\hat T_{jm}||^2$ appearing in the expression for $\mathcal I(\hat\varrho)$ can be interpreted as a mass density at distance $m$ of a token rigid body labeled by the quantum number $j$. By the analogy with the classical moment of inertia, the quantum moment of inertia of this body is given by the integral over distance of the square of $m$ times the density (the integral turns into summation since the distance is quantized): $\sum_m m^2|\langle \hat T_{jm}\rangle|^2/||\hat T_{jm}||^2$. The total moment of the state is a sum of moments of all bodies -- hence the summation over $j$ in Eq.~(\ref{eq:QMI}).

The fact that such an analogy can be established is not so surprising. The dynamical susceptibility inherits form the QFI its interpretation as a susceptibility of the state to change induced by a rotation around the $z$-axis (see Sec \ref{sec:Eff_Dist_metric}). The classical moment of inertia is also a susceptibility to rotations around given axis of the rigid body -- the angular momentum of the body is proportional to the moment of inertia.

Aside from this intuitive picture, the decomposition in the basis of spherical tensors allows to establish a relation between the dynamical entanglement of the state and its rotational symmetry. The state is considered {\it more} symmetric when the corresponding density matrix is spanned by spherical tensors with small angular momenta $j$. Indeed, if the density matrix contains only tensor with $j=0$, then it is invariant to rotations, i.e. it is a {\it scalar}. In that case the state is useless as a probe for an interferometer since no information about the rotation angle $\theta$ can be imprinted. Consequently, the dynamical susceptibility of such a state is zero. The state is {\it less} symmetric when it decomposes into spherical tensors with high $j$-s. For example, imagine that the density matrix contains a maximally polarized spherical tensor $\hat T_{jj}$ (i.e. tensor with maximal magnetic number $m=j$). When the rotation around the $z$ axis is applied to this state, the tensor transforms according to
\begin{equation}
e^{-i\theta \hat J_z}\hat T_{jj}e^{i\theta\hat J_z} = \sum_k \frac {(-i\theta)^k}{k!}\underbrace{[\hat J_z, [\hat J_z,\ldots ,}_{\text{$k$ times}}\hat T_{jj} ]\ldots ] ] = \sum_k \frac{(-i\theta j)^k}{k!}\hat T_{jj} = e^{-i j \theta}\hat T_{jj}\,.
\end{equation}
Hence, the higher the $j$ the more abrupt change due to rotation. Therefore, the less symmetric states are more susceptible to rotations which means that the speed along the path in density matrix space, as quantified by dynamical susceptibility, can potentially reach higher values. This shows that the dynamical entanglement is directly related to rotational symmetry properties of the state. 

\section[Relation between rotations and dynamical entanglement]{Relation between rotations and dynamical entanglement%
              \sectionmark{Rotations and dynamical entanglement}}
\sectionmark{Rotations and dynamical entanglement}

Throughout this thesis it was pointed out many times that rotations are coherent transformations (i.e. act on each party individually and in the same way) and as such cannot introduce entanglement between particles. Nevertheless, the example of the twin Fock state shows that rotations {\it do} affect the degree of entanglement detected by the dynamical susceptibility. In this section I explain this seemingly paradoxical behavior and provide a detailed description of the relation between the dynamical entanglement and rotations. This discussion also provides deeper insight into structure of non-classical correlations.

In the previous section I showed that a quantum state of an $N$-qubit system decomposes into parts with different symmetry properties. Each part is spanned by a set of spherical tensor operators, $\{\hat T_{jm}\}_{m=-j,\ldots,j}$, with fixed angular momentum $j$ and varying magnetic numbers $m$. For each $j$ I interpreted the expectation values of $\hat T_{jm}$ as a distribution of fictional mass over spatial dimension measured in units of $m$. This interpretation allowed for identifying an analogy between dynamical entanglement and the moment of inertia of a classical rigid body. 

\begin{figure}[h!]
\centering
\includegraphics[clip,width=\textwidth]{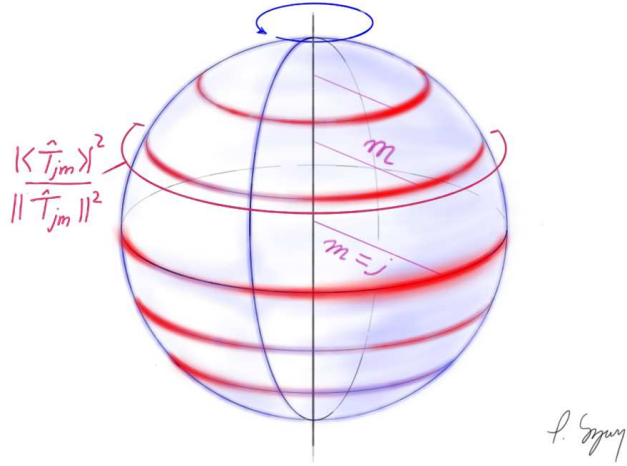}
\caption{{\it The classical rigid body modeling the part of density matrix with given symmetry, spanned by spherical tensor operators with angular momentum $j$. The sphere can rotate about vertical axis passing through its center. The mass (red) is enforcer to only occupy parallels for which the distance from the axis is quantized and measured in the units of magnetic numbers $m$.}}\label{fig:ent_model_gen}
\end{figure}

The body itself can be imagined as a massive sphere which is able to rotate around axis passing through its center. The quantum nature of the system manifests itself by forcing the mass to be distributed only over certain parallels, such that the distance from the surface to chosen rotation axis is quantized (see Fig. \ref{fig:ent_model_gen}). 

\begin{figure}[h!]
\centering
\includegraphics[clip,width=\textwidth]{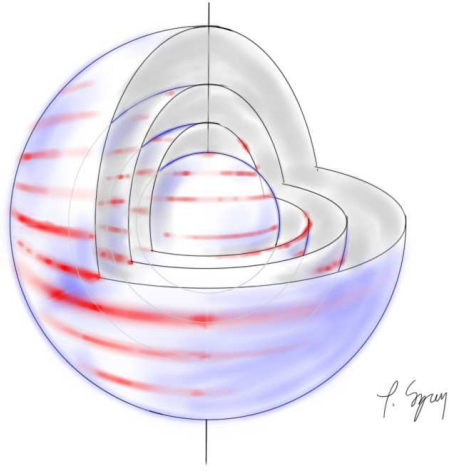}
\caption{{\it The classical model of quantum state of $N$ bosonic qubits. Each sphere represent the part of density matrix with distinguished rotational symmetry (see Fig.~\ref{fig:ent_model_gen}). The mass distribution one the surface of spheres (red) determines the moment of inertia of the system in respect to rotations about common axis. The moment of inertia is directly related to dynamical entanglement of the state.}}\label{fig:sphere_model}
\end{figure}

The total mass around parallels $|m|$ units away from the axis equals%
\footnote{{
Note that the magnetic number can be negative or positive, $m=\pm |m|$. One possible convention is to ascribe positive/negative $m$ to parallel located on northern/southern hemisphere. }\label{foot:NS_convention}}
\begin{align}
\left(\substack{\text{Total mass}\\\text{ distributed over parallels }\\\text{distanced $|m|$ units from axis}}\right)&\equiv\frac{|\langle\hat T_{jm}\rangle|^2}{||\hat T_{jm}||^2}+\frac{|\langle\hat T_{j\,-{m}}\rangle|^2}{||\hat T_{j\,-{m}}||^2} =\nonumber\\
&=\frac{|\langle\hat T_{jm}\rangle|^2+|\langle\hat T^\dagger_{j{m}}\rangle|^2}{||\hat T_{jm}||^2}=2\frac{|\langle \hat T_{jm}\rangle|^2}{||\hat T_{jm}||^2}\,.\label{eq:parallel_mass}
\end{align}
Here I used the property of spherical tensor operators that $\hat T_{j\,-m} = (-1)^m \hat T_{jm}^\dagger$. Therefore, the total mass of the sphere is given by the sum of masses occupying each allowed parallel, i.e. 
\begin{equation}\label{eq:Phys_Ent_Rot_v_Ent_total_mass}
\left(\substack{\text{ Total mass }\\\text{ of sphere }}\right) =\sum_{\text{parallels}}\left(\substack{\text{ Total mass }\\\text{ distributed over parallel }}\right)= \sum_{m=-j}^j \frac{|\langle \hat T_{jm}\rangle|^2}{||\hat T_{jm}||^2}\,.
\end{equation}
Since the maximal distance form the axis to parallel at which the mass can be found is $m=j$, the radius of the sphere is given by the quantum number $j$. The quantum state is thus analogical to collection of such spheres with radii ranging from $j=0$ to $j=N$ (see Fig.~\ref{fig:sphere_model}). 

The degree of correlations detected by the dynamical susceptibility is tied to the total moment of inertia of the spheres in respect to common rotation axis. It is important to bare in mind that according to convention adopted for the definition of the dynamical susceptibility, the axis of imagined rotation coincides with the $z$-axis of the reference frame.

The correspondence between some of the traits of the classical model and quantum states is readily apparent. For example, an obvious feature of a classical sphere is that a rotation does not change its shape, which includes total mass as well as radius. This is also the case for a quantum state. It was noted previously that rotations conserve the angular momentum $j$ of spherical tensor operators, i.e. for arbitrary rotation $\hat R$ we have
\begin{equation}
\sum_{i=x,y,z} [\hat J_i ,[\hat J_i, (\hat R\hat T_{jm}\hat R^\dagger)]] = \sum_{m'} D^j_{m',m}\sum_i [\hat J_i, [\hat J_i,\hat T_{jm'}]] = j(j+1)(\hat R\hat T_{jm}\hat R^\dagger)\,.
\end{equation}
Here 
\begin{equation}\label{eq:tensor_rotation}
\hat R\hat T_{jm}\hat R^\dagger=\sum_{m=-j}^j D^j_{m',m}\hat T_{jm'}
\end{equation}
and $D^j_{m',m}$ are elements of Wigner D-matrix corresponding to $\hat R$. Therefore, the part of the density matrix spanned by a set of tensors $\{ \hat T_{jm}\}_{m=-j,\ldots,j}$ can be represented by the sphere with the same radius $j$ even after the state has been transformed $\hat \varrho\to \hat R\hat\varrho\hat R^\dagger$.
The total ``mass'' given by Eq.~(\ref{eq:Phys_Ent_Rot_v_Ent_total_mass}) is invariant as well:
\begin{align}
&\sum_m \frac{|\langle \hat R\hat T_{jm}\hat R^\dagger\rangle|^2}{||\hat R\hat T_{jm}\hat R^\dagger||^2} = \sum_{m,m',m''}D^j_{m',m}(D^j_{m'',m})^*\frac{(2j+1)\langle \hat T_{jm'}\rangle\langle\hat T_{jm''}\rangle^*}{
\big|\big\langle \tfrac{N}2\big|\big|T_j \big|\big|\tfrac{N}2\big\rangle\big|^2}=\nonumber\\
&=\!\!\sum_{m,m',m''}\!\!\!\! e^{i\gamma(m'-m'')}\langle j m'|e^{-i\beta\hat J_y}| j ,m\rangle\langle j,m|e^{i\beta\hat J_y} |j ,m''\rangle \frac{(2j+1)\langle \hat T_{jm'}\rangle\langle\hat T_{jm''}\rangle^*}{
\big|\big\langle \tfrac{N}2\big|\big|T_j \big|\big|\tfrac{N}2\big\rangle\big|^2} \nonumber\\
&= \sum_{m',m''}\langle j,m'|e^{-i\beta\hat J_y}e^{i\beta\hat J_y}|j,m''\rangle \frac{(2j+1)\langle \hat T_{jm'}\rangle\langle\hat T_{jm''}\rangle^*}{\big|\big\langle \tfrac{N}2\big|\big|T_j \big|\big|\tfrac{N}2\big\rangle\big|^2} = \sum_{m'} \frac{|\langle\hat T_{jm'}\rangle|^2}{||\hat T_{jm'}||^2}\,.
\end{align}
Here I used the definition of the Wigner D-matrix (see footnote \ref{foot:wigner_D} below Eq.~(\ref{eq:Sym_vs_Ent_sup_of_kittens})) and the fact that the norm of a spherical tensor is invariant and depends only on $j$, 
\begin{equation}\label{eq:tensor_norm}
||\hat T_{jm}||^2=\mathrm{Tr}(\hat T_{jm}^\dagger\hat T_{jm}) =(2j+1)^{-1}\left|\left\langle \tfrac N2\right|\right|T_j\left|\left| \tfrac N2\right\rangle\right|^2\,.
\end{equation}
Hence, rotations cannot move the mass between spheres. In terms of the properties of the quantum state this means that coherent operations do not change the symmetry of the state, as they leave unaltered the distribution of the angular momenta $j$ in the decomposition of the density matrix. If the mass was not conserved, then it would be possible to transfer it to larger spheres which in turn would be equivalent to lowering the symmetry of the state. As it was shown previously, lower symmetry, i.e. the presence of higher $j$ tensors in the decomposition of the density matrix, leads to potentially higher degree of the dynamical entanglement.%
\footnote{{The mass can be transfered between spheres by, for example, non-linear evolution $e^{-i\chi \hat J_z^2}$ induced by atom-atom interactions. Such operation do not conserve angular momentum of spherical tensors. Therefore, interactions between particles alter the distribution of angular momenta in decomposition of density matrix which can decrease the symmetry of the state and generate additional dynamical entanglement.}}

However, it is not the total mass of spheres but rather the {\it mass distribution} over their surfaces that determine the degree of non-classical correlations detected by the dynamical susceptibility. In case of classical model, the rotations of spheres do not change this distribution. This can be seen from the equivalence of an {\it active} and a {\it passive} rotation transformation. The active transformation, where the point of view of an observer (represented by a choice of reference frame) is fixed while the rigid body is rotated, cannot be distinguished from a passive transformation, where the body is still and the observer is rotated. It is clear that the change of the reference frame has no influence on the dynamical properties of a body (including the mass distribution), therefore the same is true for rotations of the body itself. Hence the mass distribution over the surface of each sphere is also an invariant property of the system, independent of the choice of reference frame.

The same arguments and conclusions are also valid for quantum state, with important modification which takes into account the quantization of magnetic numbers $m$. While the distribution of the angular momenta $j$ in the decomposition of the density matrix depends only on the properties of the state, the distribution of $m$-s also depends on the choice of quantization axis (i.e the choice of reference frame). The spherical tensors $\hat T_{jm}$ are defined as the eigenoperators of the superoperator $\mathcal J_z=[\hat J_z, \,.\,]$ (see Eq.~(\ref{eq:Sym_vs_Ent_def_tensor_m})), similarly to eigenstates $|j,m\rangle$ defined as an eigenstates of $z$ component of the spin operator. When the quantization axis is changed, i.e. the operator $\hat J_z$ is rotated into $\hat J_n = \hat R\hat J_z\hat R^\dagger$, both eigenoperators and eigenstates are transformed as well:
\begin{align}
|j,m\rangle\to\hat R|j,m\rangle\,&:\,\hat J_n (\hat R|j,m\rangle)=(\hat R\hat J_z\hat R^\dagger)(\hat R|j,m\rangle) = m(\hat R|j,m\rangle),\\
\hat T_{jm}\to\hat R\hat T_{jm}\hat R^\dagger\,&:\,[\hat J_n,(\hat R\hat T_{jm}\hat R^\dagger)] = \hat R [\hat J_z,\hat T_{jm} ]\hat R^\dagger =m(\hat R\hat T_{jm}\hat R^\dagger)\, .
\end{align}
Therefore, rotation of the state, or equivalently, rotation of the reference frame causes displacement of allowed parallels on the corresponding classical sphere. This assures that the mass can be found only at quantized distances from the axis of imagined rotation (which by convention overlaps with the quantization axis), even when it shifts to new orientation. 

\begin{figure}[h!]
\centering
\includegraphics[clip,width=\textwidth]{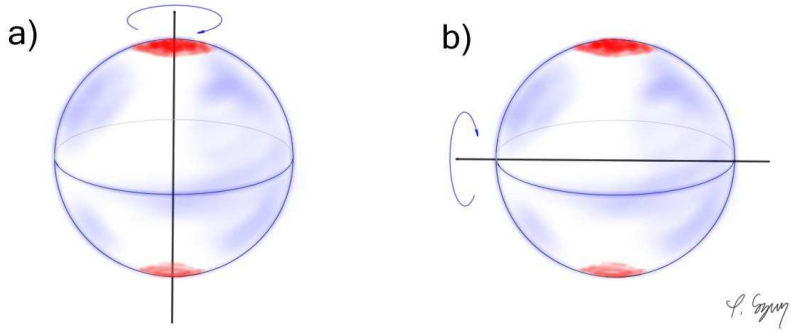}
\caption{{\it According to convention adopted for dynamical susceptibility the moment of inertia of classical model of the state is calculate in respect to $z$-axis of the reference frame. Because of that, the dynamical susceptibility is affected by rotations even though the mass distribution is not. For example, in the case presented in (a) the mass is localized on the poles of the sphere and the moment of inertia is small. The same mass distribution yields significantly greater moment of inertia when viewed in the rotate frame (b).}}\label{fig:axes_rot}
\end{figure}

Although the total mass and its distribution are invariant, the dynamical susceptibility related to the moment of inertia calculated in respect to $z$-axis must be affected by the change of the reference frame. Indeed, as the axis is rotated, the distance to a given point on the surface of the sphere changes because the latitude of the parallel occupied by the point is now different. For example, consider a case when the mass is mostly distributed on the poles of the sphere (see Fig. \ref{fig:axes_rot}, (a)). Of course, the axis passes through the poles, so that the moment of inertia of the sphere is very small. It is obvious that an appropriate rotation can reorient the $z$-axis so that in the new reference frame the mass is found on the equator instead of poles (see Fig. \ref{fig:axes_rot} (b)). The moment of inertia calculated for the new axis will be much larger then previously. Hence, even though the degree of dynamical entanglement detected for the state before and after the rotation will be different it is clear that the rotation itself did not entangle particles. Instead, it merely changed our point of view so that we could observe the entanglement which was already present in the state.

The degree of this {\it innate} dynamical entanglement existing in a given state $\hat\varrho$ can be defined by finding the optimal orientation of the axis for which the dynamical susceptibility is the largest
\begin{equation}
\left(\substack{\text{ the innate }\\\text{dynamical entanglement}\\\text{ of state $\hat\varrho$ }}\right) \equiv \max_{\hat R}\mathcal F(\hat R\hat\varrho\hat R^\dagger)=\mathcal F(\hat R_{\mathrm{max}}\hat\varrho\hat R_{\mathrm{max}}^\dagger)
\end{equation}
In case of a classical sphere the optimal axis is one of so-called {\it principle axes} for which the matrix of moment of inertia tensor has the largest eigenvalue. In this reference frame, on average, the mass is distributed closest to equator. Corresponding transformation of quantum state reorients quantization axis so that the expectation values of spherical tensors with largest magnetic numbers are maximized at the expense of those with smaller $m$-s. Note that the optimal axis for each sphere (and each part of the density matrix it represents) is in general different from the optimal axis for the whole state. The quantum state is represented by a {\it collection} of spheres, each having in general different orientation of its optimal axis. Since any rotation applied to the state affects all spheres simultaneously their relative orientation is conserved. Hence it is impossible to align the $z$-axis with all optimal axes of each sphere at the same time. The only exceptions are when the optimal axes are already aligned, or when the density matrix of the state is spanned exclusively by tensors with the same angular momentum $j>0$ ($j=0$ never contributes to value QMI).

The ability to identify transformation $\hat R_{\mathrm{max}}$ which exposes the innate dynamical entanglement of state is of particular usefulness for interferometry. Knowing the transformation allows to design interferometric sequence for a given input state, which results in the maximal precision of the parameter estimation. According to Eq.~(\ref{eq:Dyn_Ent_QFI_v_susc_reciprocal}) which establishes the connection between the dynamical entanglement and the QFI we have
\begin{equation}
F_Q\big(\hat\varrho ,\hat U_\theta(\hat R_\mathrm{max}) \big) = \mathcal F(\hat R_{\mathrm{max}}\hat\varrho\hat R^\dagger_{\mathrm{max}})
\end{equation}
Here the interferometric transformation is given by
\begin{equation}
\hat U_\theta = \hat U_\alpha\hat U_n e^{-i\theta\hat J_z}\hat U_n^\dagger\hat U_\beta\,,
\end{equation}
and the operations $\hat U_\alpha$, $\hat U_\beta$ and $\hat U_n$ which precede and follow parameter imprint phase satisfy condition
\begin{equation}
\hat R_\mathrm{max} = \hat U_\beta^\dagger\hat U_\alpha\hat U_n\,.
\end{equation}

The formulation of the classical sphere model of the quantum state of bosonic qubits finalizes my inquiry of the physical interpretation of particle entanglement found in cold-atom systems.

\section[Simple example]{Simple example%
              \sectionmark{Simple example}}
\sectionmark{Simple example}

In this section I illustrate the ideas presented above with an example of a system composed of two bososnic qubits. This system is equivalent to system consisting one pseudo-particle with spin $J=N/2=1$. Here I will focus on the two-qubit variants of previously discussed states: twin Fock state, (see Eq.~(\ref{eq:TF})) and the cat state (see Eq.~(\ref{eq:Symmetric_Cat}))
\begin{align}
&|1\rangle_L|1\rangle_R= \mathcal S\left[|\tfrac 12,\tfrac 12\rangle|\tfrac 12,-\tfrac 12\rangle\right]=|1,0\rangle\equiv |\Psi_+\rangle\\
&|\mathrm{Cat}\rangle  = \frac{|\tfrac 12,\tfrac 12\rangle|\tfrac 12,\tfrac 12\rangle+|\tfrac 12,-\tfrac 12\rangle|\tfrac 12,-\tfrac 12\rangle}{\sqrt 2} =\frac{|1,1\rangle+|1,-1\rangle}{\sqrt 2}\equiv |\Phi_+\rangle
\end{align}
Recall that entanglement of a state $|\Psi_+\rangle$ is not detected by the dynamical susceptibility, because the $z$-axis is not the optimal axis (see Eq.~(\ref{eq:Dyn_ent_Indist_TF_MZI})). 
\begin{equation}
\mathcal F(|\Psi_+\rangle) = 4\langle\Delta^2\hat J_z\rangle_{\Psi_+} = 4\left[ \langle 1,0|\hat J_z^2|1,0\rangle -(\langle 1,0|\hat J_z|1,0\rangle)^2\right] = 0\,.
\end{equation}
In order to reveal the innate dynamical entanglement of the twin Fock, the reference frame has to be rotated, for example, about the $x$-axis by an angle $\pi/2$:
\begin{align}
&\mathcal F( e^{-i\frac \pi 2\hat J_x}|\Psi_+\rangle) = \nonumber\\
&=4\left[\langle 1,0|(e^{i\frac \pi 2\hat J_x}\hat J_z e^{-i\frac \pi 2\hat J_x})^2|1,0\rangle-\big( \langle 1,0|e^{i\frac \pi 2\hat J_x}\hat J_ze^{-i\frac \pi 2\hat J_x}|1,0\rangle\big)^2\right] \nonumber\\
&= 4\left[\langle 1,0|\hat J_y^2|1,0\rangle-(\langle 1,0|\hat J_y|1,0\rangle)^2\right] = 4 = N\big(\tfrac N2+1\big)\,,
\end{align}
where I used the fact that $e^{i\alpha\hat J_x}\hat J_ze^{-i\alpha\hat J_x}=\cos\alpha \hat J_z -\sin\alpha \hat J_y$. As for state $|\Phi_+\rangle$, it is the maximally entangled cat states, so we have (see Eq.~(\ref{eq:Cat_is_max_ent}))
\begin{equation}
\mathcal F(|\Phi_+\rangle) = 4=N^2\,.
\end{equation}

Now I proceed to reproduce the above results with help of concepts introduced in the thesis and without referring to the prior experience with $N>2$ variants of the states. The first step is to decompose the density matrices of our states of interest in the basis of spherical tensors. In case of two-qubit, i.e. spin-$1$ system, the spherical tensors of angular momentum $j=1$ are given by (see Eqs.~(\ref{eq:Sym_vs_Ent_def_10_tensor}) and (\ref{def_11_tensor}) ):
\begin{align}
& \hat T_{10} = \hat J_z = \left(\begin{array}{ccc}
 1&0&0\\
 0&0&0\\
 0&0&-1\\
\end{array}\right)\,,\\
& \hat T_{11} = -\frac 1{\sqrt 2} \hat J_+ =\left(\begin{array}{ccc}
 0&-1&0\\
 0&0&-1\\
 0&0&0\\
\end{array}\right) = -\hat T^\dagger_{1\,-1}\,.
\end{align}
 The set of spherical tensor operators with $j=2$ (the maximal $j$ of basis tensors for $J=N/2=1$ system, see Eq.~(\ref{eq:Sym_vs_Ent_tensor_decomp})) can be found by employing the Wigner-Eckhart theorem (\ref{eq:Phys_Inter_Symm_v_Ent_Wigner-Eckhart}) and treating it as a definition of matrix elements of $\hat T_{jm}$. The parameter $\langle J=1||T_{2}||J=1\rangle$ sets the norm of tensor operators (see Eq,~(\ref{eq:tensor_norm})) and can be chosen at will. By setting it to equal $\sqrt{2\times j+1}=\sqrt{5}$ we get $||\hat T_{2 m}||^2 = 1$. The explicit form of $\hat T_{jm}$ matrices are given by
\begin{align}
&\hat T_{20} = \frac 1{\sqrt 6}\left(\begin{array}{ccc}
 1&0&0\\
 0&-2&0\\
 0&0&1\\
\end{array}\right)\,,\\
&\hat T_{21} =\frac 1{\sqrt{2}}\left(\begin{array}{ccc}
 0&-1&0\\
 0&0&1\\
 0&0&0\\
\end{array}\right) = -\hat T^\dagger_{2\,-1}\,,\\
&\hat T_{22}=\left(\begin{array}{ccc}
 0&0&1\\
 0&0&0\\
 0&0&0\\
\end{array}\right) = \hat T^\dagger_{2\,-2}\,.
\end{align}
The decomposition in the basis is given by the standard formula for any linear space
\begin{equation}
\hat \varrho_{\Psi_+/\Phi_+} = \sum_{j=0}^{2}\sum_{m=-j}^j \frac{(\hat T_{jm}|\hat\varrho_{\Psi_+/\Phi_+})}{||\hat T_{jm}||}\frac{\hat T_{jm}}{||\hat T_{jm}||}=
\sum_{j=0}^2\sum_{m=-j}^j \frac{\langle \hat T_{jm}\rangle^*_{\Psi_+/\Phi_+}}{||\hat T_{jm}||}\frac{\hat T_{jm}}{||\hat T_{jm}||}\,. 
\end{equation}
with the scalar product defined as $(\hat A |\hat B)=\mathrm{Tr}(\hat A^\dagger\hat B)$ (see Eq.~(\ref{eq:Sym_vs_Ent_tensor_decomp})). The last equality comes from the direct relation between the scalar product and the expectation value 
\begin{equation}
(\hat T_{jm}|\hat\varrho)=\mathrm{Tr}(\hat T_{jm}^\dagger\hat\varrho)=\langle\hat T_{jm}\rangle_{\hat\varrho}^*\,.
\end{equation}
A straightforward calculation gives us the desired decompositions
\begin{align}
&\hat\varrho_{\Psi_+} = \frac 13\hat{\mathbf 1} -\sqrt{\frac  23}\hat T_{20}\,,\label{eq:tf_decomp}\\
&\hat \varrho_{\Phi_+} =\frac 13\hat {\mathbf 1} +\frac 1{\sqrt 6}\hat T_{20} + \frac 12\left(\hat T_{22}+\hat T_{2\,-2}\right)\,.\label{eq:cat_decomp}
\end{align}
Note that the unit matrix is spanned by $j=0$ tensor, $\hat{\mathbf 1} \propto \hat T_{00}$.

From Eqs.~(\ref{eq:tf_decomp}) and (\ref{eq:cat_decomp}) we see that the density matrices of the states are spanned exclusively by unity and a set of tensors with angular momentum $j=2$. Hence the states can be represented by a single sphere of radius two. This implies that the states have only one optimal axis, and it is possible to align it with the $z$-axis of the reference frame.

The decompositions give us a general idea about the distribution of the mass around the spheres representing the states. The state $|\Psi_+\rangle$ is spanned only by $m=0$ tensor, so the totality of its mass is located at the poles of the sphere. In case of $|\Phi_+\rangle$ state, a portion of its mass is also located the the poles. However, it is smaller amount then for $|\Psi_+\rangle$, which can be seen by comparing masses of parallels distanced $m=0$ units away from axis:
\begin{equation}
\frac{|\langle\hat T_{20}\rangle_{\Phi_+}|^2}{||\hat T_{20}||^2}=\frac{1}6<\frac{|\langle \hat T_{20}\rangle_{\Psi_+}|^2}{||\hat T_{20}||^2}=\frac 23
\end{equation}
(recall I set the norm of $j=2$ tensors to one). The rest of the mass of the sphere is located at the equator, which is evident from the presence of $\hat T_{2\pm2}$ tensors in the decomposition.

We can gain a better insight into detailed distribution of mass by examining transformation properties of the states. For example, the quantity $|\langle\hat T_{20}\rangle_{\Psi_+}|^2 /\allowbreak {||\hat T_{20}||^2}$ only tells us that certain amount of mass can be found at the poles of sphere representing state $|\Psi_+\rangle$. Just from this one number we cannot say whether this mass occupies only one of the poles or it split between them in some manner. However, when we examine the distribution from different point of view, i.e. we rotate the state, the details of the structure can become more apparent. In order to track the changes due to transformations more easily I adopt a following notation: the mass distribution on a sphere of radius $j$ is represented by a vector where each row is a mass accumulated at the allowed parallel (with the convention that negative magnetic number refers to parallel at the southern hemisphere, see footnote \ref{foot:NS_convention} just before Eq.~(\ref{eq:parallel_mass})):
\begin{equation}
\hat\varrho \leftrightarrow \left[\begin{array}{c}
\tfrac{|\langle \hat T_{jj}\rangle|^2}{||\hat T_{jj}||^2}\\[.25cm]
\tfrac{|\langle \hat T_{jj-1}\rangle|^2}{||\hat T_{jj-1}||^2}\\
\vdots\\
\tfrac{|\langle \hat T_{j\,-j}\rangle|^2}{||\hat T_{j\,-j}||^2}\\
\end{array}\right]^{j}_{\hat\varrho}\,.
\end{equation}
In this notation the states $\Psi_+$ and $\Phi_+$ are described by
\begin{equation}
\hat\varrho_{\Psi_+} \leftrightarrow \left[\begin{array}{c} 0\\0\\{\tfrac 23}\\0\\0\\\end{array}\right]^{j=2}_{\Psi_+}\quad,\quad
\hat\varrho_{\Phi_+} \leftrightarrow \left[\begin{array}{c} \tfrac 14\\0\\\tfrac 1{ 6}\\0\\\tfrac 14\\\end{array}\right]^{j=2}_{\Phi_+}
\end{equation}
Now I shall consider a sequence of two rotations about $x$-axis by $\pi/4$ angle applied to state $|\Psi_+\rangle$. Due to transformation $\hat\varrho_{\Psi_+} \to e^{-i\alpha\hat J_x}\hat\varrho_{\Psi_+}e^{i\alpha\hat J_x}$ the mass distribution changes according to Eq.~(\ref{eq:tensor_rotation}) which is reflected in my notation as
\begin{equation}\label{eq:Phys_Inter_Example_sequence}
\left[\begin{array}{c} 0\\0\\\tfrac 23\\0\\0\\\end{array}\right]^{2}_{\Psi_+}\!\!\!\!\!\!\xrightarrow{e^{-i\frac{\pi}{4}\hat J_x}}
\left[\begin{array}{c} \tfrac 1{16}\\[.1cm]\tfrac 14\\[.1cm]\tfrac 1{ 24 }\\[.1cm]\tfrac 14\\[.1cm]\tfrac 1{16}\\\end{array}\right]^{2}_{\Psi_+'}
\!\!\!\!\!\!\xrightarrow{e^{-i\frac{\pi}{4}\hat J_x}} \left[\begin{array}{c} \tfrac 14\\0\\\tfrac 1{ 6}\\0\\\tfrac 14\\\end{array}\right]^2_{\Psi_+''}\!\!\!\!\!\leftrightarrow \hat\varrho_{\Phi_+}
\end{equation}
\begin{figure}[h]
	\centering
	\includegraphics[clip,width=\textwidth]{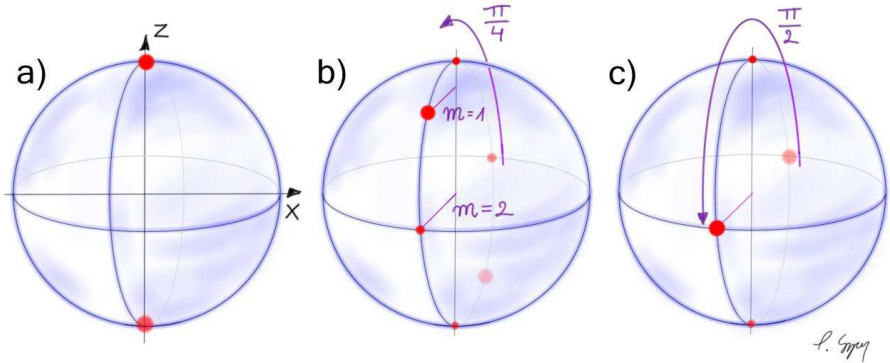}
	\caption{{\it Transformation of $|\Psi_+\rangle$ state as described in Eq.~(\ref{eq:Phys_Inter_Example_sequence}) illustrated on the massive sphere model: a) The initial state, b) Initial state is rotated through $\pi/4$ angle about $x$-axis. c) The state is rotated through additional $\pi/4$ angle which transforms it into $|\Phi_-\rangle$.}}\label{fig:rot_sequence}
\end{figure}
By tracing the sequence (\ref{eq:Phys_Inter_Example_sequence}), illustrated in Fig.~\ref{fig:rot_sequence}, we see that initially the mass of state $|\Psi_+\rangle$ must have been equally distributed among the poles of the sphere. If this was not the case, then the first rotation could not produce equally occupied northern and southern parallels distanced $m=1$ units away from the axis. Also, since the rotation is performed about the $x$-axis, the mass is kept during the whole sequence in the $z-y$ plane, instead of being, for example, ``smeared'' around allowed parallels.  Finally, we discover that the second transformation, which composes with the first one to rotation by $\pi/2$ angle, brought state $|\Psi_+\rangle$ into $|\Phi_+\rangle$. Thanks to this occurrence, we also gained an insight into details of mass distribution of the cat state $|\Phi_+\rangle$: it has an equal fractions of its mass placed on both of the poles and what is remaining is split evenly on equator, at the point where the $y$-axis intersects the surface of the sphere.

In the last step I shall utilize the decompositions (\ref{eq:tf_decomp}) and (\ref{eq:cat_decomp}) of the states to calculate quantum moment of inertia (see Eq.~(\ref{eq:QMI})), which in the case of pure states equals the dynamical susceptibility. I start with the cat state $|\Phi_+\rangle$:
\begin{equation}
\mathcal F(\hat \varrho_{\Phi_+}) = 4\mathcal I(\hat \varrho_{\Phi_+}) = 4\left( 2^2\times \frac{|\langle \hat T_{22}\rangle|^2}{||\hat T_{22}||^2}\right) = 4 =N^2\,.
\end{equation}
In case of $|\Psi_+\rangle$ state we immediately get
\begin{equation}
\mathcal F(\hat \varrho_{\Psi_+}) =4\mathcal I(\hat\varrho_{\Psi_+}) = 0\,,
\end{equation}
because it is spanned only by $m=0$ tensors which do not contribute to QMI. However, we learned form examining the sequence (\ref{eq:Phys_Inter_Example_sequence}) that state $|\Psi_+\rangle$ possesses large degree of innate dynamical entanglement: 
\begin{equation}
\max_{\hat R}\mathcal F(\hat R\hat\varrho_{\Psi_+}\hat R^\dagger) =\mathcal F(e^{-i\frac\pi 2\hat J_x}\hat\varrho_{\Psi_+}e^{i\frac\pi 2\hat J_x}) =\mathcal F(\hat \varrho_{\Phi_+}) =4 = N^2\,.
\end{equation}
Therefore, the $|\Psi_+\rangle$ state contains the same amount of innate dynamical entanglement as the cat state $|\Phi_+\rangle$. The fact that these states are equally entangled is unique feature of two-qubit systems. For higher numbers of particles twin Fock states are still significantly entangled, but to the lesser degree then the maximally entangled cat states.

\chapter{Conclusions}

The main purpose of the thesis was to better understand the non-classical correlations in ultra-cold atom systems. The main difficulty in achieving this goal lies in the general and discriminative definition of entanglement. One of the solutions to this problem, which was presented in the thesis, is to examine the entanglement form the point of view of its usefulness for certain tasks, like precise metrology. The efficiency at which this task is performed, when the quantum state is used as a resource, can serve as a measure of the degree of the entanglement.

The atomic interferometry, a branch of quantum metrology, exploits the wave nature of matter and the particle entanglement to attain precise measurements of the phase beyond capabilities of classical devices. It turn out that ultra-cold atom systems are most well suited for this type of task. 

In modern experimental realizations of such interferometers the Bose-Einstein condensate (BEC) used as a probe can be described as a collection of qubits. Due to their bosonic nature, the state of the BEC can be described as a state of a single pseudo-particle with the spin equal to the half of its particle number $N$. It follows that any (unitary) interferometric transformation is mapped onto a sequence of rotations of the spin, with one of the angles being the unknown phase. Moreover, it is important to note that all possible unitary transformations of a single qubit are equivalent to rotation, and it follows that any coherent transformation of the collection of qubits (i.e. the transformation which acts on each qubit in the same way) is also mapped onto global rotation of the system. Since the indistinguishable qubits cannot be addressed individually, coherent rotations constitute the entirety of possible local operations -- transformations which do not introduce or destroy the entanglement between particles.

The efficiency of the interferometer is tied to the precision of the phase estimation. According to the Cram\'{e}r-Rao theorem, which is known from classical theory and was later supplemented by Braunstein and Caves with quantum considerations, the precision is bounded by the quantum Fisher information (QFI). The QFI quantifies the susceptibility of the state to change induced by a given transformation, which in this case is the interferometric sequence. 

When the transformation is coherent, the QFI cannot surpass the threshold of the shot-noise limit unless the state is entangled. Therefore, for a given interferometric sequence, the QFI can serve as a criterion for the entanglement that is useful for enhancing the precision of this particular interferometer. For another choice of the interferometer, the QFI would deem different set of states as usefully entangled. 

On one hand, the freedom of choice of the transformation is essential for embedding the question of entanglement nature in the context of a real experiment. On the other hand, this ambiguity might obscure some of the important structures and relations which could help in understanding the non-classical correlations, which is the main aim of this thesis. This problem was solved by introducing the dynamical susceptibility, which is defined as the QFI for ``standard'' interferometer and is a function of the state alone. The dynamical susceptibility allowed to show that states which are usefully entangled for one interferometric sequence are related to states useful for different interferometer through coherent transformation, which conserves the degree of particle entanglement.

By exploiting the equivalence between bosonic qubits and the spin system, it has been found how particle indistinguishability can enhance the overall degree of the entanglement. The susceptibility to change due to coherent transformation (quantified by the dynamical susceptibility) of the spin $N/2$ system is in general greater than the susceptibility of the ensemble of $N$ individuals with spins $1/2$. If the qubits constituting the state were distinguishable, the spins of the individuals could add-up to a whole range of total angular momenta, thus diminishing the state's susceptibility. For identical qubits, the only possibility is to add the individual spins to the maximal total spin of $N/2$ -- allowing for potentially maximal degree of entanglement.

The entanglement criteria based on the QFI, including the dynamical susceptibility, establish a direct correspondence between the particle entanglement and the susceptibility of the state to rotations. In classical physics such susceptibility is measured by the moment of inertia of the system, which is given by the distribution of the mass in respect to the axis of rotation. It was shown that the dynamical susceptibility is indeed closely related to the quantum analog of the classical moment of inertia. In the quantum case the axis of rotation is given by the quantization axis and the classical mass is replaced by the components of the density matrix of the state decomposed in the basis of spherical tensor operators. 

The final result of the thesis is the formulation of the classical model of the quantum state and its entanglement. Within this model the state of $N$ bosonic qubits is represented by nested massive spheres of increasing radii. The quantum nature of the system is manifested by enforcing the mass of each sphere to only occupy parallels located at quantized distances from the common rotation axis. The model embraces the analogy between the dynamical susceptibility and the classical moment of inertia of a rigid body and identifies the degree of correlations between qubits with the distribution of the fictional mass on the surfaces of the spheres. The most entangled states are represented by the system with the majority of its mass located at the equator of the largest sphere so that the moment of inertia is the greatest.

I hope that the existence of the model such as this can convince the Reader that at least certain type of particle entanglement can be understood in terms of concrete physical properties of the system. That the difference between the classically-correlated state and the entangled one is only quantitative not qualitative. There is no sharp border between the two cases: separable states are characterized by a lesser ``intensity'' of some special physical property then the states that are considered as entangled. I believe that it is only natural to expect such smeared border rather then the sharp one. After all, in the quantum world there are no ``classical correlations'' because everything is just ``quantum''. Hence, the distinction between separable and entangled state is in most cases artificial, and so no spectacular physically effects should accompany the passing of the border.

\newpage



\newpage
\pagestyle{empty}
\vspace*{14cm}
\noindent
{This work was supported by the Foundation for Polish Science International PhD Projects Programme co-financed by the EU European Regional Development Fund.}\\[.5cm]
\includegraphics[width=\textwidth]{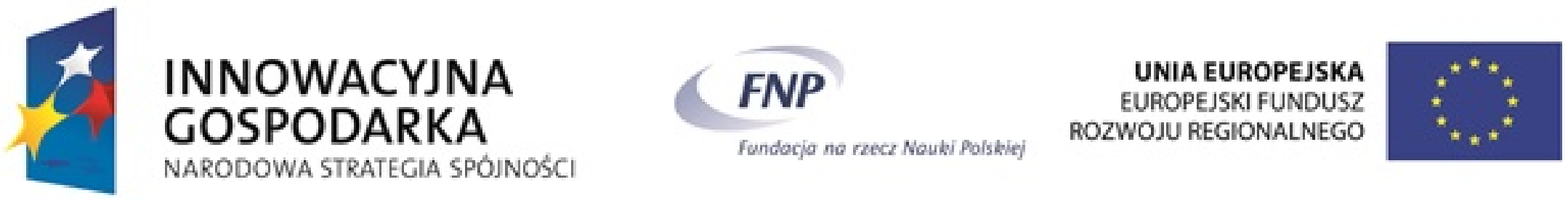}\\[-.5cm]
\begin{center}
\includegraphics[width=.2\textwidth]{uw3.eps}\hspace{.5cm}
\includegraphics[width=.3\textwidth]{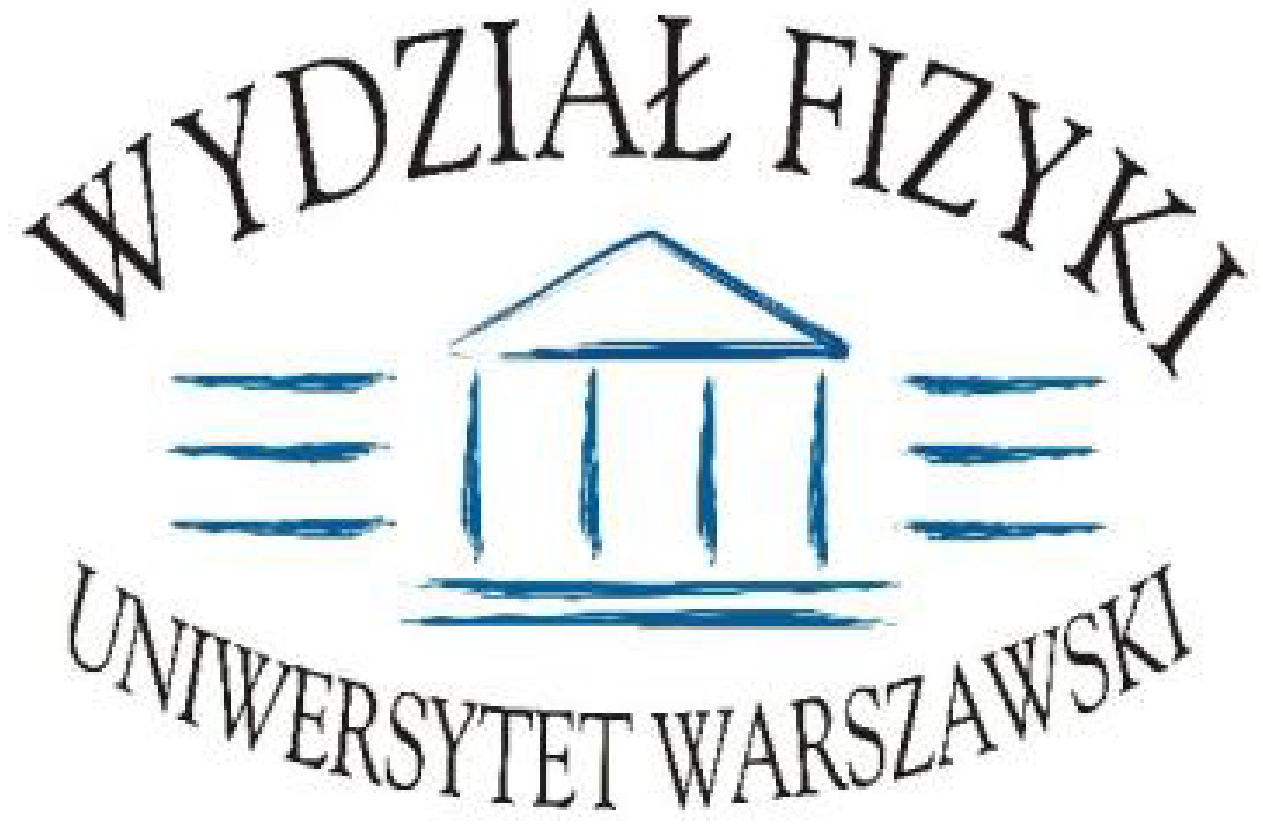}
\end{center}


\addcontentsline{toc}{chapter}{Bibliography}
\begin{thebibliography}{99}
\bibitem{Young_double_slit_88}
	A. Zeilinger, G. Roland, C.G. Shull, W. Treimer and W. Mampe, 
	Rev. Mod. Phys. {\bf 60}, 1067 (1988)
\bibitem{Young_double_slit_91}
	O.Carnal and J. Mlynek, 
	Phys. Rev. Lett. {\bf 66}, 2689 (1991)
\bibitem{Einstein_05}
	 A. Einstein, 
	 Annalen der Physik {\bf 17}, 132 (1905)
 \bibitem{EPR_35}
	 A. Einstein, B. Podolsky and N. Rosen,
	Phys. Rev. {\bf 47}, 777 (1935)
\bibitem{Schroedinger_35}
	E. Schr\"{o}dinger,
	Mathematical Proceedings of the Cambridge Philosophical Society {\bf 31}, 555 (1935)
\bibitem{Bell_64}
	J. Bell,
	Physics {\bf 1}, 195 (1964)
\bibitem{Aspect_81}
	A. Aspect, P. Grangier and G. Roger,
	Phys. Rev. Lett. {\bf 47}, 460 (1981)
\bibitem{Aspect_82}
	A. Aspect, P. Grangier and G. Roger,
	Phys. Rev. Lett. {\bf 49}, 91 (1982)
\bibitem{Aspect_82_2}
	A. Aspect, J. Dalibard and G. Roger, 
	Phys. Rev. Lett. {\bf 49}, 1804 (1982)
\bibitem{CSI}
	T. Wasak, P. Sza\'{n}kowski, P. Zi\'{n}, M. Trippenbach and J. Chwede\'{n}czuk,
	Phys. Rev. A {\bf 90}, 033616
\bibitem{Sudarshan_63}
	E. C. G. Sudarshan,
	Phys. Rev. Lett. {\bf 10}, 277 (1963)
\bibitem{Glauber_63}
	R. J. Glauber,
	Phys. Rev. {\bf 131}, 2766 (1963)
\bibitem{Terhal_00}
	B. M. Terhal,
	Physics Letters A {\bf 271}, 319 (2000)
\bibitem{Toth_09}
	O. Guhne, G. Toth,
	Phys. Rep. {\bf 474}, 1 (2009)
\bibitem{Ueda_93}
	M. Kitagawa and M. Ueda, 
	Phys. Rev. A {\bf 47}, 5138 (1993)
\bibitem{Sorensen_01}
	A. S\o{}rensen, L.-M. Duan, J. I. Cirac and P. Zoller, Nature {\bf 409}, 63 (2001)
\bibitem{Toth_07}
	G. Toth, C. Knapp, O. G\"{u}hne, and H. J. Briegel, Phys. Rev. Lett. {\bf 99}, 250405 (2007)
\bibitem{Wang_03}
	X. Wang and B. C. Sanders, Phys. Rev. A {\bf 68}, 012101 (2003)
\bibitem{Schmiedmayer_RevModPhys_2009}
	A. D. Cronin, J. Schmiedmayer and D. E. Pritchard,
	Rev. Mod. Phys. {\bf 81}, 1051 (2009)
\bibitem{Schmiedmayer_95}
	C. Ekstrom, J. Schmiedmayer, M. Chapman, T. Hammond and D. Pritchard,
	Phys. Rev. A. {\bf 51}, 3883 (1995)
\bibitem{Hoff_08}
	S. Hofferberth, I. Lesanovsky, T. Schumm, A. Imambekov, V. Gritsev, E. Demler and J. Schmiedmayer,  
	Nat. Phys. {\bf 4}, 489 (2008).
\bibitem{Gring_12}
	M. Gring, M. Kuhnert, T. Langen, T. Kitagawa, B. Rauer, M. Schreitl, I. Mazets, D. Adu Smith, E. Demler and J. Schmiedmayer,
	Science {\bf 337}, 1318 (2012)
\bibitem{Chu_93}
	D. Weiss, B. Young and S. Chu,
	Phys. Rev. Lett. {\bf 70}, 2706 (1993)
\bibitem{Clauser_88}
	J.  F. Clauser, 
	Physica B {\bf 151}, 262 (1988)
\bibitem{Chu_99}
	A. Peters, K. Y. Chung and S. Chu,
	Nature {\bf 400}, 849 (1999)
\bibitem{Gustavson_00}
	T. L. Gustavson, A. Landragin and M. A. Kasevich,
	Classical and Quantum Gravity {\bf 17}, 2385 (2000)
\bibitem{Perrin_12}
	A. Perrin, R. B\"{u}cker, S. Manz, T. Betz, C. Koller, T. Plisson, T. Schumm and J. Schmiedmayer, 
	Nat. Phys. {\bf  8}, 195 (2012)
\bibitem{McDonald_13}
	G. D. McDonald, H. Keal, P. A. Altin, J. E. Debs, S. Bennetts, C. C. N. Kuhn, K. S. Hardman, M. T. Johnsson, J. D. Close, and N. P. Robins
	Phys. Rev. A. {\bf 87}, 013632 (2013)
\bibitem{micrograv_13}
	H. M\"{u}ntinga, M. Saba, T. A. Pasquini, W. Ketterle, D. E. Pritchard, and A. E. Leanhardt, 
	Phys. Rev. Lett. {\bf 110}, 093602 (2013)
\bibitem{Shin_04}
	Y. Shin, M. Saba, T. A. Pasquini, W. Ketterle, D. E. Pritchard, and A. E. Leanhardt, 
	Phys. Rev. Lett. {\bf 92}, 050405 (2004)
\bibitem{Schumm_05}
	T. Schumm, S. Hofferberth, L. M. Andersson, S. Wildermuth, S. Groth, I. Bar-Joseph, J. Schmiedmayer and P. Kr\"{u}ger, 
	Nat. Phys. {\bf 1}, 57–62 (2005)
\bibitem{Baumg_10}
	F. Baumg\"{a}rtner, R. J. Sewell, S. Eriksson, I. Llorente-Garcia, Jos Dingjan, J. P. Cotter, and E. A. Hinds,
	Phys. Rev. Lett. {\bf 105}, 243003 (2010)
\bibitem{RMP_feshbach_10}
	C. Chin, R. Grimm, P. Julienne and E. Tiesinga, 
	Rev. Mod. Phys. {\bf 82}, 1225 (2010)
\bibitem{Oberthaler_08}
	J. Est\'{e}ve, C. Gross, A. Weller, S. Giovanazzi and M. K. Oberthaler,
	Nature {\bf 455}, 1216 (2008)
\bibitem{Schmiedmayer_13}
	T. Berrada, S. van Frank, R. B\"{u}cker, T. Schumm, J.-F. Schaff and J. Schmiedmayer,
	Nat. Comm. {\bf 4}, 2077 (2013)
\bibitem{Smerzi_exp_11}
	B. L\"{u}cke, M. Scherer, J. Kruse, L. Pezz\'{e}, F. Deuretzbacher, P. Hyllus, O. Topic, J. Peise, W. Ertmer, J. Arlt, L. Santos, A. Smerzi and C. Klempt, 
	Science {\bf 334}, 773 (2011)
\bibitem{Bucker_11}
	R. B\"ucker, J. Grond, S. Manz, T. Berrada, T. Betz, C. Koller, U. Hohenester, T. Schumm, A. Perrin and J. Schmiedmayer, 
	Nat. Phys. {\bf 7}, 608 (2011)
\bibitem{Wasak_twin_beam_14}
	T. Wasak, P. Sza\'{n}kowski, R. B\"{u}cker, J. Chwede\'{n}czuk and M. Trippenbach, 
	New J. Phys. {\bf 16}, 013041 (2014)
\bibitem{Caves_QFI}
	S. L. Braunstein and C. M. Caves, 
	Phys. Rev. Lett. {\bf 72}, 3439 (1994)
\bibitem{Wootters_81}
	W. K. Wootters, 
	Phys. Rev. D {\bf 23}, 357 (1981)
\bibitem{Bures_69}
	D. J. C. Bures, 
	Trans. Am. Math. Soc. {\bf 135}, 199 (1969)
\bibitem{Uhlmann_76}
	A. Uhlmann, 
	Rep. Math. Phys. {\bf 9}, 273 (1976)
\bibitem{Hubner_92}
	M. H\"{u}bner, 
	Phys. Lett. A {\bf 168}, 239 (1992).
\bibitem{Cramer_book}
	H. Cram\'{e}r, 
	{\it Mathematical Methods of Statistics}
	(Princeton University, Princeton, NJ, 1946)
\bibitem{Fisher_1925}
	R. A. Fisher, 
	Proc. Camb. Soc. {\bf 22}, 700 (1925)
\bibitem{Braunstein_92}
	S. L. Braunstein, 
	J. Phys. A {\bf 25}, 3813 (1992)
\bibitem{Braunstein_92_2}
	S. L. Braunstein, 
	Phys. Rev. Lett. {\bf 69}, 3598 (1992)
\bibitem{Smerzi_QFI}
	L. Pezz\'e and A. Smerzi, 
	Phys. Rev. Lett. {\bf 102}, 100401 (2009)
\bibitem{Plenio_ent_measures} 
	M. B. Plenio and S. Virmani, 
	Quant. Inf. Comput. {\bf 7}, 1 (2007)
\bibitem{Smerzi_measured_QFI}
	H. Strobel, W. Muessel, D. Linnemann, T. Zibold, D. B. Hume, L. Pezz\'{e}, A. Smerzi, M. K. Oberthaler, 
	Science {\bf 345}, 424 (2014)
\bibitem{N-qubit_noise}
	P. Sza\'{n}kowski, M. Trippenbach, J. Chwede\'{n}czuk, 
	arXiv:1212.2528 (2014)
\bibitem{2-qubits_noise}
	P. Sza\'{n}kowski, M. Trippenbach, \L{}. Cywi\'{n}ski, Y. B. Band, 
	arXiv:1408.4254 (2014)
\bibitem{sakurai}
	J. J. Sakurai, 
	{\it Modern Quantum Mechanics} 
	(Addison-Wesley, 1994)
\bibitem{Toth_Petz_2013}
	G. Toth and D. Petz, 
	Phys. Rev. A {\bf 87}, 032324 (2013)
\bibitem{QFI_is_a_roof}
	S. Yu,
	arXiv:1302.5311 (2013)

\end{thebibliography}
\end{document}